\documentclass[a4paper,10pt]{article}

\usepackage{latexsym}
\usepackage{graphics}
\usepackage{graphicx}
\usepackage{subfigure}
\usepackage{algorithmic}
\usepackage{algorithm}
\usepackage{longtable}
\usepackage{lscape}

\usepackage{rotating}
\usepackage{pst-tree}
\usepackage{pst-node}
\usepackage{pst-grad} 
\usepackage{pst-plot}
\usepackage{pst-3d}
\usepackage{pstricks}
\psset{arrows=->,levelsep=5ex,nodesep=3pt,arrowsize=4pt}

\usepackage{amsmath}
\usepackage{amsthm}

\theoremstyle{remark}
\newtheorem*{example}{Example}

\theoremstyle{plain}

\newtheorem{lemma}{Lemma}         
\newtheorem{cor}{Corollary}
\newtheorem{property}{Property}

\newcommand{\qleft}{Q_{\text{left}}}
\newcommand{\qright}{Q_{\text{right}}}
\newcommand{\pleft}{P_{\text{left}}}
\newcommand{\hleft}{H_{\text{left}}}

\newcommand{\sigmaleft}{\Sigma_{\text{left}}}

\newcommand{\proj}{\boldsymbol\pi}
\newcommand{\sel}{\boldsymbol\sigma}

\newcommand{\cantab}[2]{{\it CanTab}_{#1,#2}}
\newcommand{\freqtab}[2]{{\it FreqTab}_{#1,#2}}
\newcommand{\freqtabPattern}[1]{{\it FreqTab}_{#1}}

\newcommand{\Pinst}[1]{P_{#1}^{\alpha_{#1}}}
\newcommand{\PinstPrime}[1]{{P'}_{#1}^{{\alpha'}_{#1}}}
\newcommand{\Qinst}[1]{Q_{#1}^{\alpha_{#1}}}
\newcommand{\Freq}[0]{\mathit{Freq}}
\newcommand{\Conf}[0]{\mathit{Conf}}

\title{Mining Tree-Query Associations in Graphs}
\date{}
\author{Eveline Hoekx and Jan Van den Bussche\\ Hasselt University and transnational University of Limburg\\ Agoralaan D, 3590 Diepenbeek, Belgium}

\begin{document}

\maketitle

\begin{abstract}
New applications of data mining, such as in biology, bioinformatics, or sociology, are faced with large datasets structured as graphs. We introduce a novel class of tree-shaped patterns called tree queries, and present algorithms for mining tree queries and tree-query associations in a large data graph.  Novel about our class of patterns is that they can contain constants, and can contain existential nodes which are not counted when determining the number of occurrences of the pattern in the data graph.  Our algorithms have a number of provable optimality properties, which are based on the theory of conjunctive database queries.  We propose a practical, database-oriented implementation in SQL, and show that the approach works in practice through experiments on data about food webs, protein interactions, and citation analysis.
\end{abstract}

\section{Introduction}
The problem of mining patterns in graph-structured data has received
considerable attention in recent years, as it has many interesting
applications in such diverse areas as biology, the life sciences,
the World Wide Web, or social sciences. In the present work we introduce a novel class of patterns, called tree queries, and we present algorithms for mining these tree queries and tree-query associations in a large data graph. This article is based on two earlier conference papers \cite{tqmKDD2005, tqmICDM2006}.

Tree queries are powerful tree-shaped patterns, inspired by
conjunctive data\-base queries \cite{gettingwarmer}.  In comparison to the kinds of patterns used in most other
graph mining approaches, tree queries have some extra features:

\begin{itemize}
\item

Patterns may have ``existential'' nodes: any occurrence of the pattern
must have a copy of such a node, but existential nodes
are not counted when determining the number of occurrences.

\item

Moreover, patterns may have ``parameterized'' nodes, labeled by constants, which
must map to fixed designated nodes of the data graph.  

\item

An ``occurrence'' of the pattern in a data graph $G$ is defined as any homomorphism from
the pattern in $G$.  When counting the number of occurrences, two occurrences that differ only on existential nodes are identified.

\end{itemize}

Past work in graph mining has dealt with node labels, but only with non-unique ones: such labels are easily simulated
by constants, but the converse is not obvious. It is also possible to simulate edge labels using constants. To simulate a node label $a$, add a
special node $a$, and express that node $x$ has label $a$ by drawing an edge
from $x$ to $a$.  For an edge $x\to y$ labeled $b$, introduce an intermediate node $x.y$ with $x\to
x.y \to y$, and label node $x.y$ by $b$.

\begin{figure}
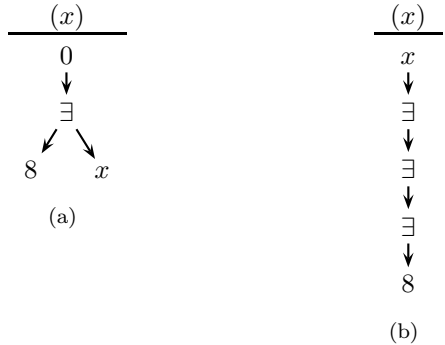

\hspace*{\fill}
\subfigure[]
{
\label{figpattern}
\begin{tabular}[b]{c}
$(x)$\\
\hline
\pstree{\TR{$0$}}
{
	\pstree{\TR{$\exists$}}
	{
		\TR{8}
		\TR{$x$}
	}
}
\end{tabular}
}
\hfill
\subfigure[]
{\label{figpattern2}
\begin{tabular}[b]{c}
$(x)$\\
\hline	
\pstree{\TR{$x$}}
	{
		\pstree{\TR{$\exists$}}
		{
			\pstree{\TR{$\exists$}}
			{
				\pstree{\TR{$\exists$}}
				{
					\TR{$8$}
				}
			}
		}
	}
\end{tabular}
}
\hspace*{\fill}
\label{figpatterns}
\caption{Simple examples of tree-query patterns}
\end{figure}

A simple example of a tree query is shown in Figure~\ref{figpattern}; when
applied to a food web: a data graph of organisms, where there is an edge $x \rightarrow y$ if $y$ feeds on $x$, it describes all organisms $x$ that compete with organism~\#8 for some organism as food, that itself feeds on
organism~\#0.  This pattern has one existential node, two parameters, and
one distinguished node $x$.
Figure~\ref{figpattern2} shows another example of a tree query; when applied to a food web, it describes all organisms $x$ that have a path of length four beneath them that ends in organism~\#8.

\subfigcapskip=10pt

\begin{figure}
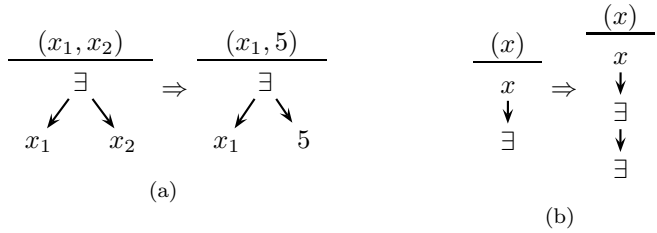

\hspace*{\fill}
\subfigure[]{\label{intro_1}
\begin{tabular}{c}
$(x_1,x_2)$\\
\hline
\pstree{\TR{$\exists$}}
{
	\TR{$x_1$}
	\TR{$x_2$}  
}   
\end{tabular}
\hfill
$\Rightarrow$
\hfill
\begin{tabular}{c}
$(x_1,5)$\\
\hline
\pstree{\TR{$\exists$}}
{
	\TR{$x_1$}
	\TR{$5$}  
}   
\end{tabular}
}
\hfill
\subfigure[]{\label{intro_2}
\begin{tabular}{c}
$(x)$\\
\hline
\pstree{\TR{$x$}}
{
	\TR{$\exists$} 
}   
\end{tabular}
\hfill
$\Rightarrow$
\hfill
\begin{tabular}{c}
$(x)$\\
\hline
\pstree{\TR{$x$}}
{
	\pstree{\TR{$\exists$}}
	{
		\TR{$\exists$}
	}  
}   
\end{tabular}
}
\hspace*{\fill}
\caption{Simple examples of association rules over tree queries.}
\label{fig_intro}
\end{figure}

Effectively, tree queries are what is known in database research as
\emph{conjunctive queries} \cite{cm,ullmanII,ahv_book}; these are the queries we could
pose to the data graph (stored as a two-column table) in the core fragment of SQL
where we do not use aggregates or subqueries, and use only conjunctions of
equality comparisons as where-conditions.  For example, the pattern of
Figure~\ref{figpattern} amounts to the following SQL query on a table
\texttt{G(from,to)}:

\begin{verbatim}
select distinct G3.to as x
from G G1, G G2, G G3
where G1.from=0 and G1.to=G2.from
  and G2.to=8 and G3.from=G2.from
\end{verbatim}

In the present work we also introduce association rules over tree queries. By mining for tree-query associations we can discover quite subtle properties of the data graph.  Figure~\ref{intro_1} shows a very simple example of an
association that our algorithm might find in a social network: a data graph of
persons where there is an edge $x \to y$ if $x$ considers $y$ to be a close
friend.  The tree query on the left matches all pairs $(x_1,x_2)$ of
``co-friends'': persons that are friends of a common person (represented by an
existential variable).  The query on the right matches all co-friends $x_1$ of
person~\#5 (represented by a parameterized node), and pairs all those
co-friends to person~\#5.  Now were the association from the left to the right to be
discovered with a confidence of $c$, with $0 \leq c \leq 1$, then this would
mean that the pairs retrieved by the right query actually constitute a
fraction of $c$ of all pairs retrieved by the left query, which indicates (for
nonnegligible $c$) that
5 plays a special role in the network.\footnote{Note that this does not just
  mean that 5 has many co-friends; if we only wanted to express that, just a
  frequent pattern in the form of the right query would suffice.  For
  instance, imagine a data graph consisting of $n$ disjoint 2-cliques (pairs of
  persons who have each other as a friend), where additionally all these
  persons also consider 5 to be an extra friend (but not vice versa).  In such
  a data graph,
5 is a co-friend of everybody, and the association has a
  rather high confidence of more than $2/7$.  If, however, we would now add to
  the data graph a separate $n$-clique, then still 2/3rds of all persons are a
  co-friend of 5, which is still a lot, but the confidence drops to below
  $2/n$.}
  
Figure~\ref{intro_2} shows quite a different, but again simple, example
of a tree-query association that our algorithm might discover in a food web.
With confidence $c$, this association means that of all organisms that are not
on top of the food chain (i.e., they are fed upon by some other organism),
a fraction of $c$ is actually at least two down in the food chain.  

The examples of tree queries and associations we just saw are didactical examples, but
in Section~\ref{sec_exp} we will see more complicated examples of tree queries and associations
mined in real-life datasets.

In this paper we present algorithms for mining tree queries and associations rules over tree queries in a large data graph. Some important features of these algorithms are the following:

\begin{enumerate}

\item
Our algorithms belong to the group of graph mining algorithms where the input is a single large graph, and the task is to discover patterns that occur sufficiently often in the single data graph. We will refer to this group of algorithms as the single graph category. There is also a second category of graph mining algorithms, called the transactional category, which is explained in Section~\ref{relwork}.

\item 
We restrict to patterns that are trees, such as the example in Figure~\ref{figpatterns}.  Tree patterns have formed an important special case in the transactional category (Section~\ref{relwork}), but have not yet received special attention in the single-graph literature.  Note that the data graph that is being mined is not restricted in any way.

\item

The tree-query-mining algorithm is incremental in the number of nodes of the pattern. So, our algorithm systematically considers ever larger trees, and can be stopped any time it has run long enough or has produced
enough results.  Our algorithm does not need any space beyond what is
needed to store the mining results. Thanks to the restriction to tree shapes the duplicate-free generation of trees can be done efficiently.


\item

For each tree, all conjunctive queries based on that tree are generated in the tree-query-mining algorithm.
Here, we work in a levelwise fashion in the sense of Mannila and Toivonen \cite{levelwise}.  

\item
As in classical association rules over itemsets \cite{apriori}, our
association rule generation phase comes after the generation of frequent
patterns and does not require access to the original dataset.  


\item
We apply the theory of conjunctive database queries
\cite{cm,ullmanII,ahv_book} to formally define and to correctly generate association rules 
over tree queries.  The conjunctive-query approach to pattern matching allows
for an efficiently checkable notion of frequency, whereas in the 
subgraph-based approach, determining whether a pattern is frequent is NP-complete
(in that approach the frequency of a pattern is the maximal
number of disjoint subgraphs isomorphic to the pattern
\cite{kukajourn2005}).

\item
There is a notion of equivalence among tree queries and association rules over tree queries. We carefully and efficiently avoid the generation of equivalent tree queries and associations, by using and adapting what is known from the theory of conjunctive database queries. Due to the restriction to tree shapes, equivalence and redundancy (which are normally NP-complete) are efficiently checkable.

\item 
Last but not least, our algorithms naturally suggest a database-oriented
implementation in SQL\@.  This is useful for several reasons.  First, the
number of discovered patterns can be quite large, and it is important to keep
them available in a persistent and structured manner, so that they can be
browsed easily, and so that association rules can be derived
efficiently.  Moreover, we will show how the use of SQL allows us to generate
and check large numbers of similar patterns in parallel, taking advantage of
the query processing optimizations provided by modern relational database
systems.  Third, a database-oriented implementation does not require us to
move the dataset out of the database before it can be mined.  In classical
itemset mining, database-oriented implementations have received serious
attention \cite{queryflocks,shibythomas}, but less so in graph mining, a
recent exception being an implementation in SQL of the seminal SUBDUE
algorithm \cite{dbsubdue}.
\end{enumerate}

The purpose of this paper is to introduce tree queries and tree-query associations and to present algorithms for mining tree queries and tree-query associations. Concrete applications to discover new knowledge about scientific datasets are the topic of current research. Yet, the algorithms are fully implemented and we can already show that our approach works in practice, by showing some concrete results mined from a food web, a protein interactions graph, and a citation graph. We will also give performance results on random data graphs (as a worst-case scenario).

\section{Related Work}
\label{relwork}

Approaches to graph mining, especially mining for frequent patterns or association rules, can be divided in two major categories which are not to be
confused.  
\begin{enumerate}
\item
In transactional graph mining, e.g., \cite{warmer,ramon_outerplanar,hwp_ffsm,iwm_agm_journ,kuka_fsg_journ,yaha_gspan, zaki_journ_forest}, the dataset
consists of many small data graphs which we call transactions, and the task is to discover patterns that occur at least once in a sufficient number of
transactions.  (Approaches from machine learning or inductive logic
programming usually call the small data graphs ``examples'' instead of
transactions.) 
\item
In single-graph mining the dataset is a single large data graph, and the task is to discover patterns that occur
sufficiently often in the dataset.  
\end{enumerate}

Note that single-graph mining is more difficult than transactional mining, in the sense that transactional graph mining can be simulated by single-graph mining, but the converse is not obvious. 

Since our approach falls squarely within the single-graph category, we will focus on that category in this section. Most work in this category has been done on frequent pattern mining, and less attention has been spend on association rules. We briefly review the work in this category next:

\begin{itemize}
	
	\item 
	
	Cook and Holder \cite{subdue} apply in their SUBDUE system the minimum description length (MDL) principle to discover substructures in a labeled data graph. The MDL principle states that the best pattern, is that pattern that minimizes the description length of the complete data graph. Hence, in SUBDUE a pattern is evaluated on how well it can compress the entire dataset. The input for the SUBDUE system is a labeled data graph; nodes and edges are labeled with non-unique labels. This is in contrast with the unique labels (`constants') in our system. But as we already noted, non-unique node labels and edge-labels can easily be simulated by constants, but the converse is not obvious. The SUBDUE system only mines patterns, no association rules. 	

	\item
	
	Ghazizadeh and Chawathe \cite{ghch_seus} mine in their SEuS system for connected subgraphs in a labeled, directed data graph, as in the SUBDUE system.  Instead of generating candidate patterns using the input data graph, SEuS uses a summary of the data graph. This summary gives an upper bound for the support of the patterns, and the user can then select those patterns of which he wants to know the exact support. SEuS also only mines for frequent patterns and not for associations.
	
	\item
	
	Vanetik, Gudes, and Shimony \cite{vgs_freqpat_journ} propose an Apriori-like \cite{apriori} algorithm for mining subgraphs from a labeled data graph.  The support of a graph pattern is defined as the maximal number of edge-disjoint instances of the pattern in the data graph. By reducing the support counting problem to the maximal independent set problem on graphs, they show that in worst case, computing the support of a graph pattern is NP-hard. They propose an Apriori-like algorithm to minimize the number of patterns for which the support needs to be computed. The major idea of their approach is using edge-disjoint paths as building blocks instead of items in classical itemset mining. Vanetik, Gudes, and Shimony also only mine for frequent patterns in the data graph.	
	
	\item
	
	Kuramochi en Karypis \cite{kukajourn2005} use the same support measure for graph patterns as Vanetik, Gudes and Shimony \cite{vgs_freqpat_journ}. They also note that computing the support of a graph pattern is NP--hard in worst case, since it can be reduced to finding the maximum independent set (MIS) in a graph. Kuramochi and Karypis quickly compute the support of a graph pattern using approximate MIS-algorithms. The number of candidate patterns is restricted using canonical labeling. As the majority of algorithms, Kuramochi and Karypis only mine for frequent patterns. 
		
	\item
	
	Jeh and Widom \cite{jw_graphprop} consider patterns that are, like our tree queries, inspired by conjunctive database queries, and they also emphasize the tree-shaped case. A severe restriction, however, is that their patterns can be matched by single nodes only, rather than by tuples of nodes. Still their work is interesting in that it presents a rather nonstandard approach to graph mining, quite different from the standard incremental, levelwise approach, and in that it incorporates ranking. Jeh and Widom mention association rules as an example of an application of their mining framework.

\end{itemize}

The related work that was most influential for us is Warmr
\cite{warmer,AR_warmr}, although it belongs to the transactional category. 
Based on inductive logic programming, patterns in Warmr also feature
existential variables and parameters.  While not restricted to tree shapes,
the queries in Warmr are restricted in another sense so that only
transactional mining can be supported.
Association rules in Warmr are defined
in a naive manner through pattern extension, rather than being founded upon
the theory of conjunctive query containment.  The Warmr system is also
Prolog-oriented, rather than database-oriented, which we believe is
fundamental to mining of single large data graphs, and which allows a more uniform
and parallel treatment of parameter instantiations, as we will show in this
paper.  Finally, Warmr does not seriously attempt to avoid the generation of
duplicates.  Yet, Warmr remains a
pathbreaking work, which did not receive sufficient follow-up in the data
mining community at large.  We hope our present work represents an improvement
in this respect.  Many of the improvements we make to Warmr were already
envisaged (but without concrete algorithms) in 2002 by Goethals and the second
author \cite{gettingwarmer}.

Finally, we note that parameterized conjunctive
database queries have been used in data mining quite early,
e.g., \cite{queryflocks,zan_metamining}, but then in the setting of ``data
mining query languages'', where a \emph{single}
such query serves to specify a family of patterns to be mined or queried
for, rather than the mining for such queries themselves, let alone
associations among them.

\section{Problem Statement}

In this section we define some concepts formally. In the appendix an overview of all notations used in this paper is given.

We basically assume a set $U$ of \emph{data constants} from which the nodes of the data graph to be mined will be
taken.

\subsection{Graph-theoretic concepts}

Let $N \subseteq U$ be any finite set of \emph{nodes}; nodes can be any data objects such
as numbers or strings.
For our purposes, we define a (directed) \emph{graph}
on $N$ as a subset of $N^2$, i.e., as a finite set of ordered pairs of nodes.
These pairs are called \emph{edges}.  We assume familiarity with the notion of
a \emph{tree} as a special kind of graph, and with standard
graph-theoretic concepts such as \emph{root} of a tree; \emph{children}, 
\emph{descendants}, \emph{parent}, and \emph{ancestors} of a node;
and \emph{path} in a graph.  Any good algorithms
textbook will supply the necessary background.

In this paper all trees we consider are rooted and unordered,
unless stated otherwise.

\subsection{Tree Pattern}

\paragraph*{Tree Patterns} A \emph{parameterized tree pattern} $P$ is a tree whose nodes are called \emph{variables}, and where additionally:
\begin{itemize}
	\item Some variables may be marked as being \emph{existential};
	\item Some other variables may be marked as \emph{parameters}; 
	\item The variables of $P$ that are neither existential nor parameters are called \emph{distinguished}.
\end{itemize}
We will denote the set of existential variables by $\Pi$, the set of parameters by $\Sigma$, and the set of distinguished variables by $\Delta$.
To make clear that these sets belong to some parameterized tree pattern $P$ we will use a
subscript as in $\Pi_P$ or $\Sigma_P$.

A \emph{parameter assignment} $\alpha$, for a parameterized tree pattern $P$, is a mapping $\Sigma \rightarrow U$ which assigns data constants to the parameters. 

An \emph{instantiated} tree pattern is a pair $(P, \alpha)$, with $P$ a parameterized tree pattern and $\alpha$ a parameter assignment for $P$. We will also denote this by $P^{\alpha}$.

When depicting parameterized tree patterns, existential nodes are indicated by labeling them with the symbol `$\exists$' and parameters are indicated by labeling them with the symbol `$\sigma$'.  When depicting instantiated tree patterns, parameters are indicated by directly writing down their parameter assignment.

\begin{figure}
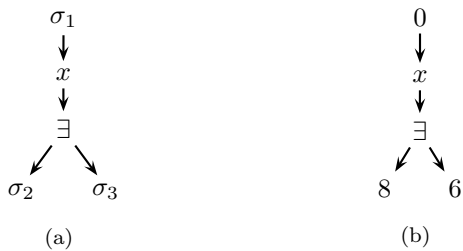

\hspace*{\fill}
\subfigure[]{\label{fig_tree_patterna}
\begin{tabular}{c}
\pstree{\TR{$\sigma_1$}}
{
	\pstree{\TR{$x$}}
	{
		\pstree{\TR{$\exists$}}
		{
			\TR{$\sigma_2$}
			\TR{$\sigma_3$}
		}
	}
}
\end{tabular}
}
\hfill
\subfigure[]{\label{fig_tree_patternb}
\begin{tabular}{c}
\pstree{\TR{$0$}}
{
	\pstree{\TR{$x$}}
	{
		\pstree{\TR{$\exists$}}
		{
			\TR{$8$}
			\TR{$6$}
		}
	}
}
\end{tabular}
}
\hspace*{\fill}
\caption{(a) is a parameterized tree pattern, and (b) is an instantiation of (a).}
\label{fig_tree_pattern_def}
\end{figure}

Figure~\ref{fig_tree_pattern_def} shows an illustration.

\paragraph*{Matching} Recall that a \emph{homomorphism} from a graph $G_1$ to
a graph $G_2$ is a mapping $\mu$ from the nodes of $G_1$ to the nodes of $G_2$
that preserves edges, i.e., if $(i,j) \in G_1$ then $(\mu(i),\mu(j)) \in G_2$. We
now define a \emph{matching} of an instantiated tree pattern $P^{\alpha}$ in a
data graph $G$ as a homomorphism $\mu$ from the underlying tree of $P$ to $G$, with
the constraint that for any parameter $\sigma$, if $\alpha(\sigma)=a$, then
$\mu(\sigma)$ must be the node $a$. We denote the set $\{\mu|_\Delta:\ \mu \text{ is a matching of } P^{\alpha} \text{ in } G\}$ by $P^{\alpha}(G)$.

\paragraph*{Frequency of a tree pattern} The \emph{frequency} of an
instantiated tree pattern $P^{\alpha}$ in a data graph $G$, is formally defined as the
cardinality of $P^{\alpha}(G)$. So, we count the number of matchings of $P^{\alpha}$ in $G$, with the important provision that \emph{we identify any two matchings that agree on the distinguished variables.} Indeed, two
matchings that differ only on the existential nodes need not be distinguished,
as this is precisely the intended semantics of existential nodes.  Note that
we do not need to worry about selected nodes, as all matchings will agree on
those by definition. For a given threshold
$k$ (a natural number) we say that $P^{\alpha}$ is $k$-\emph{frequent} if its
frequency is at least $k$. Often the threshold is understood implicitly, and
then we talk simply about ``frequent'' patterns and denote the threshold by
\emph{minsup}.

\begin{figure}
\hspace*{\fill}
\subfigure[]{\label{figgraaf1}
\begin{tabular}{c}
\includegraphics{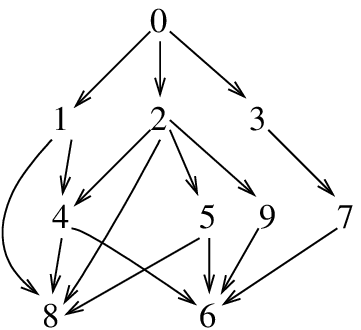}
\end{tabular}
}
\hfill
\subfigure[]{\label{figgraaf2}
\begin{tabular}{c}
\includegraphics{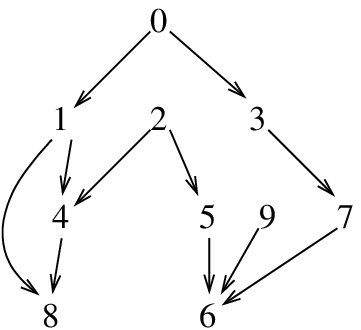}
\end{tabular}
}
\hspace*{\fill}
\label{fig_input_graphs}
\caption{Two data graphs.}
\end{figure}

\begin{example}
Take again the instantiated tree pattern $P^{\alpha}$ shown in Figure~\ref{fig_tree_patternb}.
Let us name the existential node by $y$; let us name the parameter labeled 0 by $z_1$; the parameter labeled 8 by $z_2$; and the parameter labeled 6 by $z_3$. The distinguished node already has the name $x$.
Now let us apply $P^{\alpha}$ to the simple example data graph $G$ shown in
Figure~\ref{figgraaf1}.  The following table lists all matchings of $P^{\alpha}$ in $G$:
$$
\begin{array}{c|ccccc}
& z_1 & x & y & z_2 & z_3 \\
\hline
h_1& 0  & 1 & 4 & 8 & 6\\
h_2& 0  & 2 & 4 & 8 & 6\\
h_3& 0  & 2 & 5 & 8 & 6
\end{array}
$$
As required by the definition, all matchings match $z_1$ to 0, $z_2$ to 8, and $z_3$ to 6.
Although there are three matchings, when determining the frequency of $P^{\alpha}$ in $G$, we only look at their value on $x$ to distinguish them, as $y$ is existential.  So, $h_2$ and $h_3$ are identified as identical matchings when counting the number of matchings.  In conclusion, the frequency of $P^{\alpha}$ in $G$ is two, as $x$ can be matched to the two different nodes 1 and 2.
\qed
\end{example}

\subsection{Tree Query}
\label{sec_treequery}
\paragraph*{Tree Queries} A \textit{parameterized tree query} $Q$ is a pair $(H, P)$ where:

\begin{enumerate}
	\item $P$ is a parameterized tree pattern, called the \emph{body}
	of $Q$;

	\item $H$ is a tuple of distinguished variables and parameters coming
	from $P$.  All distinguished variables of $P$ must appear at least
	once in $H$. We call $H$ the \emph{head} of $Q$.

\end{enumerate}

A parameter assignment for $Q$ is simply a parameter assignment for its body,
and an \emph{instantiated} tree query is then again a pair $(Q,\alpha)$ with
$Q$ a parameterized tree query and $\alpha$ a parameter assignment for $Q$.
We will again also denote this by $Q^{\alpha}$.

\begin{figure}
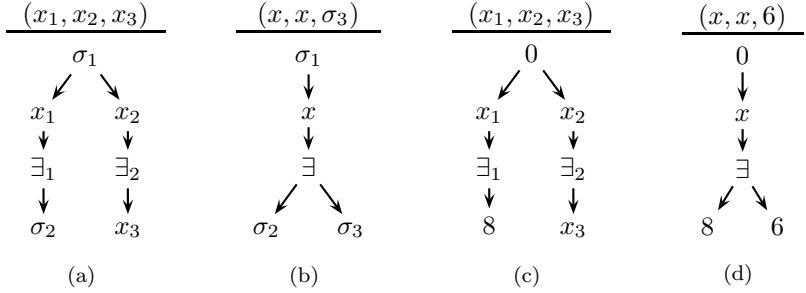

\hspace*{\fill}
\subfigure[]{\label{fig_tree_query_a}
\begin{tabular}{c}
$(x_1,x_2,x_3)$\\
\hline
\pstree{\TR{$\sigma_1$}}
{
	\pstree{\TR{$x_1$}}
	{  
		 \pstree{\TR{$\exists_1$}}
		    {
		    	\TR{$\sigma_2$}
		    }
	} 
	  \pstree{\TR{$x_2$}}
	  {
		    \pstree{\TR{$\exists_2$}}
		    {
		    	\TR{$x_3$}
		    }
	  }
}
\end{tabular}
}
\hfill
\subfigure[]{\label{fig_tree_query_b}
\begin{tabular}{c}
$(x,x,\sigma_3)$\\
\hline
\pstree{\TR{$\sigma_1$}}
{
	\pstree{\TR{$x$}}
	{
		\pstree{\TR{$\exists$}}
		{
			\TR{$\sigma_2$}
			\TR{$\sigma_3$}
		}
	}
}
\end{tabular}
}
\hfill
\subfigure[]{\label{fig_tree_query_c}
\begin{tabular}{c}
$(x_1,x_2, x_3)$\\
\hline
\pstree{\TR{$0$}}
{
	\pstree{\TR{$x_1$}}
	{  
		 \pstree{\TR{$\exists_1$}}
		    {
		    	\TR{$8$}
		    }
	} 
	  \pstree{\TR{$x_2$}}
	  {
		    \pstree{\TR{$\exists_2$}}
		    {
		    	\TR{$x_3$}
		    }
	  }
}
\end{tabular}
}
\hfill
\subfigure[]{\label{fig_tree_query_d}
\begin{tabular}{c}
$(x,x,6)$\\
\hline
\pstree{\TR{$0$}}
{
	\pstree{\TR{$x$}}
	{
		\pstree{\TR{$\exists$}}
		{
			\TR{$8$}
			\TR{$6$}
		}
	}
}
\end{tabular}
}
\hspace*{\fill}
\caption{(a) and (b) are parameterized tree queries; (c) is an instantiation of (a); (d) is an instantiation of (b); and query (b) is $\rho$-contained in query (a)}
\label{fig_tree_query}
\end{figure}

When depicting tree queries, the head is given above a horizontal line, and
the body below it.  Two illustrations are given in
Figure~\ref{fig_tree_query}.

\paragraph*{Frequency of a tree query} The \emph{frequency} of an instantiated tree query $Q^{\alpha} = ((H, P), \alpha)$ in a data graph $G$, is defined as the frequency of the body $P^{\alpha}$ in $G$. When $G$ is understood, we denote the frequency by $\Freq(P^{\alpha})$. For a given threshold $k$ (a natural number) we say that $Q^{\alpha}$ is \emph{k-frequent} if its frequency is at least $k$. Again, this threshold is often understood implicitly, and then we talk simply about ``frequent'' queries and denote the threshold by \emph{minsup}.

\paragraph*{Containment of tree queries}
An important step towards our formal definition of tree-query association is the
notion of \emph{containment} among queries. Since queries are parameterized, a variation of the classical notion of containment \cite{cm,ullmanII,ahv_book} is needed in that we now need to specify a parameter correspondence. 

First, we define the \emph{answer set} of an instantiated tree query
$Q^{\alpha}$, with $Q=(H,P)$, in a data graph $G$ as follows: $$Q^\alpha(G) := \{
\mu(H) \mid \mu \text{ is a matching of } P^{\alpha} \text{ in } G\}$$ 

Consider two parameterized tree queries $Q_1$ and $Q_2$, with $Q_i = (H_i,P_i)$ for $i=1,2$.  A \emph{parameter correspondence} from $Q_1$ to $Q_2$ is any mapping $\rho: \Sigma_1 \rightarrow \Sigma_2$. We then say that a parameterized tree query $Q_2$ is \emph{$\rho$-contained} in a parameterized tree query $Q_1$, if for every $\alpha_2$, a parameter assignment for $Q_2$, $Q_2^{\alpha_2}(G) \subseteq Q_1^{\alpha_{2} \circ \rho}(G)$ for all data graphs $G$.  In shorthand notation we write this as $Q_2 \subseteq_{\rho} Q_1$.

Containment as just defined is a semantical property, referring to all
possible data graphs, and it is not immediately
clear how one could decide this property syntactically.  The required
syntactical notion for this is that of \emph{$\rho$-containment mapping}, which we
next define in two steps.  For the tree queries $Q_1$ and $Q_2$ as above, and $\rho$ a parameter correspondence from $Q_1$ to $Q_2$:

\begin{enumerate}

\item
A $\rho$-containment mapping from $P_1$ to $P_2$ is a homomorphism $f$ from the
underlying tree of $P_1$ to the underlying tree of $P_2$, with the properties:
\begin{enumerate}
 \item $f$ maps the distinguished nodes of $P_1$ to distinguished nodes or
parameters of $P_2$; and
 \item $f|_{\Sigma_1} = \rho$, i.e., for each $z \in \Sigma_1$ we have $f(z)=\rho(z)$.
\end{enumerate}

\item
Finally, a $\rho$-containment mapping from $Q_1$ to $Q_2$ is a $\rho$-containment mapping $f$ from $P_1$ to $P_2$ such that $f(H_1)=H_2$.

\end{enumerate}

For later use, we note:

\begin{lemma}
\label{lemma_comp}
Consider three parameterized tree patterns $P_1$, $P_2$, and $P_3$, a parameter correspondence $\rho_1: \Sigma_1 \rightarrow \Sigma_2$, a parameter correspondence $\rho_2: \Sigma_2 \rightarrow \Sigma_3$, a $\rho_1$-containment mapping $f_1$ from $P_1$ to $P_2$, and a $\rho_2$-containment mapping $f_2$ from $P_2$ to $P_3$. Then $f_2 \circ f_1$ is a $(\rho_2 \circ \rho_1)$-containment mapping from $P_1$ to $P_3$.
\end{lemma}
\begin{proof}
We will show that:
\begin{enumerate}
	\item $f_2 \circ f_1$ is homomorphism;
	\item $f_2 \circ f_1$ maps distinguished nodes of $P_1$ to distinguished nodes or parameters of $P_3$; and
	\item $(f_2 \circ f_1)|_{\Sigma_1} = \rho_2 \circ \rho_1$.
\end{enumerate}

$(1)$ Clearly $f_2 \circ f_1$ is a homomorphism since both $f_1$ and $f_2$ are homomorphisms, and it is already known that a composition of homomorphisms is a homomorphism. 

$(2)$ Consider a $x_1 \in \Delta_1$, then there are two possibilities for $f_1(x_1)$:
\begin{enumerate}
	\item $f_1(x_1) = x_2$, with $x_2 \in \Delta_2$. Then we know, since $f_2$ is a $\rho_2$-containment mapping, that $f_2(x_2)$ is either a distinguished node $x_3 \in \Delta_3$, or a parameter $z_3 \in \Sigma_3$. 
	\item $f_1(x_1) = z_2$, with $z_2 \in \Sigma_2$. Then we know, since $f_2|_{\Sigma_2} = \rho_2$, that $f_2(z_2) = z_3$, with $z_3 \in \Sigma_3$.
\end{enumerate}
Hence, we can conclude that $f_2 \circ f_2$ maps distinguished nodes of $P_1$ to distinguished nodes or parameters of $P_3$. 

$(3)$ For each $z_1 \in \Sigma_1$, we have $f_2(f_1(z_1)) = \rho_2(\rho_1(z_1))$. Hence, $(f_2 \circ f_1)|_{\Sigma_1} = \rho_2 \circ \rho_1$.
\quad
\end{proof}

\begin{figure}
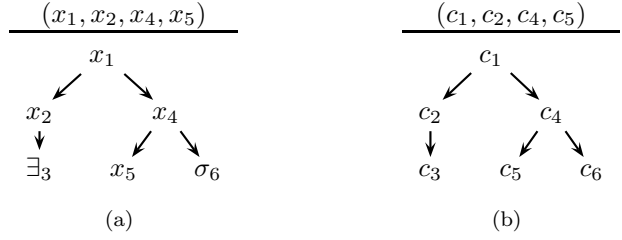

\hspace*{\fill}
\subfigure[]{\label{fig_frozen_a}
\begin{tabular}{c}
$(x_1,x_2, x_4, x_5)$\\
\hline
\pstree{\TR{$x_1$}}
{
	\pstree{\TR{$x_2$}}
	{  
		\TR{$\exists_3$}
	} 
	\pstree{\TR{$x_4$}}
	  {
		\TR{$x_5$}
		\TR{$\sigma_{6}$}
	  }
}
\end{tabular}
}
\hfill
\subfigure[]{\label{fig_frozen_b}
\begin{tabular}{c}
$(c_1,c_2, c_4, c_5)$\\
\hline
\pstree{\TR{$c_1$}}
{
	\pstree{\TR{$c_2$}}
	{  
		\TR{$c_3$}
	} 
	\pstree{\TR{$c_4$}}
	  {
		\TR{$c_5$}
		\TR{$c_{6}$}
	  }
}
\end{tabular}
}
\hspace*{\fill}
\caption{(b) is a freezing of the parameterized tree query in (a)}
\label{fig_frozen}
\end{figure}

From the theory of conjunctive database queries \cite{cm,ullmanII,ahv_book} we can derive the following:

\begin{lemma}
\label{containment_lemma}
 Consider two parameterized tree queries $Q_{1}$ and $Q_{2}$, with $Q_{1}=(H_{1}, P_{1})$ and $Q_{2}=(H_{2}, P_{2})$ and a parameter correspondence $\rho: \Sigma_{1} \rightarrow \Sigma_{2}$. Then $Q_{2}$ is $\rho$-contained in $Q_{1}$ ($Q_{2} \subseteq_{\rho} Q_{1}$), if and only if there exists a $\rho$-containment mapping from $Q_{1}$ to $Q_{2}$.
\end{lemma}
\begin{proof}
Let us start with the `only if' direction. We first introduce the concept of a \emph{freezing} of a parameterized tree query $Q=(H,P)$. Recall that $U$ is the set of data constants from which the nodes of the data graph to be mined will be taken. A freezing $\beta$ of $P$ is then a one-to-one mapping from the nodes of $P$ to $U$. We denote by $\text{freeze}_{\beta}(P)$ the data graph constructed from $P$ by replacing each node $n$ of $P$ by $\beta(n)$, and we denote by $\text{freeze}_{\beta}(H)$ the tuple constructed from $H$ by replacing each node $n$ in $H$ by the data constant $\beta(n)$.

For example, consider the parameterized tree query $Q=(H,P)$ in Figure~\ref{fig_frozen_a}. Figure~\ref{fig_frozen_b}, shows $\text{freeze}_{\beta}(P)$ and $\text{freeze}_{\beta}(H)$ for the freezing $\beta$ given as follows: $x_{1} \rightarrow c_{1}$; $x_{2} \rightarrow c_{2}$;
$\exists_{3} \rightarrow c_{3}$; $x_{4} \rightarrow c_{4}$; $x_{5} \rightarrow c_{5}$; $\sigma_{6} \rightarrow c_{6}$.

We can now continue with the proof of the `only if' direction. Consider a freezing $\beta$ from the nodes of $P_2$ to $U$. Note that $\beta|_{\Sigma_{2}}$ is a parameter assignment for $Q_{2}$, and $\text{freeze}_{\beta}(H_{2}) \in Q_{2}^{\beta|_{\Sigma_{2}}}(\text{freeze}(P_{2}))$. Since $Q_{2} \subseteq_{\rho} Q_{1}$, also $\text{freeze}_{\beta}(H_{2}) \in Q_{1}^{\beta|_{\Sigma_{2}} \circ \rho}(\text{freeze}(P_{2}))$. Hence, there must be a matching $\mu$ from $P_{1}^{\beta|_{\Sigma_{2}} \circ \rho}$ in $\text{freeze}_{\beta}(P_{2})$ such that $\mu(H_1) = \text{freeze}_{\beta}(H_2)$. Now consider the function $g: \beta^{-1} \circ \mu$. We show that $g$ is $\rho$-containment mapping from $Q_1$ to $Q_2$: 
\begin{enumerate}
 \item Clearly, $g$ is a homomorphism from $P_1$ to $P_2$ since $\mu$ is a homomorphism and $\beta^{-1}$ is an isomorphism. Also the following properties hold for $g$:
	\begin{enumerate}
	 \item $g$ maps distinguished nodes of $P_1$ to distinguished nodes or parameters of $P_2$ since $g(H_1) = H_2$ (as shown in (2)); and
	 \item for each $z \in \Sigma_1$: $g(z)=\beta^{-1}(\mu(z)) = \beta^{-1}(\beta(\rho(z))) = \rho(z)$, hence $g|_{\Sigma_1} = \rho$
	\end{enumerate}
\item $g(H_1) = \beta^{-1}(\mu(H_1)) = \beta^{-1}(\text{freeze}_{\beta}(H_2)) = H_2$.
\end{enumerate}
Hence, we conclude that $g$ is a $\rho$-containment mapping from $Q_1$ to $Q_2$. 

Let us then look at the `if' direction. Let $h$ be the $\rho$-containment mapping from $Q_{1}$ to $Q_{2}$. Consider an arbitrary parameter assignment $\alpha_{2}$ for $Q_{2}$. We must prove that for every data graph $G$, if $a \in \Qinst{2}(G)$, then also $a \in Q_{1}^{\alpha_{1} \circ \rho}(G)$. Consider such an arbitrary data graph $G$. Since, $a \in \Qinst{2}(G)$, we know that there exists a matching $\mu$ of $P_{2}^{\alpha_{2}}$ in $G$ such that $a=\mu(H_{2})$. Now consider the function $g= \mu \circ h$. We show that $g$ is a matching from $P_1^{\alpha_2 \circ \rho}$ in $G$ and $a = g(H_1)$:
\begin{enumerate}
 \item $g$ is a homomorphism since both $\mu$ and $g$ are homomorphisms; and
 \item for each $z \in \Sigma_1$ we have $g(z)= \mu(h(z)) = \mu(\rho(z))= \alpha_2(\rho(z))$.
\end{enumerate}
So, $g$ is indeed a matching of $P_1^{\alpha_2 \circ \rho}$ in $G$. Finally, we observe that $g(H_1) = \mu(h(H_1)) = \mu(H_2) =a$, as desired.
\quad
\end{proof}

Checking for a containment mapping is evidently computable, and
although the problem for general database conjunctive queries is NP-complete,
our restriction to tree shapes allows for efficient checking, as we will
see later.

\begin{example}
Consider the parameterized and instantiated tree queries shown in Figure~\ref{fig_tree_query}. In the example data graph in Figure~\ref{figgraaf1} the frequency of query~(c) is 10 and that of query~(d) is 2. Let $\Sigma_{a}$ be the set of parameters of query~(a), and let $\Sigma_{b}$ be the set of parameters of query~(b); then let the parameter correspondence $\rho: \Sigma_{a} \rightarrow \Sigma_{b}$ be as follows: $\sigma_{1} \rightarrow \sigma_{1}$; $\sigma_{2} \rightarrow \sigma_{2}$. A moment's reflection should convince the reader that (b) is $\rho$-contained in (a), and indeed a $\rho$-containment mapping $f$ from (a) to (b) can be found as follows: \\

\begin{center}
\begin{tabular}{|c|c|}
\hline \multicolumn{2}{|c|}{$f$}  \\ \hline
$\sigma_{1}$ & $\sigma_{1}$ \\  
$x_1$ & $x$ \\ 
$x_2$ & $x$ \\ 
$\exists_1$ & $\exists$ \\ 
$\exists_2$ & $\exists$ \\ 
$\sigma_{2}$ & $\sigma_{2}$ \\ 
$x_3$ & $\sigma_{3}$ \\ \hline
\end{tabular}
\end{center}

\end{example}

\subsection{Tree-Query Association}
\label{subsec_def_association}

\paragraph*{Association Rules}

A \emph{parameterized association rule (pAR)} is of the form $Q_1 \Rightarrow_{\rho} Q_2$, with $Q_1$ and $Q_2$
parameterized tree queries and $\rho$ a parameter correspondence from $\Sigma_1$ to $\Sigma_2$. 
We call a pAR \emph{legal} if $Q_2 \subseteq_\rho Q_1$. We call $Q_1$ the left-hand side (lhs),
and $Q_2$ the right-hand side (rhs).
A \emph{parameter assignment} $\alpha$, for a pAR, is a mapping $\Sigma_2 \rightarrow U$ which assigns data constants to the parameters. 
An \emph{instantiated association rule (iAR)} is a pair $(Q_1 \Rightarrow_{\rho} Q_2, \alpha)$, with $Q_1 \Rightarrow_{\rho} Q_2$ a pAR and $\alpha$ a parameter assignment for $Q_1 \Rightarrow_{\rho} Q_2$.
Note that while $\alpha$ is only defined on the rhs, we can also apply it to the lhs by using $\rho$ first.
\paragraph*{Confidence}

The \emph{confidence} of an iAR in a data graph $G$ is defined as the frequency of $Q_2^{\alpha_2}$ divided by the frequency of  $Q_1^{\alpha_2 \circ \rho}$.
If the AR is legal, we know that the answer set of $Q_2^{\alpha_2}$ is a subset of the answer set of  $Q_1^{\alpha_2 \circ \rho}$, and hence the confidence equals precisely the proportion that the $Q_2^{\alpha_2}$ answer set
takes up in the $Q_1^{\alpha_2 \circ \rho}$ answer set.  Thus, our notions of a legal
pAR and confidence are very intuitive and natural.

For a given threshold $c$ (a rational number, $0 \leq c\leq 1$) we say that
the iAR is \emph{$c$-confident} in $G$ if its confidence in $G$ is at least
$c$.  Often the threshold is understood implicitly, and then we
talk simply about ``confident'' iARs and denote the threshold by \emph{minconf}.

Furthermore, the iAR is called \emph{frequent} in $G$ if $Q_2^{\alpha_2}$ is frequent in $G$.  Note that if the iAR is legal and frequent, then also $Q_1^{\alpha_2 \circ \rho}$ is frequent, since the rhs is $\rho$-contained in the lhs.

\begin{example}
Continuing the previous example, we can see that we can form a legal pAR from
the queries of Figure~\ref{fig_tree_query}, with (a) the lhs and (b) the rhs and $\rho$ as follows: $\sigma_{1} \rightarrow \sigma_{1}$; $\sigma_{2} \rightarrow \sigma_{2}$. We can also form an iAR with the tree queries in Figure~\ref{fig_tree_query_c} and Figure~\ref{fig_tree_query_d}; the confidence of this iAR in the data graph of
Figure~\ref{figgraaf1} is $2/10$.
Many more examples of ARs are given in Section~\ref{sec_algo_associations}.
\end{example}

\subsection{Mining Problems}

We are now finally ready to define the graph mining problems we want to solve.

\subsubsection{Mining Tree Queries} 
\begin{description}
\item[Input:]
A data graph $G$; a threshold \emph{minsup}.
\item[Output:]
All frequent instantiated tree queries $Q=((H,P),\alpha)$.
\end{description}

In theory, however, there are infinitely many $k$-frequent tree queries, and even if we
set an upper bound on the size of the patterns, there may be exponentially
many.  As an extreme example, if $G$ is the complete graph on the set of nodes
$\{1,\dots,n\}$, and $k \leq n$, then \emph{any} instantiated pattern with all parameters assigned to values in $\{1,\dots,n\}$, and with at least one distinguished variable, is frequent.

Hence, in practice, we want an algorithm that runs incrementally, and that can be stopped any time it has run long
enough or has produced enough results. We introduce such an algorithm in Section~\ref{sec_algo_patterns}.

\subsubsection{Association Rule Mining}
\begin{description}
\item[Input: ]  
A data graph $G$; a threshold \emph{minsup}; a parameterized tree query $\qleft$; and a threshold \emph{minconf}.
\item[Output: ]
All iARs $(\qleft \Rightarrow_\rho Q_{\text{right}}, \alpha)$ that are legal, frequent and confident in $G$
\end{description}

In theory, however, there are infinitely many legal, frequent and confident association rules for a fixed lhs, and even if we set an upper bound on the size of the rhs, there may be exponentially many. Hence, in practice, we want an algorithm that runs incrementally, and that can be stopped any time it has run long enough or has produced enough results. We introduce such an algorithm in Section~\ref{sec_algo_associations}.

\section{Mining Tree Queries}
\label{sec_algo_patterns}

In this Section we present an algorithm for mining frequent instantiated tree queries in a large data graph. But first we show that we do not need to tackle the problem in its full generality.

\subsection{Problem Reduction}
\label{sec_probred}

In this subsection we show that, without loss of generality, we can focus on parameterized tree queries that are `pure'. 

\paragraph*{Pure Tree Queries} To define this formally, assume that all possible variables (nodes of tree patterns) have been arranged in some fixed but arbitrary order. We then call a parameterized tree query $Q=(H,P)$ \emph{pure} when $H$ consists of the enumeration, in order and without repetitions, of all the distinguished variables of $P$. In particular $H$ cannot contain parameters. We call $H$ the \emph{pure head} for $P$. As an illustration, the parameterized tree query in Figure~\ref{fig_tree_query_a} is pure, while the parameterized tree query in Figure~\ref{fig_tree_query_b} is not pure.

\begin{figure}
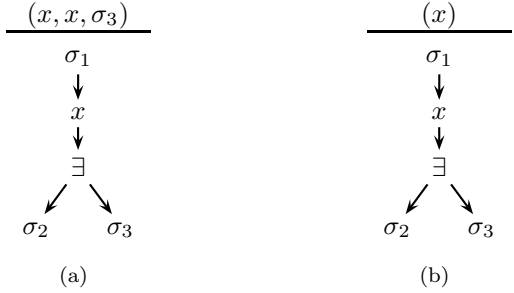

\hspace*{\fill}
\subfigure[]{\label{fig_pure_a}
\begin{tabular}{c}
$(x,x,\sigma_3)$\\
\hline
\pstree{\TR{$\sigma_1$}}
{
	\pstree{\TR{$x$}}
	{
		\pstree{\TR{$\exists$}}
		{
			\TR{$\sigma_2$}
			\TR{$\sigma_3$}
		}
	}
}
\end{tabular}
}
\hfill
\subfigure[]{\label{fig_pure_b}
\begin{tabular}{c}
$(x)$\\
\hline
\pstree{\TR{$\sigma_1$}}
{
	\pstree{\TR{$x$}}
	{
		\pstree{\TR{$\exists$}}
		{
			\TR{$\sigma_2$}
			\TR{$\sigma_3$}
		}
	}
}
\end{tabular}
}
\hspace*{\fill}
\caption{(a) is an impure parameterized tree query. The parameterized tree query in (b) is the purification of the parameterized tree query (a), and expresses precisely the same information.}
\label{fig_pure}
\end{figure}

A parameterized tree query that is not pure can always be rewritten to a parameterized tree query that is pure, in such a way that all instantiations of the impure query correspond to instantiations of the pure query, with the same frequency. Indeed, take a parameterized tree query $Q=(H,P)$. We can purify $Q$ by removing all parameters and repetitions of distinguished variables from $H$, and sort $H$ by the order on the variables. An illustration of this is given in Figure~\ref{fig_pure}.

\begin{figure}
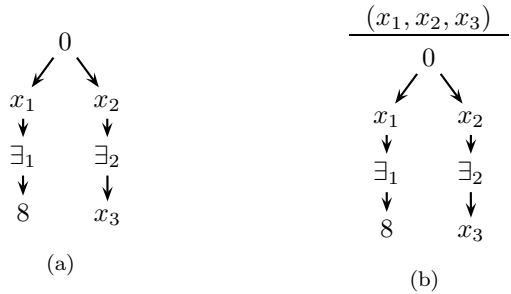

\hspace*{\fill}
\subfigure[]{\label{fig_pure_pattern_a}
\begin{tabular}{c}
\pstree{\TR{$0$}}
{
	\pstree{\TR{$x_1$}}
	{  
		 \pstree{\TR{$\exists_1$}}
		    {
		    	\TR{$8$}
		    }
	} 
	  \pstree{\TR{$x_2$}}
	  {
		    \pstree{\TR{$\exists_2$}}
		    {
		    	\TR{$x_3$}
		    }
	  }
}
\end{tabular}
}
\hfill
\subfigure[]{\label{fig_pure_pattern_b}
\begin{tabular}{c}
$(x_1,x_2, x_3)$\\
\hline
\pstree{\TR{$0$}}
{
	\pstree{\TR{$x_1$}}
	{  
		 \pstree{\TR{$\exists_1$}}
		    {
		    	\TR{$8$}
		    }
	} 
	  \pstree{\TR{$x_2$}}
	  {
		    \pstree{\TR{$\exists_2$}}
		    {
		    	\TR{$x_3$}
		    }
	  }
}
\end{tabular}
}
\hspace*{\fill}
\caption{(b) is the pure instantiated tree query constructed from the instantiated tree pattern in (a)}
\label{fig_pure_pattern}
\end{figure}

We can conclude that it is sufficient to only consider pure instantiated tree queries. As a consequence, rather than mining tree \emph{queries}, it suffices to mine for tree \emph{patterns}, because the frequency of a query is nothing else then the frequency of his body, i.e., a pattern.  An illustration is given in Figure~\ref{fig_pure_pattern}.

\subsection{Overall Approach}
\label{sec_overall}

An overall outline of our tree-query mining algorithm is the following:

\begin{description}
\item[Outer loop:]
Generate, incrementally, all possible trees $T$ of increasing sizes.  Avoid
trees that are isomorphic to previously generated ones.
\item[Inner loop:]
For each $T$, generate all instantiated tree patterns $P^{\alpha}$ based on $T$,
and test their frequency.  
\end{description}

The algorithm is incremental in the number of nodes of the pattern.  We
generate canonically ordered rooted trees of increasing sizes, avoiding the generation
of isomorphic duplicates.  It is well known how to do this efficiently
\cite{scions,ruskey,zaki_journ_forest,muntz_trees}.
Note that this generation of trees is in no way
``levelwise'' \cite{levelwise}.  Indeed, under the way we count pattern
occurrences, a subgraph of a pattern might be less frequent than the pattern
itself (this was already pointed out by Kuramochi and Karypis
\cite{kukajourn2005}).  So, our algorithm systematically considers ever larger
trees, and can be stopped any time it has run long enough or has produced
enough results.  Our algorithm does not need any space beyond what is
needed to store the mining results. The outer loop of our algorithm will be explained in more detail in Section~\ref{outerLoop}.

For each tree, all conjunctive queries based on that tree are generated.
Here, we do work in a levelwise fashion.  This aspect of our algorithm has
clear similarities with ``query flocks'' \cite{queryflocks}.  A query flock is
a user-specified conjunctive query, in which some constants are left
unspecified and viewed as parameters.  A levelwise algorithm was proposed for
mining all instantiations of the parameters under which the resulting query
returns enough answers.  We push that approach further by also mining the
query flocks themselves.  Consequently, the specialization relation on queries
used to guide the levelwise search is quite different in our approach. The inner loop of our algorithm will be explained in more detail in Section~\ref{innerLoop}.

A query based on some tree may be equivalent to a query based on a previously
seen tree.  Furthermore, two queries based on the same tree may be equivalent.
We carefully and efficiently
avoid the counting of equivalent queries, by using and adapting
what is known from the theory of conjunctive database queries. This will be discussed in Section~\ref{equivalentQueries}.

\subsection{Outer Loop}
\label{outerLoop}
\begin{figure}
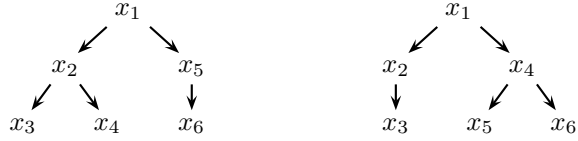

\begin{center}
\hspace*{\fill}
\pstree{\TR{$x_1$}}
{
  \pstree[levelsep=5ex]{\TR{$x_2$}}
  {
    \TR{$x_3$}
    \TR{$x_4$}
  }
  \pstree[levelsep=5ex]{\TR{$x_5$}}
  {
    \TR{$x_6$}
  }
}
\hfill
\pstree{\TR{$x_1$}}
{
  \pstree[levelsep=5ex]{\TR{$x_2$}}
  {
    \TR{$x_3$}
  }
  \pstree[levelsep=5ex]{\TR{$x_4$}}
  {
    \TR{$x_5$}
    \TR{$x_6$}
  }
}
\hspace*{\fill}
\end{center}
\caption{Two orderings of the same tree.  The left one is canonical.}
\label{figcanonical}
\end{figure}

In the outer loop we generate all possible trees of increasing sizes and we avoid trees that are isomorphic to previously generated ones.  In fact, it is well known how to do this \cite{scions,ruskey,zaki_journ_forest,muntz_trees}.
What these procedures typically do is generating trees that are \emph{canonically ordered}
in the following sense.  Given an (unordered) tree $T$, we can order the children of every
node in some way, and call this an \emph{ordering} of $T$.  For example, Figure~\ref{figcanonical} shows two orderings of the same tree. 
From the different orderings of a tree $T$, we want to uniquely select one, to be the canonical ordering of $T$.  For each such possible ordering of $T$, we can write down the \emph{level sequence} of the
resulting tree. This is actually a string representation of the resulting tree. This level sequence is as follows: if the tree has $n$ nodes then this is a sequence of $n$ numbers, where the $i$th number is the depth of the $i$th node in preorder.
Here, the depth of the root is 0, the depth of its children is 1, and so on.
The canonical ordering of $T$ is then the ordering of $T$ that yields the
lexicographically maximal level sequence among all possible orderings of $T$.

For example, in Figure~\ref{figcanonical}, the left one is the canonical one.

\subsection{Inner Loop}
\label{innerLoop}

Let $G$ be the data graph being mined, and let $U$ be its set of nodes.
In this section, we fix a tree $T$, and we want to find all instantiated tree patterns $P^{\alpha}$ based on $T$ whose frequency in $G$ is at least \emph{minsup}. 

This tasks lends itself naturally to a levelwise approach \cite{levelwise}.
A natural choice for the specialization relation is suggested by an
alternative notation for the patterns under consideration.
Concretely, since the underlying tree $T$ is fixed, any parameterized tree pattern $P$ based on $T$ is characterized by two parameters:
\begin{enumerate}
\item
The set $\Pi$ of existential nodes;
\item
The set $\Sigma$ of parameters.
\end{enumerate}
Note that $\Pi$ and $\Sigma$ are disjoint. 

Thus, a parameterized tree pattern $P$ is completely characterized by the pair $(\Pi,\Sigma)$. An instantiation $P^{\alpha}$ of $P$ is then represented by the triple $(\Pi,\Sigma, \alpha)$. For two parameterized tree patterns $P_1 = (\Pi_1, \Sigma_1)$ and $P_2 = (\Pi_2, \Sigma_2)$ we now say that $P_{1}$ \emph{specializes} $P_{2}$ if $\Pi_1 \supseteq \Pi_2$ and $\Sigma_1 \supseteq \Sigma_2$; and $\alpha_2 = \alpha_1|_{\Sigma_2}$. We also say that $P_{2}$ \emph{generalizes} $P_{1}$.

\paragraph*{Parent} An immediate generalization of a tree pattern is called a \emph{parent}. Formally, let $P = (\Pi, \Sigma)$ and $P'= (\Pi', \Sigma')$ be parameterized tree patterns based on $T$. We say that $P'$ is a \emph{parent} of $P$ if:
\begin{itemize}
	\item [(i)] $\Sigma = \Sigma'$ and $\Pi = \Pi' \cup \{y\}$ for some node $y \not\in \Pi'$ ; or
	\item [(ii)] $\Pi = \Pi'$ and $\Sigma = \Sigma' \cup \{z\}$ for some node $z \not\in \Sigma'$.
\end{itemize}

From the following lemma, it follows that specialized patterns have a lower frequency, as expected for a specialization relation:

\begin{lemma}
\label{lemma_freq}
Let $P$ and $P'$ be parameterized tree patterns such that $P'$ is a parent of $P$. Let $\Pinst{}$ be an instantiation of $P$, and let $\alpha'=\alpha|_{\Sigma'}$. Then $\Freq(\Pinst{}) \leq \Freq(\PinstPrime{})$.
\end{lemma}
\begin{proof}
We will show that $\#P^{\alpha}(G) \leq \#P'^{\alpha'}(G)$ by defining an injection $I: P^{\alpha}(G) \rightarrow P'^{\alpha'}(G)$. \\Since $P'$ is a parent of $P$, we know that $\Delta'=\Delta \cup \{u\}$ where $u$ is either an existential node or a parameter of $P$. Note that each $\mu \in P^{\alpha}(G)$ is of the form $\overline{\mu}|_{\Delta}$ for some matching $\overline{\mu}$ of $P^{\alpha} \in G$. For each $\mu$ in $P^{\alpha}(G)$, we fix arbitrarily $\overline{\mu}$. Now we define $I(\mu):= {\overline{\mu}}|_{\Delta'}$. To see that $I$ is an injection, let $\mu_1, \mu_2 \in \Pinst{}(G)$ and suppose that $I(\mu_1)=I(\mu_2)$. In other words, ${\overline{\mu_1}}|_{\Delta'} = {\overline{\mu_2}}|_{\Delta'}$. In particular, $\mu_1 = {\overline{\mu_1}}|_{\Delta} = {\overline{\mu_2}}|_{\Delta} = \mu_2$, as desired.\\Hence, we can conclude that $\#\Pinst{}(G) \leq \#\PinstPrime{}(G)$ and that $\Freq(\Pinst{}) \leq \Freq(\PinstPrime{})$. \quad
\end{proof}

The above lemma suggests the following definition of specialization among \emph{instantiated} tree patterns: we say that $(\Pi_1, \Sigma_1, \alpha_1)$ is a specialization of \\$(\Pi_2, \Sigma_2, \alpha_2)$ if the parameterized tree pattern $(\Pi_1, \Sigma_1)$ is a specialization of the parameterized tree pattern $(\Pi_2, \Sigma_2)$, and $\alpha_2 = \alpha_1|_{\Sigma_2}$. 

Intuitively, the previous lemma then expresses that the frequency of an instantiated tree pattern is always at most the frequency of any of its instantiated parents.

\subsubsection{Candidate generation}
\label{secgeneration}

\paragraph*{Candidate pattern} A \emph{candidate pattern} is an instantiated tree pattern whose frequency is not yet determined, but all whose generalizations are known to be frequent. 

Using the specialization relation and the definition for a candidate pattern we explain how the levelwise search for frequent instantiated tree patterns will go. 

\paragraph*{Levelwise search}
We start with the most general instantiated tree pattern $P=(\emptyset,\emptyset,\emptyset)$, and we progressively consider more specific patterns.  The search has the typical property that, in each new iteration,
new \emph{candidate} patterns are generated; the frequency of all newly discovered candidate patterns is determined, and the process repeats.

There are many different instantiations to consider for each parameterized tree pattern. Hence, to generate candidate patterns in an efficient manner, we propose the use of \emph{candidacy tables} and \emph{frequency tables}.  These candidacy and frequency tables allow us to generate all frequent instantiations for a particular parameterized tree pattern in parallel. A frequency table contains all frequent instantiations for a particular parameterized tree pattern.

Formally, for any parameterized pattern $P=(\Pi,\Sigma)$, we define:
\begin{align*}
\cantab \Pi \Sigma & = \{\alpha \mid \text{$\Pinst{}$ is a candidate instantiated tree pattern}\} \\
\freqtab \Pi \Sigma & = \{\alpha \mid \text{$\Pinst{}$ is a frequent instantiated tree pattern}\}
\end{align*}

Technically, the table has columns for the different parameters, plus a column \verb"freq". Note that when $\Sigma = \emptyset$, i.e., $P$ has no parameters, this is a single-column, single-row table containing just the frequency of $P$. This still makes sense and can be interpreted as boolean values; for example, if $\freqtab \Pi \emptyset$ contains the empty tuple, then the pattern $(\Pi,\emptyset,\emptyset)$ is frequent; if the table is empty, the
pattern is not frequent.  Of course in practice, all frequency tables for parameterless patterns can be combined into a single table. All frequency tables are kept in a relational database.

The following crucial lemma shows these tables can be populated efficiently.

\newtheorem*{joinlemma}{Join Lemma}
\begin{joinlemma}
A parameter assignment $\alpha$ is in $\cantab \Pi \Sigma$ if and only if the following
conditions are satisfied for every parent $(\Pi',\Sigma')$ of $(\Pi,\Sigma)$:
\begin{itemize}
\item[(i)]
If $\Pi=\Pi'$, then $\alpha|_{\Sigma'} \in \freqtab{\Pi'}{\Sigma'}$;
\item[(ii)]
If $\Sigma=\Sigma'$, then $\alpha \in \freqtab{\Pi'}{\Sigma'}$.
\end{itemize}
\end{joinlemma}
\begin{proof}
For the `only-if' direction: By definition of a candidacy table, if $\alpha \in \cantab{\Pi}{\Sigma}$, then all generalizations of $(\Pi, \Sigma, \alpha)$ are frequent. In particular, for all parents $(\Pi', \Sigma')$ of $(\Pi, \Sigma)$, we know that $(\Pi', \Sigma', \alpha|_{\Sigma'})$ is frequent, since parents are generalizations.

For the `if' direction, we must show that all generalizations of $(\Pi, \Sigma, \alpha)$ are frequent. Consider such a generalization $(\Pi_{g_1}, \Sigma_{g_1}, \alpha|_{\Sigma_{g_1}})$. Let us denote the parent relation by $\geq_{p}$. Then there is a sequence of parent patterns: $(\Pi_{g_1}, \Sigma_{g_1}) \geq_{p} (\Pi_{g_2}, \Sigma_{g_2}) \geq_{p} ... \geq_{p} (\Pi', \Sigma')$. And we have: $\Freq(\Pi_{g_1}, \Sigma_{g_1}, \alpha|_{\Sigma_{g_1}})$ \\$\geq \Freq(\Pi_{g_2}, \Sigma_{g_2}, \alpha|_{\Sigma_{g_2}}) \geq ... \geq \Freq(\Pi', \Sigma', \alpha|_{\Sigma'}) \geq \text{minsup}$. The last inequality is given by (i) or (ii), the other inequalities are given by Lemma~\ref{lemma_freq}.  
\quad
\end{proof}

The Join Lemma has its name because, viewing the tables as relational database tables,
it can be phrased as follows:
\begin{quote}
\emph{Each candidacy table can be computed by
taking the natural join of its parent frequency tables.}
\end{quote}

The only exception is when $\Pi = \emptyset$ and $\Sigma = \{z\}$ is a
singleton; this is the initial iteration of the search process, when there are
no constants in the parent tables to start from.  In that case, we define
$\cantab \emptyset {\{z\}}$ as the table with a single column $z$, holding all
nodes of the data graph $G$ being mined.

\subsubsection{Frequency counting using SQL}

The search process starts by determining the frequency of the
underlying tree $T = (\emptyset,\emptyset)$; indeed,
formally this amounts to computing $\freqtab \emptyset \emptyset$.
Similarly, for each parameterized tree pattern $P=(\Pi,\emptyset)$ with $\Pi \neq
\emptyset$, all we can do is determine its frequency, except that here, we do
this only on condition that its parent patterns are frequent.

We have seen above that, if the frequency tables are viewed as relational
database tables, we can compute each candidacy table by a single database
query, using the Join Lemma.  Now suppose the data graph $G$ that is being mined is
stored in the relational database system as well, in the form of a table
\texttt{G(from,to)}.  Then also each frequency table can be computed by a
single SQL query.

Indeed, in the cases where $\Sigma=\emptyset$ this simply
amounts to formulating the pattern in SQL, and determining its count
(eliminating duplicates). Since our patterns are in fact conjunctive queries (or datalog rules) known from database research \cite{ahv_book,ullmanII}. They can easily be translated in SQL:
\begin{itemize}
 \item The \verb"FROM"-clause consists of all table references of the form \verb"G as Gij", for all edges $x_i \rightarrow x_j$ in $T$.
 \item The \verb"WHERE"-clause consists of all equalities of the form \verb"Gij.from = "\\ \verb"Gik.from" as well of equalities of the form \verb"Gij.to = Gjh.from".
 \item The \verb"SELECT"-clause is of the from \verb"SELECT DISTINCT" and consists of all column references of the form \verb"Gij.to" when $x_{ij}$ is a distinguished node in $P$, plus one reference of the form \verb"G1k.from" if the root node is distinguished.
\end{itemize}
The SQL query for the tree in Figure~\ref{fig_sql} with $\Pi = \{x_2\}$ and $\Sigma = \emptyset$ is as follows: 
\begin{verbatim}
E = SELECT G12.from, G23.to, G24.to
FROM G as G12, G as G23, G as G24
WHERE G12.to = G23.from AND G12.to = G24.from
\end{verbatim}

\begin{figure}
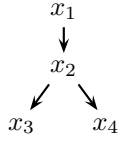

\begin{center}
\pstree{\TR{$x_1$}}
{
	\pstree{\TR{$x_2$}}
	{
		\TR{$x_3$}
		\TR{$x_4$}
	}
}
\end{center}
\caption{Illustration on translating a tree pattern without parameters in SQL.}
\label{fig_sql}
\end{figure}

But also when $\Sigma \neq \emptyset$, we can compute $\freqtab \Pi \Sigma$ by
a single SQL query.  Note that we thus compute the frequency of a large number
of instantiated tree patterns in parallel!  We proceed as follows:
\begin{enumerate}
\item 
we formulate the pattern $(\Pi,\emptyset)$ in SQL; call the resulting expression $E$
\item
We then take the natural join of $E$ and $\cantab \Pi \Sigma$, group by
$\Sigma$, and count each group.  
\end{enumerate}

The join with the candidacy table ensures that only candidate patterns are counted.

For instance, the SQL query to compute the frequency table for the tree in Figure~\ref{fig_sql}, with $\Pi = \{ x_2 \}$ and $\Sigma = \{ x_1, x_3\}$, with \verb"E" as above, is as follows:
{\ttfamily
\begin{tabbing}
SELECT E.x1, E.x3, COUNT(*)\\
FROM E, $\cantab{\{x_2\}}{\{x_1, x_3\}}$ CT\\
WHERE E.x1 = CT.x1 AND E.x3 = CT.x3\\
GROUP BY E.x1, E.x3 HAVING COUNT(*) >= minsup
\end{tabbing}}

It goes without saying that, whenever the frequency table of a tree pattern is found
to be empty, the search for more specialized patterns is pruned at that point.

\subsubsection{The algorithm}

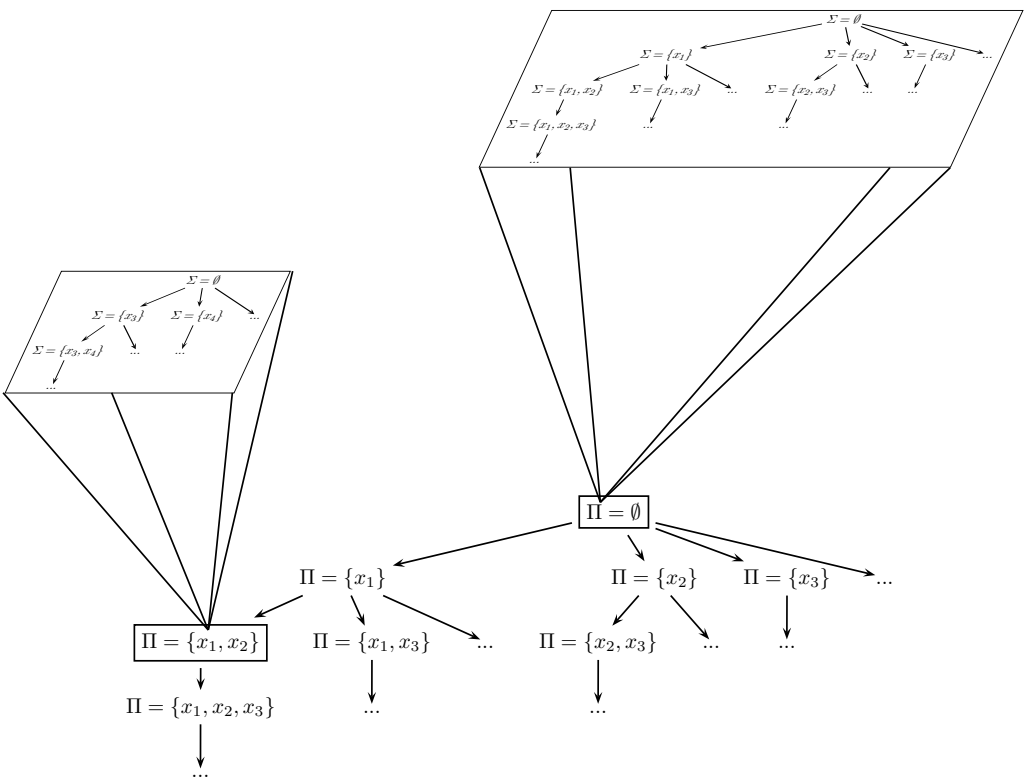
\begin{figure}
\psset{arrows=->,levelsep=7ex,nodesep=3pt,arrowsize=4pt}
\scalebox{0.8}
{
\rotatebox{90}
{
\begin{pspicture}(8,-5)(26,9)
\rput(17,-2)
{
	\pstree{\TR{\psframebox{$\Pi = \emptyset$}}}
	{
		\pstree{\TR{$\Pi = \{x_1\}$}}
		{
			\pstree{\TR{\psframebox{$\Pi = \{ x_1, x_2\}$}}}
			{
				\pstree{\TR{$\Pi = \{ x_1, x_2, x_3\}$}}
				{
					\TR{...}
				}
			}
			\pstree{\TR{$\Pi = \{ x_1, x_3\}$}}
			{
				\TR{...}
			}
			\TR{...}
		}
		\pstree{\TR{$\Pi = \{x_2\}$}}
		{
			\pstree{\TR{$\Pi = \{ x_2, x_3\}$}}
			{
				\TR{...}
			}
			\TR{...}
		}
		\pstree{\TR{$\Pi = \{x_3\}$}}
		{
			\TR{...}
		}
		\TR{...}	
	}
}
\psline[arrowsize=0pt](18.5,0.2)(16.5,5.7)
\psline[arrowsize=0pt](18.5,0.2)(18,5.7)
\psline[arrowsize=0pt](18.5,0.2)(23.3,5.7)
\psline[arrowsize=0pt](18.5,0.2)(24.3,5.7)
\rput(21,7){
\scalebox{0.6}
{
\pstilt{65}{
\psframebox{
	\pstree{\TR{$\Sigma = \emptyset$}}
	{
		\pstree{\TR{$\Sigma = \{x_1\}$}}
		{
			\pstree{\TR{$\Sigma = \{ x_1, x_2\}$}}
			{
				\pstree{\TR{$\Sigma = \{ x_1, x_2, x_3\}$}}
				{
					\TR{...}
				}
			}
			\pstree{\TR{$\Sigma = \{ x_1, x_3\}$}}
			{
				\TR{...}
			}
			\TR{...}
		}
		\pstree{\TR{$\Sigma = \{x_2\}$}}
		{
			\pstree{\TR{$\Sigma = \{ x_2, x_3\}$}}
			{
				\TR{...}
			}
			\TR{...}
		}
		\pstree{\TR{$\Sigma = \{x_3\}$}}
		{
			\TR{...}
		}
		\TR{...}	
	}
}
}
}
}
\rput(11,3){
\scalebox{0.6}
{
\pstilt{65}{
\psframebox{
	\pstree{\TR{$\Sigma = \emptyset$}}
	{
		\pstree{\TR{$\Sigma = \{x_3\}$}}
		{
			\pstree{\TR{$\Sigma = \{ x_3, x_4\}$}}
			{
				\TR{...}
			}
			\TR{...}
		}
		\pstree{\TR{$\Sigma = \{x_4\}$}}
		{
			\TR{...}
		}
		\TR{...}	
	}
}
}
}
}
\psline[arrowsize=0pt](12,-1.9)(8.6,2)
\psline[arrowsize=0pt](12,-1.9)(10.4,2)
\psline[arrowsize=0pt](12,-1.9)(12.4,2)
\psline[arrowsize=0pt](12,-1.9)(13.4,4)
\end{pspicture}
}
}
\caption{Illustration of the algorithm in Algorithm~\ref{figalgo}}
\label{illust_algo}
\end{figure}

Putting everything together so far, the algorithm is given in
Algorithm~\ref{figalgo}.  In outline it is a double Apriori algorithm
\cite{apriori}, where the sets $\Pi$ form one dimension of itemsets, and the
sets $\Sigma$ another. A graphical illustration of the algorithm is given in Figure~\ref{illust_algo}. In this illustration we use \emph{tries} (or prefix-trees) to store the itemsets. A trie \cite{trie, brin_trie, bayardo_trie} is commonly used in implementations of the Apriori algorithm.

\begin{algorithm}
\caption{Levelwise search for frequent tree patterns.}
\label{figalgo}
\begin{algorithmic}[1]
\FOR{each unordered, rooted tree $T$} 
\STATE $X := \text{set of nodes of $T$}$
\STATE $p:=0$; $\mathcal{P}_0:=\{\emptyset\}$
\REPEAT
\FOR{each $\Pi \in \mathcal{P}_p$}
\STATE Compute $\freqtab \Pi \emptyset$ in SQL
\IF{$\freqtab \Pi \emptyset \neq \emptyset$}
\STATE $s := 1$
\STATE $\mathcal{S}_1 := \{\{z\} \mid z \in X-\Pi\}$ 
\REPEAT
\FOR{each $\Sigma \in \mathcal{S}_s$}
\IF{$p=0$ and $s=1$}
\STATE $\cantab \Pi \Sigma := \text{set of nodes of $G$}$
\ELSE
\STATE $\cantab \Pi \Sigma := \: \Join\{\freqtab{\Pi'}{\Sigma'} \mid (\Pi',\Sigma')$ parent of $(\Pi,\Sigma)\}$
\ENDIF
\STATE Compute $\freqtab \Pi \Sigma$ in SQL
\IF{$\freqtab \Pi \Sigma = \emptyset$} 
\STATE remove $\Sigma$ from $\mathcal{S}_s$ \COMMENT{$\Sigma$ is pruned away}
\ENDIF
\ENDFOR
\STATE $\mathcal{S}_{s+1}:=\{\Sigma\subseteq X-\Pi \mid \text{$\#\Sigma=s+1$}$
\STATE \hspace{4cm} $\text{and each $s$-subset of $\Sigma$ is in $\mathcal{S}_s\}$}$
\STATE $s := s+1$
\UNTIL{$\mathcal{S}_s=\emptyset$}
\ELSE 
\STATE remove $\Pi$ from $\mathcal{P}_p$ \COMMENT{$\Pi$ is pruned away}
\ENDIF
\ENDFOR
\STATE $\mathcal{P}_{p+1}:=\{\Pi\subseteq X \mid \text{$\#\Pi=p+1$ and each $p$-subset of $\Pi$ is in $\mathcal{P}_p\}$}$
\STATE $p := p+1$
\UNTIL{$\mathcal{P}_p=\emptyset$}
\ENDFOR 
\end{algorithmic}
\end{algorithm}

\subsubsection{Example run}
\label{sec_example}

\begin{figure}
\hspace*{\fill}
\subfigure[]{\label{fig_exampledatagraph_a}
\begin{pspicture}(0,0)(2,2)
\rput(1,2){\rnode{0}{0}}
\rput(0,1){\rnode{1}{1}}
\rput(1,1){\rnode{2}{2}}
\rput(2,1){\rnode{3}{3}}
\rput(0,0){\rnode{4}{4}}
\rput(1,0){\rnode{5}{5}}
\rput(2,0){\rnode{6}{6}}
\ncline{0}{1}
\ncline{0}{2}
\ncline{0}{3}
\ncline{1}{4}
\ncline{2}{4}
\ncline{2}{5}
\ncline{3}{6}
\end{pspicture}
}
\hfill
\subfigure[]{\label{fig_exampledatagraph_b}
\pstree{\TR{$x_1$}}
{
  \TR{$x_2$}
  \TR{$x_3$}
}
}
\hspace*{\fill}
\caption{Data graph $G$ and unordered rooted tree $T$ for the example run in Section \ref{sec_example}.}
\label{fig_exampledatagraph}
\end{figure}
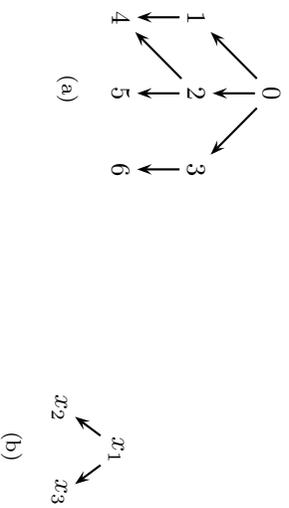

In this Section we give an example run of the proposed algorithm in Algorithm~\ref{figalgo}. Consider the example data graph $G$ in Figure~\ref{fig_exampledatagraph_a}; the unordered rooted tree $T$ in Figure~\ref{fig_exampledatagraph_b}; and let the minimum support threshold be $3$. 

The example run then looks as follows:


\newlength{\iets}
\setlength{\iets}{0.5ex}
\newlength{\struthoogte}
\newcommand{\mystrut}[1]{\settoheight{\struthoogte}{#1}%
\addtolength{\struthoogte}{\iets}%
#1\rule[-\struthoogte]{0pt}{2\struthoogte}}

\newcommand{\FreqEnv}[1]{\mystrut{%
\begin{minipage}{25ex}%
\begin{multline*}%
#1%
\end{multline*}%
\end{minipage}}}

{\footnotesize
\begin{longtable}{|c|c|c|c|c|}
\hline \multicolumn{1}{|c|}{$\Pi$} & \multicolumn{1}{|c|}{$\Sigma$} & \multicolumn{1}{|c|}{$P$} & \multicolumn{1}{|c|}{${\it CanTab}$} & \multicolumn{1}{|c|}{${\it FreqTab}$}\\ \hline
\endfirsthead

\hline
\endlastfoot
$\emptyset$
&
$\emptyset$
&
\mystrut{\begin{tabular}{c}
\pstree{\TR{$x_1$}}
{
  \TR{$x_2$}
  \TR{$x_3$}
}
\end{tabular}}
&
&
\mystrut{\begin{tabular}{|c|}
\hline $\Freq$  \\ \hline
$15$ \\ \hline
\end{tabular}}
\\ \hline
$\emptyset$
&
$\{x_1\}$
&\mystrut{\begin{tabular}{c}
\pstree{\TR{$\sigma_1$}}
{
	\TR{$x_2$}
	\TR{$x_3$}
}
\end{tabular}}
&
\FreqEnv{\text{ nodes of }G}
&
\mystrut{\begin{tabular}{|c|c|}
\hline $\sigma_1$ & $\Freq$  \\ \hline
0 & 9 \\  
2 & 4 \\ \hline
\end{tabular}}
\\ \hline
$\emptyset$
&
$\{x_2\}$
&\mystrut{\begin{tabular}{c}
\pstree{\TR{$x_1$}}
{
\TR{$\sigma_2$}
\TR{$x_3$}
}
\end{tabular}}
&
\FreqEnv{\text{ nodes of }G}
&
\mystrut{\begin{tabular}{|c|c|}
\hline $\sigma_2$ & $\Freq$  \\ \hline
1 & 3 \\
2 & 3 \\
3 & 3 \\  
4 & 3 \\ \hline
\end{tabular}}
\\ \hline
$\emptyset$
&
$\{x_3\}$
&\mystrut{\begin{tabular}{c}
\pstree{\TR{$x_1$}}
{
\TR{$x_2$}
\TR{$\sigma_3$}
}
\end{tabular}}
&
\FreqEnv{\text{ nodes of }G}
&
\mystrut{\begin{tabular}{|c|c|}
\hline $\sigma_3$ & $\Freq$  \\ \hline
1 & 3 \\
2 & 3 \\
3 & 3 \\  
4 & 3 \\ \hline
\end{tabular}}
\\ \hline
$\emptyset$
&
$\{x_1, x_2\}$
&
\mystrut{\begin{tabular}{c}
\pstree{\TR{$\sigma_1$}}
{
\TR{$\sigma_2$}
\TR{$x_3$}
}
\end{tabular}}
&
\FreqEnv{\freqtab{\emptyset}{\{x_1\}} \\ \Join \freqtab{\emptyset}{\{x_2\}}}
&
\mystrut{\begin{tabular}{|c|c|c|}
\hline $\sigma_1$ & $\sigma_2$ & $\Freq$  \\ \hline
0 & 1 & 3 \\
0 & 2 & 3 \\
0 & 3 & 3 \\ \hline
\end{tabular}}
\\ \hline
$\emptyset$
&
$\{x_1, x_3\}$
&\mystrut{\begin{tabular}{c}
\pstree{\TR{$\sigma_1$}}
{
\TR{$x_2$}
\TR{$\sigma_3$}
}
\end{tabular}}
&
\FreqEnv{\freqtab{\emptyset}{\{x_1\}} \\ \Join \freqtab{\emptyset}{\{x_3\}}}
&
\mystrut{\begin{tabular}{|c|c|c|}
\hline $\sigma_1$ & $\sigma_3$ & $\Freq$  \\ \hline
0 & 1 & 3 \\
0 & 2 & 3 \\
0 & 3 & 3 \\ \hline
\end{tabular}}
\\ \hline
$\emptyset$
&
$\{x_2, x_3\}$
&\mystrut{\begin{tabular}{c}
\pstree{\TR{$x_1$}}
{
\TR{$\sigma_2$}
\TR{$\sigma_3$}
}
\end{tabular}}
&
\FreqEnv{\freqtab{\emptyset}{\{x_2\}} \\ \Join \freqtab{\emptyset}{\{x_3\}}}
&
$\emptyset$ 
\\ \hline
$\emptyset$
&
$\{x_1, x_2, x_3\}$
& 
\mystrut{
\begin{tabular}{c}
\pstree{\TR{$\sigma_1$}}
{
\TR{$\sigma_2$}
\TR{$\sigma_3$}
}
\end{tabular}}
&
\multicolumn{2}{|c|}{Pruned}
\\ \hline
$\{x_1\}$
&
$\emptyset$
&
\mystrut{
\begin{tabular}{c}
\pstree{\TR{$\exists$}}
{
\TR{$x_2$}
\TR{$x_3$}
}
\end{tabular}}
&
&
\mystrut{
\begin{tabular}{|c|}
\hline $\Freq$  \\ \hline
$14$ \\  \hline
\end{tabular}}
\\ \hline
$\{x_1\}$ 
& 
$\{x_2\}$
&
\mystrut{\begin{tabular}{c}
\pstree{\TR{$\exists$}}
{
	\TR{$\sigma_2$}
	\TR{$x_3$}
}
\end{tabular}}
&
\FreqEnv{\freqtab{\emptyset}{\{ x_2 \}} \\ \Join \freqtab{\{x_1\}}{\{ \emptyset \}}}
&
\mystrut{\begin{tabular}{|c|c|}
\hline $\sigma_2$ & $\Freq$  \\ \hline
1 & 3 \\
2 & 3 \\
3 & 3 \\ \hline
\end{tabular}}
\\ \hline
$\{x_1\}$ & $\{x_3\}$
&
\mystrut{\begin{tabular}{c}
\pstree{\TR{$\exists$}}
{
\TR{$x_2$}
\TR{$\sigma_3$}
}
\end{tabular}}
&
\FreqEnv{\freqtab{\emptyset}{\{ x_3 \}} \\ \Join \freqtab{\{x_1\}}{\{ \emptyset \}}}
&
\mystrut{\begin{tabular}{|c|c|}
\hline $\sigma_3$ & $\Freq$  \\ \hline
1 & 3 \\
2 & 3 \\
3 & 3 \\ \hline
\end{tabular}}
\\ \hline
$\{x_1\}$ & $\{x_2, x_3\}$
&
\mystrut{\begin{tabular}{c}
\pstree{\TR{$\exists$}}
{
\TR{$\sigma_2$}
\TR{$\sigma_3$}
}
\end{tabular}}
& 
\FreqEnv{\freqtab{\emptyset}{\{ x_2, x_3 \}} \\ \Join \freqtab{\{x_1\}}{\{ x_2 \}} \\ \Join \freqtab{\{x_1\}}{\{ x_3 \}}}
& 
$\emptyset$ 
\\ \hline
$\{x_2\}$
&
$\emptyset$
&
\mystrut{\begin{tabular}{c}
\pstree{\TR{$x_1$}}
{
\TR{$\exists$}
\TR{$x_3$}
}
\end{tabular}}
&
&
\mystrut{ \begin{tabular}{|c|}
\hline $\Freq$  \\ \hline
7 \\ \hline
\end{tabular}}
\\ \hline
$\{x_2\}$
&
$\{x_1\}$
&
\mystrut{\begin{tabular}{c}
\pstree{\TR{$\sigma_1$}}
{
\TR{$\exists$}
\TR{$x_3$}
}
\end{tabular}}
&
\FreqEnv{\freqtab{\emptyset}{\{ x_1 \}} \\ \Join \freqtab{\{x_2\}}{\{ \emptyset \}}}
&
\mystrut{\begin{tabular}{|c|c|}
\hline $\sigma_1$ & $\Freq$  \\ \hline
0 & 3 \\ \hline
\end{tabular}}
\\ \hline
$\{x_2\}$
&
$\{x_3\}$
&
\mystrut{\begin{tabular}{c}
\pstree{\TR{$x_1$}}
{
\TR{$\exists$}
\TR{$\sigma_3$}
}
\end{tabular}}
&
\FreqEnv{\freqtab{\emptyset}{\{ x_3 \}} \\ \Join \freqtab{\{x_2\}}{\{ \emptyset \}}}
&
$\emptyset$
\\ \hline
$\{x_2\}$
&
$\{x_1, x_3\}$
&
\mystrut{\begin{tabular}{c}
\pstree{\TR{$\sigma_1$}}
{
\TR{$\exists$}
\TR{$\sigma_3$}
}
\end{tabular}}
&
\multicolumn{2}{|c|}{Pruned}
\\ \hline
$\{x_3\}$
&
$\emptyset$
&
\mystrut{\begin{tabular}{c}
\pstree{\TR{$x_1$}}
{
\TR{$x_2$}
\TR{$\exists$}
}
\end{tabular}}
&
&
\mystrut{
\begin{tabular}{|c|}
\hline $\Freq$  \\ \hline
7 \\ \hline
\end{tabular}}
\\ \hline
$\{x_3\}$
&
$\{x_1\}$
&
\mystrut{\begin{tabular}{c}
\pstree{\TR{$\sigma_1$}}
{
\TR{$\exists$}
\TR{$x_3$}
}
\end{tabular}}
&
\FreqEnv{\freqtab{\emptyset}{\{ x_1 \}} \\ \Join \freqtab{\{x_3\}}{\{ \emptyset \}}}
&
\mystrut{\begin{tabular}{|c|c|}
\hline $\sigma_1$ & $\Freq$  \\ \hline
0 & 3 \\ \hline
\end{tabular}}
\\ \hline
$\{x_3\}$
&
$\{x_2\}$
&
\mystrut{\begin{tabular}{c}
\pstree{\TR{$x_1$}}
{
\TR{$\sigma_2$}
\TR{$\exists$}
}
\end{tabular}}
&
\FreqEnv{\freqtab{\emptyset}{\{ x_2 \}} \\ \Join \freqtab{\{x_3\}}{\{ \emptyset \}}}
&
$\emptyset $
\\ \hline
$\{x_3\}$
&
$\{x_1, x_2\}$
&\mystrut{\begin{tabular}{c}
\pstree{\TR{$\sigma_1$}}
{
\TR{$\sigma_2$}
\TR{$\exists$}
}
\end{tabular}}
&
\multicolumn{2}{|c|}{Pruned}
\\ \hline
$\{x_1, x_2\}$
&
$\emptyset$
&
\mystrut{\begin{tabular}{c}
\pstree{\TR{$\exists$}}
{
\TR{$\exists$}
\TR{$x_3$}
}
\end{tabular}}
&
&
\mystrut{\begin{tabular}{|c|}
\hline $\Freq$  \\ \hline
6 \\ \hline
\end{tabular}}
\\ \hline
$\{x_1, x_2\}$
&
$\{x_3\}$
&
\mystrut{\begin{tabular}{c}
\pstree{\TR{$\exists$}}
{
\TR{$\exists$}
\TR{$\sigma_3$}
}
\end{tabular}}
&
\FreqEnv{\freqtab{\{x_1\}}{\{ x_3 \}} \\ \Join \freqtab{\{x_2\}}{\{ x_3 \}} \\ \Join \freqtab{\{x_1, x_2\}}{\emptyset}}
&
$\emptyset$
\\ \hline
$\{x_1, x_3\}$
&
$\emptyset$
&
\mystrut{\begin{tabular}{c}
\pstree{\TR{$\exists$}}
{
\TR{$x_2$}
\TR{$\exists$}
}
\end{tabular}}
&
&
\mystrut{\begin{tabular}{|c|}
\hline $\Freq$  \\ \hline
6 \\ \hline
\end{tabular}}
\\ \hline
$\{x_1, x_3\}$
&
$\{x_2\}$
&
\mystrut{\begin{tabular}{c}
\pstree{\TR{$\exists$}}
{
\TR{$\sigma_2$}
\TR{$\exists$}
}
\end{tabular}}
&
\FreqEnv{\freqtab{\{x_1\}}{\{ x_2 \}} \\ \Join \freqtab{\{x_3\}}{\{ x_2 \}} \\ \Join \freqtab{\{x_1, x_3\}}{\emptyset}}
&
$\emptyset$
\\ \hline
$\{x_2, x_3\}$
&
$\emptyset$
&
\mystrut{\begin{tabular}{c}
\pstree{\TR{$x_1$}}
{
\TR{$\exists$}
\TR{$\exists$}
}
\end{tabular}}
&
&
\mystrut{\begin{tabular}{|c|}
\hline $\Freq$  \\ \hline
4 \\ \hline
\end{tabular}}
\\ \hline
$\{x_2, x_3\}$
&
$\{x_1\}$
&
\mystrut{\begin{tabular}{c}
\pstree{\TR{$\sigma_1$}}
{
\TR{$\exists$}
\TR{$\exists$}
}
\end{tabular}}
&
\FreqEnv{\freqtab{\{x_2, x_3\}}{\emptyset} \\ \Join \freqtab{\{x_2\}}{\{ x_1 \}} \\ \Join \freqtab{\{x_3\}}{\{ x_2 \}}}
&
$\emptyset$
\\ \hline
\end{longtable}
}

\subsection{Equivalence among Tree Patterns}
\label{equivalentQueries}
In this Section we make a number of modifications to the algorithm described so far,  so as to avoid duplicate work. 

As an example of duplicate work, consider the parameterized tree pattern $P_1$ from the example run in Section~\ref{sec_example} ($\Pi=\{x_2\}$ and $\Sigma = \emptyset$): 
\begin{center}
\pstree{\TR{$x_1$}}
{
	\TR{$\exists$}
	\TR{$x_3$}
}
\end{center} 
and the parameterized tree pattern $P_2$: 
\begin{center}
\pstree{\TR{$x_1$}}
{
	\TR{$x_2$}
}
\end{center} 
Clearly, $P_1$ and $P_2$ have the same answer set for all data graphs $G$, up to renaming of the distinguished variables ($x_2$ by $x_3$).  However, these patterns have different underlying trees, and hence Algorithm~\ref{figalgo} will compute the answer set for both patterns (line~$6$). The answer set of $P_1$ is computed before the answer set of $P_2$, since our algorithm is incremental in the number of nodes of $T$. Hence, we can conclude that our algorithm performs some duplicate work which we want to avoid.

Another example of duplicate work our algorithm performs: Consider the parameterized tree pattern $P_3$ from the example run in Section~\ref{sec_example} ($\Pi=\{x_1\}$ and $\Sigma = \{x_2\}$): 
\begin{center}
\pstree{\TR{$\exists$}}
{
	\TR{$\sigma_2$}
	\TR{$x_3$}
}
\end{center} 
and the parameterized tree pattern $P_4$ also from the example run in Section~\ref{sec_example} ($\Pi=\{x_1\}$ and $\Sigma = \{x_3\}$): 
\begin{center}
\pstree{\TR{$\exists$}}
{
	\TR{$x_2$}
	\TR{$\sigma_3$}
}
\end{center} 
As one can see in Section~\ref{sec_example}, these two parameterized patterns have the same instantiations for all data graphs $G$, up to renaming of the parameters ($\sigma_2$ by $\sigma_3$), and for each instantiantion, the same answer set for all data graphs $G$, up to renaming of the distinguished variables ($x_3$ by $x_2$).  However, when we look at the outline of our algorithm in Algorithm~\ref{figalgo}, we see that for both patterns the candidacy and frequency tables are computed between line~$11$ and line~$17$. Hence, we can conclude again that our algorithm performs duplicate work that we want to avoid.

In the rest of this Section we formalize the duplicate work our algorithm performs, and we make a number of modifications to the algorithm described so far, so as to avoid the duplicate work. 

\subsubsection{Equivalency}

Intuitively we call two parameterized tree patterns \emph{equivalent} if they have the same answer sets and the same parameter assignments for all data graphs $G$, up to renaming of the parameters and the distinguished variables. For instance, the parameterized tree patterns $P_1$ and $P_2$ from above we call equivalent, as the tree patterns $P_3$ and $P_4$ from above. 

To define equivalent parameterized tree patterns formally we introduce the notion of $\mathit{(\delta, \rho)}$\textit{-equivalence}.

\paragraph*{$\mathbf{(\delta, \rho)}$-Equivalence} Let $P_1$ and $P_2$ be two parameterized tree patterns and $\rho$ a parameter correspondence from $P_1$ to $P_2$ (recall Section~\ref{sec_treequery}). We define an \emph{answer set correspondence} from $P_1$ to $P_2$ as any mapping $\delta: \Delta_1 \rightarrow \Delta_2$. Furthermore, assume that $\delta$ and $\rho$ are bijections. We then say that $P_1$ and $P_2$ are $\mathbf{(\delta, \rho)}$-\emph{equivalent}, denoted by $P_1 \equiv_{\rho}^{\delta} P_2$, if for all data graphs $G$, and all parameter assignments $\alpha_2: \Sigma_2 \rightarrow U$, we have $P_2^{\alpha_{2}}(G) \circ \delta  = P_1^{\alpha_2 \circ \rho}(G)$, where $P_2^{\alpha_{2}}(G) \circ \delta$ denotes the set $\{f \circ \delta :\ f \in \Pinst{2}(G)\}$. 

\begin{figure}
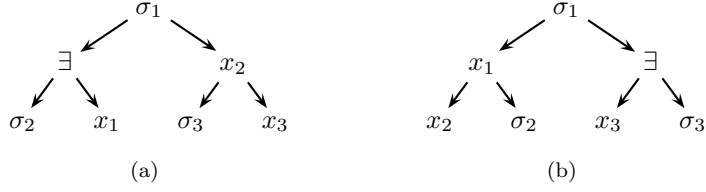

\hspace*{\fill}
\subfigure[]{\label{figisomorf1}
\begin{tabular}{c}
\pstree{\TR{$\sigma_1$}}
{
  \pstree{\TR{$\exists$}}
  {
    \TR{$\sigma_2$}
    \TR{$x_1$}
  }
  \pstree{\TR{$x_2$}}
  {
    \TR{$\sigma_3$}
    \TR{$x_3$}
  }
}
\end{tabular}
}
\hfill
\subfigure[]{\label{figisomorf2}
\begin{tabular}{c}
\pstree{\TR{$\sigma_1$}}
{
  \pstree{\TR{$x_1$}}
  {
    \TR{$x_2$}
    \TR{$\sigma_2$}
  }
  \pstree{\TR{$\exists$}}
  {
    \TR{$x_3$}
    \TR{$\sigma_3$}
  }
}
\end{tabular}
}
\hspace*{\fill}
\caption{Two equivalent parameterized tree patterns.}
\label{figisomorf}
\end{figure}

For example, consider the two parameterized tree patterns in Figure~\ref{figisomorf}, and let $\rho: \Sigma_a \rightarrow \Sigma_b$ be as follows:

\begin{center}
\begin{tabular}{|c|c|}
\hline \multicolumn{2}{|c|}{$\rho$}  \\ \hline
$\sigma_{1}$ & $\sigma_{1}$ \\  
$\sigma_{2}$ & $\sigma_{3}$ \\ 
$\sigma_{3}$ & $\sigma_{2}$ \\ \hline
\end{tabular}
\end{center} and let $\delta: \Delta_a \rightarrow \Delta_b$ be as follows: 

\begin{center}
\begin{tabular}{|c|c|}
\hline \multicolumn{2}{|c|}{$\delta$}  \\ \hline
$x_{1}$ & $x_{3}$ \\  
$x_{2}$ & $x_{1}$ \\ 
$x_{3}$ & $x_{2}$ \\ \hline
\end{tabular}
\end{center}

\begin{figure}
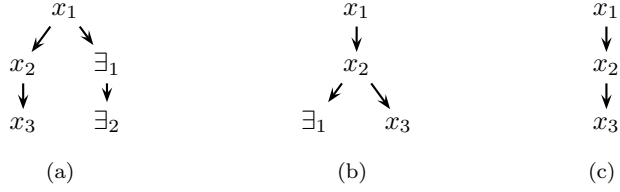

\begin{center}
\hspace*{\fill}
\subfigure[]{\label{figredundant1}
\begin{tabular}{c}
\pstree{\TR{$x_1$}}
{
  \pstree{\TR{$x_2$}}{\TR{$x_3$}}
  \pstree{\TR{$\exists_1$}}{\TR{$\exists_2$}}
}
\end{tabular}
}
\hfill
\subfigure[]{\label{figredundant2}
\begin{tabular}{c}
\pstree{\TR{$x_1$}}
{
  \pstree{\TR{$x_2$}}
  {
    \TR{$\exists_1$}
    \TR{$x_3$}
  }
}
\end{tabular}
}
\hfill
\subfigure[]{\label{figredundant3}
\begin{tabular}{c}
\pstree{\TR{$x_1$}}
{
  \pstree{\TR{$x_2$}}
  {
    \TR{$x_3$}
  }
}
\end{tabular}
}
\hspace*{\fill}
\end{center}
\caption{Three equivalent parameterized tree patterns.}
\label{figredundant}
\end{figure}

The two parameterized tree patterns are clearly $(\delta, \rho)$-equivalent, as are the three parameterized tree patterns shown in Figure~\ref{figredundant} with an empty parameter correspondence $\rho$ and $\delta$ the identity.

The parameter correspondence $\rho$ is a bijection in the definition of $\rho$-equivalen\-ce, since intuitively we want equivalent parameterized tree patterns to have essentially the same set of instantiations. Hence it is necessary that the tree patterns have the same number of parameters. Intuitively we also want equivalent tree patterns to have the same answer sets up to renaming of the distinguished variables. That is the reason why an answer set correspondence is introduced that is a bijection.

We then define equivalent parameterized tree patterns as follows:

\paragraph*{Equivalent parameterized tree patterns} We call two parameterized tree patterns $P_1$ and $P_2$ \emph{equivalent} if $P_1$ is $(\delta, \rho)$-equivalent with $P_2$ for some bijective parameter correspondence $\rho$ and some bijective answer set correspondence $\delta$.

\begin{figure}
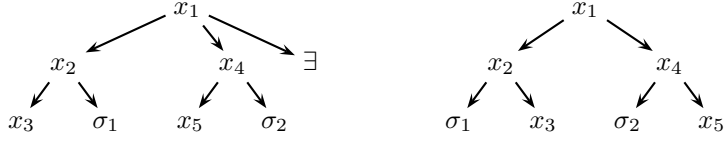

\hspace*{\fill}
\subfigure{\label{figequivalent1}
\begin{tabular}{c}
\pstree{\TR{$x_1$}}
{
  \pstree{\TR{$x_2$}}
  {
    \TR{$x_3$}
    \TR{$\sigma_1$}
  }
  \pstree{\TR{$x_4$}}
  {
    \TR{$x_5$}
    \TR{$\sigma_2$}
  }
  \TR{$\exists$}
}
\end{tabular}
}
\hfill
\subfigure{\label{figequivalent2}
\begin{tabular}{c}
\pstree{\TR{$x_1$}}
{
  \pstree{\TR{$x_2$}}
  {
    \TR{$\sigma_1$}
    \TR{$x_3$}
  }
  \pstree{\TR{$x_4$}}
  {
    \TR{$\sigma_2$}
    \TR{$x_5$}
  }
}
\end{tabular}
}
\hspace*{\fill}
\caption{Two equivalent parameterized tree patterns with more than one possibility for a parameter and answer set correspondence.}
\label{figequivalent}
\end{figure}

Note that there can exist more than one parameter correspondence $\rho$ and more than one answer set correspondence $\delta$ for which the two parameterized tree patterns are $(\delta,\rho)$-equivalent. An illustration of this is given in Figure~\ref{figequivalent}. Let $\rho_1: \Sigma_a \rightarrow \Sigma_b$, $\delta_1: \Delta_a \rightarrow \Delta_b$, $\rho_2: \Sigma_a \rightarrow \Sigma_b$ and $\delta_2: \Delta_a \rightarrow \Delta_b$ be as follows: $\rho_1$ is the identity; $\delta_1$ is the identity and

\begin{center}
\begin{tabular}{|c|c|}
\hline \multicolumn{2}{|c|}{$\rho_2$}  \\ \hline
$\sigma_{1}$ & $\sigma_{2}$ \\  
$\sigma_{2}$ & $\sigma_{1}$ \\ \hline
\end{tabular}\hspace*{1cm}
\begin{tabular}{|c|c|}
\hline \multicolumn{2}{|c|}{$\delta_2$}  \\ \hline
$x_{1}$ & $x_{1}$ \\  
$x_{2}$ & $x_{4}$ \\ 
$x_{3}$ & $x_{5}$ \\
$x_{4}$ & $x_{2}$ \\
$x_{5}$ & $x_{3}$ \\ \hline
\end{tabular}
\end{center}

Then the two tree patterns in Figure~\ref{figequivalent} are clearly $(\delta_1, \rho_1)$-equivalent and $(\delta_2, \rho_2)$-equivalent.

Equivalence as just defined is a semantical property, referring to all possible data graphs, and it is not immediately clear how one could decide this property syntactically. The required syntactical notion is given by the following Lemma and Corollary.

\begin{lemma}
\label{lemma_equiv1}
Consider two parameterized tree patterns $P_1$ and $P_2$, $\delta: \Delta_1 \rightarrow \Delta_2$ a bijective answer set correspondence, and $\rho: \Sigma_1 \rightarrow \Sigma_2$ a bijective parameter correspondence. Then $P_1 \equiv_{\rho}^{\delta} P_2$ if and only if we have the following containment relations among the tree queries $(H_1, P_1)$ and $(\delta(H_1), P_2)$, with $H_1$ the pure head of $P_1$ (cfr. Section~\ref{sec_probred}):
\begin{enumerate}
	\item $(\delta(H_1), P_2) \subseteq_{\rho} (H_1, P_1)$; and
 	\item $(H_1, P_1) \subseteq_{\rho^{-1}} (\delta(H_1), P_2)$
\end{enumerate}
\end{lemma}
\begin{proof}
Let us start with the if direction. We need to prove that for every parameter assignment $\alpha_2$ for $P_2$, and every data graph $G$ that $\Pinst{2}(G) \circ \delta = P_1^{\alpha_2 \circ \rho}(G)$.  We know that $(\delta(H_1), P_2)^{\alpha_2}(G) \subseteq (H_1, P_1)^{\alpha_2 \circ \rho}(G)$ since $(\delta(H_1), P_2) \subseteq_{\rho} (H_1, P_1)$. We may rewrite this as: $\Pinst{2}(G) \circ \delta \subseteq P_1^{\alpha_2 \circ \rho}(G)$ since $H_1$ is an enumeration of $\Delta_1$. 

We also know that $(H_1, P_1)^{\alpha_1}(G) \subseteq (\delta(H_1), P_2)^{\alpha_1 \circ \rho^{-1}}(G)$ for every parameter assignment $\alpha_1$ for $P_1$ since $(H_1, P_1) \subseteq_{\rho^{-1}} (\delta(H_1), P_2)$. Now take $\alpha_1 = \alpha_2 \circ \rho$. We then have $(H_1, P_1)^{\alpha_2 \circ \rho}(G) \subseteq (\delta(H_1), P_2)^{\alpha_2}(G)$. Again since $H_1$ is an enumeration of $\Delta_1$ we may rewrite this as: $P_1^{\alpha_2 \circ \rho}(G) \subseteq P_2^{\alpha_2}(G) \circ \delta$. Hence we can conclude that $\Pinst{2}(G) \circ \delta = P_1^{\alpha_2 \circ \rho}(G)$.

Let us then look at the only-if direction. To prove that $(\delta(H_1), P_2) \subseteq_{\rho} (H_1, P_1)$, we will show that for every $\alpha_2$ parameter assignment for $P_2$, and every data graph $G$, we have $(\delta(H_1), P_2)^{\alpha_2}(G) \subseteq (H_1, P_1)^{\alpha_2 \circ \rho}(G)$. Since $\Pinst{2}(G) \circ \delta = P_1^{\alpha_2 \circ \rho}(G)$, we have $(\delta(H_1), P_2)^{\alpha_2}(G) = (H_1, P_1)^{\alpha_2 \circ \rho}(G)$, and hence clearly $(\delta(H_1), P_2)^{\alpha_2}(G) \subseteq (H_1, P_1)^{\alpha_2 \circ \rho}(G)$.

To prove that $(H_1, P_1) \subseteq_{\rho^{-1}} (\delta(H_1), P_2)$, we will show that for every $\alpha_1$ parameter assignment for $P_1$, and every data graph $G$, we have $(H_1, P_1)^{\alpha_1}(G) \subseteq (\delta(H_1), P_2)^{\alpha_1 \circ \rho^{-1}}(G)$. We know that for every $\alpha_2$ parameter assignment for $P_2$, we have $\Pinst{2}(G) \circ \delta = P_1^{\alpha_2 \circ \rho}(G)$. Now take $\alpha_2 = \alpha_1 \circ \rho^{-1}$. We then have $P^{\alpha_1 \circ \rho}_{2}(G) \circ \delta = P_1^{\alpha_1}(G)$, hence $(\delta(H_1), P_2)^{\alpha_1 \circ \rho^{-1}}(G) = (H_1, P_1)^{\alpha_1}(G)$, hence clearly $(H_1, P_1)^{\alpha_1}(G) \subseteq (\delta(H_1), P_2)^{\alpha_1 \circ \rho^{-1}}(G)$. \quad
\end{proof}

\begin{cor}
\label{lemma_equivCont}
Consider two parameterized tree patterns $P_1$ and $P_2$. Then $P_1$ is equivalent with $P_2$ if and only if there exist:
\begin{enumerate}
	\item a bijective answer set correspondence $\delta: \Delta_1 \rightarrow \Delta_2$;
 	\item a bijective parameter correspondence $\rho: \Sigma_1 \rightarrow \Sigma_2$;
	\item a $\rho$-containment mapping $f_1: (H_1, P_1) \rightarrow (\delta(H_1), P_2)$; and
	\item a $\rho^{-1}$-containment mapping $f_2: (\delta(H_1), P_2) \rightarrow (H_1, P_1)$.
\end{enumerate}
with $H_1$ the pure head for $P_1$.
\end{cor}

Of course, we want to avoid that our algorithm considers some parameterized tree pattern $P_{2}$ if
it is equivalent to an earlier considered parameterized tree pattern $P_{1}$.
Since our algorithm generates trees in increasing sizes, there are two cases to consider:
\begin{description}
\item[Case A:] $P_{1}$ has fewer nodes than $P_{2}$.
\item[Case B:] $P_{1}$ and $P_{2}$ have the same number of nodes.
\end{description}

Armed with the above Lemma and Corollary, we can now analyze the above two cases.

\subsubsection{Case A: Redundancy checking}

Let us start by defining the notion of a redundancy.

\paragraph*{Redundant subtree}

A \emph{redundant subtree} $R$, is a subtree of a parameterized tree pattern $P$, such that removing $R$ from $P$ yields a parameterized tree pattern $P'$ that is equivalent with $P$. 

For example, the first two parameterized tree patterns in Figure~\ref{figredundant} indeed contain a redundant subtree.

The following lemma shows that two parameterized tree patterns with different numbers of nodes can only be equivalent if the largest one contains redundant subtrees:

\begin{lemma}
Consider two parameterized tree patterns $P$ and $P'$, and the number of nodes of $P'$ is smaller than the number of nodes of $P$. Then $P$ can only be equivalent with $P'$ if $P$ contains redundant subtrees.
\end{lemma}
\begin{proof}
Since $P$ and $P'$ are equivalent we know from Corollary~\ref{lemma_equivCont} that the following exist:
\begin{enumerate}
	\item an answer set correspondence $\delta: \Delta \rightarrow \Delta'$ that is a bijection;
	\item a parameter correspondence $\rho: \Sigma \rightarrow \Sigma'$ that is a bijection;
	\item a $\rho$-containment mapping $f_1: (H,P) \rightarrow (\delta(H), P')$; and
	\item a $\rho^{-1}$-containment mapping $f_2: (\delta(H), P') \rightarrow (H,P)$.
\end{enumerate}
with $H$ the pure head for $P$.
Since the number of nodes of $P'$ is smaller than the number of nodes of $P$, we know that some subtrees $R$ of $P$ are not in the range of $f_2$. We will prove that these subtrees $R$ are redundant subtrees, by showing that $P$ and $P-R$ are equivalent. 

Since the containment mappings $f_1$ and $f_2$ exist, we know that in particular the following containment mappings will exist:
\begin{enumerate}
	\item $g_1 = f_1|_{P-R}$, a $\rho$-containment mapping from $(H, P-R)$ to $(\delta(H), P')$, and
	\item $g_2 = f_2$, a $\rho^{-1}$-containment mapping from $(\delta(H), P')$ to $(H, P-R)$.
\end{enumerate}
with $\delta$ and $\rho$ as above.

Let us now look at the following mappings: $h_1 = g_2 \circ f_1$ and $h_2 = f_2 \circ g_1$. 
By Lemma~\ref{lemma_comp}, $h_1 = g_2 \circ f_1$ and $h_2 = f_2 \circ g_1$ are identity-containment mappings. 

Using Corollary~\ref{lemma_equivCont} we can now conclude that $P$ and $P-R$ are (identity, identity)-equivalent and thus $R$ is a redundant subtree. \quad

\end{proof}

From this lemma follows that Case A can only happen if $P_2$ contains redundant subtrees. Hence, if we can avoid redundancies, Case A will never occur. 

The following lemma provides us with an efficient check for redundancies.

\newtheorem*{redundancylemma}{Redundancy Lemma}
\begin{redundancylemma}
Let $P$ be a parameterized tree pattern.  Then $P$
has a redundancy if and only it contains a subtree $C$ in the form of a linear
chain of existential nodes (possibly just a single node),
such that the parent of $C$ has another subtree
that is at least as deep as $C$.
\end{redundancylemma}

\begin{figure}
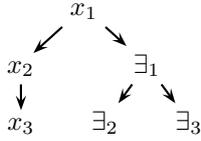

\begin{center}
\pstree{\TR{$x_1$}}
{
  \pstree{\TR{$x_2$}}
  {
    \TR{$x_3$}
  }
  \pstree{\TR{$\exists_1$}}
  {
	\TR{$\exists_2$}
	\TR{$\exists_3$}
  }
}
\end{center}
\caption{A tree pattern that contains a linear chain of existential nodes that is redundant.}
\label{figlinearchain}
\end{figure}

Before we prove this Lemma, let us see some examples. For instance the parameterized tree patterns in Figure~\ref{figredundant1} and Figure~\ref{figredundant2} contain a linear chain of existential nodes that is redundant. In both tree patterns this linear chain is rooted in $\exists_1$. Another example of such a redundancy is given in Figure~\ref{figlinearchain}. Here the linear chain is rooted in $\exists_3$.  Note that when we remove the linear chain rooted in $\exists_3$, we have a new linear chain rooted in $\exists_1$ that is redundant.

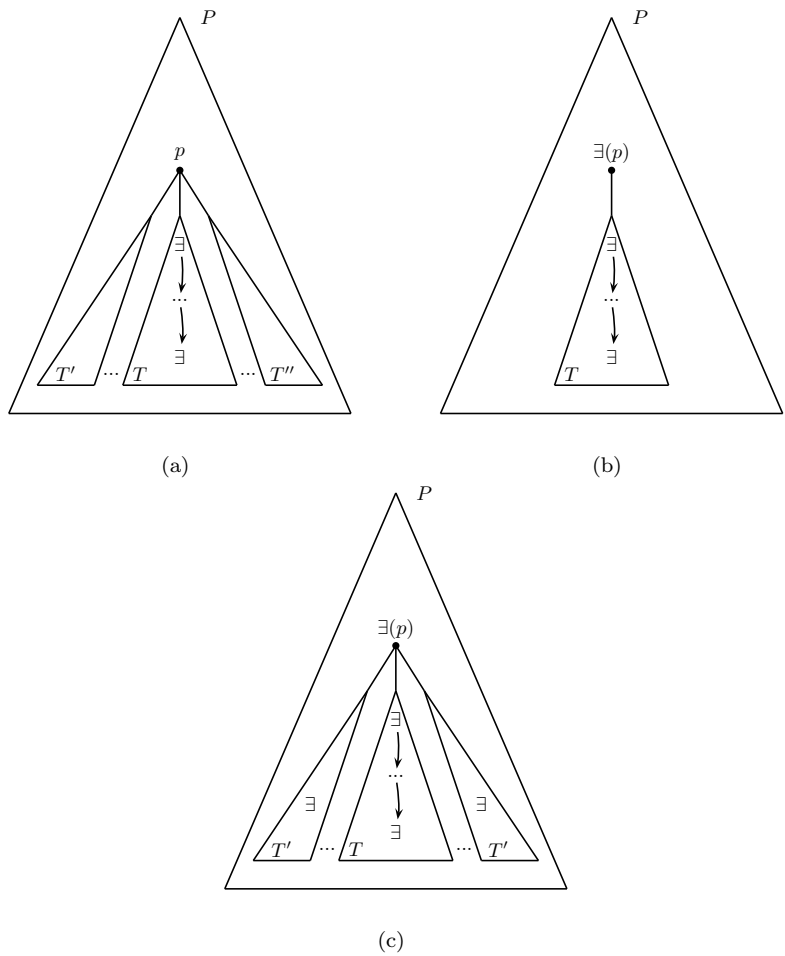
\begin{figure}
\hspace*{\fill}
\subfigure[]{\label{redLemma_case1}
\scalebox{0.75}
{
\begin{pspicture}(0,0)(6,7)
\psline{-}(0,0)(3,7)
\psline{-}(3,7)(6,0)
\psline{-}(0,0)(6,0)
\psline{-}(3.0,3.5)(2, 0.5)
\psline{-}(3.0,3.5)(4, 0.5)
\psline{-}(4,0.5)(2,0.5)
\psline{-}(3,4.3)(3,3.5)
\psline{-}(3,4.3)(2.5,3.5)
\psline{-}(3,4.3)(3.5,3.5)
\psline{-}(2.5,3.5)(1.5,0.5)
\psline{-}(2.5,3.5)(0.5,0.5)
\psline{-}(0.5,0.5)(1.5,0.5)
\psline{-}(3.5,3.5)(4.5,0.5)
\psline{-}(3.5,3.5)(5.5,0.5)
\psline{-}(4.5,0.5)(5.5,0.5)

\rput(3.5,7){\rnode{E}{$P$}}
\rput(2.3,.7){\rnode{F}{$T$}}
\rput(1.0,.7){\rnode{G}{$T'$}}
\rput(4.8,.7){\rnode{H}{$T''$}}
\rput(4.2,.7){\rnode{H}{$...$}}
\rput(1.8,.7){\rnode{H}{$...$}}
\rput(3.0,4.6){\rnode{A}{$p$}}
\dotnode(3.0,4.3){I}
\rput(3.0,3.0){\rnode{B}{$\exists$}}
\rput(3.0,2.0){\rnode{C}{$...$}}
\rput(3.0,1.0){\rnode{D}{$\exists$}}
\ncarc{->}{B}{C}
\ncarc{->}{C}{D}
\end{pspicture}
}
}
\hfill
\subfigure[]{\label{redLemma_case2}
\scalebox{0.75}
{
\begin{pspicture}(0,0)(6,7)
\psline{-}(0,0)(3,7)
\psline{-}(3,7)(6,0)
\psline{-}(0,0)(6,0)
\psline{-}(3.0,3.5)(2, 0.5)
\psline{-}(3.0,3.5)(4, 0.5)
\psline{-}(4,0.5)(2,0.5)
\psline{-}(3,4.3)(3,3.5)
\dotnode(3.0,4.3){I}
\rput(3.0,4.6){\rnode{A}{$\exists (p)$}}

\rput(3.5,7){\rnode{E}{$P$}}
\rput(2.3,.7){\rnode{F}{$T$}}
\rput(3.0,3.0){\rnode{B}{$\exists$}}
\rput(3.0,2.0){\rnode{C}{$...$}}
\rput(3.0,1.0){\rnode{D}{$\exists$}}
\ncarc{->}{B}{C}
\ncarc{->}{C}{D}
\end{pspicture}
}
}
\hspace*{\fill}
\\
\hspace*{\fill}
\subfigure[]{\label{redLemma_case3}
\scalebox{0.75}
{
\begin{pspicture}(0,0)(6,7)
\psline{-}(0,0)(3,7)
\psline{-}(3,7)(6,0)
\psline{-}(0,0)(6,0)
\psline{-}(3.0,3.5)(2, 0.5)
\psline{-}(3.0,3.5)(4, 0.5)
\psline{-}(4,0.5)(2,0.5)
\psline{-}(3,4.3)(3,3.5)
\psline{-}(3,4.3)(2.5,3.5)
\psline{-}(3,4.3)(3.5,3.5)

\psline{-}(2.5,3.5)(1.5,0.5)
\psline{-}(2.5,3.5)(0.5,0.5)
\psline{-}(0.5,0.5)(1.5,0.5)
\psline{-}(3.5,3.5)(4.5,0.5)
\psline{-}(3.5,3.5)(5.5,0.5)
\psline{-}(4.5,0.5)(5.5,0.5)

\rput(3.5,7){\rnode{E}{$P$}}
\rput(2.3,.7){\rnode{F}{$T$}}
\rput(1.0,.7){\rnode{G}{$T'$}}
\rput(1.5,1.5){\rnode{J}{$\exists$}}
\rput(4.5,1.5){\rnode{K}{$\exists$}}
\rput(4.8,.7){\rnode{H}{$T'$}}
\rput(4.2,.7){\rnode{H}{$...$}}
\rput(1.8,.7){\rnode{H}{$...$}}
\rput(3.0,4.6){\rnode{A}{$\exists(p)$}}
\dotnode(3.0,4.3){I}
\rput(3.0,3.0){\rnode{B}{$\exists$}}
\rput(3.0,2.0){\rnode{C}{$...$}}
\rput(3.0,1.0){\rnode{D}{$\exists$}}
\ncarc{->}{B}{C}
\ncarc{->}{C}{D}
\end{pspicture}
}
}
\hspace*{\fill}
\caption{Figures to illustrate the proof of the Redundancy Lemma.}
\label{fig_redundancyLemma}
\end{figure}

\begin{proof}
Let us refer to a subtree $C$ as described in the lemma as an
``eliminable path''.  An eliminable path is clearly redundant, so we only need
to prove the only-if direction.  Let $T$ be a redundant subtree of $P$ that
is maximal, in the sense that it is not the subtree of another redundant
subtree.  Then following Corollary~\ref{lemma_equivCont}, there must be a $\rho$-containment mapping $h$ from $(H,P)$ to $(\delta(H),P-T)$ with $\rho$ and $\delta$ bijections and $H$ the pure head for $P$.  All distinguished variables of $P$ must be in $P-T$, since $\delta$ is a bijection. Also all parameters of $P$ must be in $P-T$, since $\rho$ is also a bijection. So $T$ consists entirely of existential nodes. 

Furthermore, note that $h$ must fix the root of $P$, since the height of $P$ is at least that of $P-T$.

Any iteration $h^n$ of $h$ is a $\rho^{n}$-containment mapping from  $(H, P)$ to $(\delta^n(H),P-T)$ by Lemma~\ref{lemma_comp}. Moreover, each $h^n|_{\Delta \cup \Sigma}$ induces a permutation on the set $\Delta \cup \Sigma$ of distinguished variables and parameters.  Since $\Delta \cup \Sigma$ is finite, there are only a finite number of possible
permutations of $\Delta \cup \Sigma$, namely $|\Delta \cup \Sigma|!$.  Hence, there will be an iteration $h^k|_{\Delta \cup \Sigma}$ and an iteration $h^{(k+l)}|_{\Delta \cup \Sigma}$ such that $h^k|_{\Delta \cup \Sigma} = h^{(k+l)}|_{\Delta \cup \Sigma}$. Hence, $h^l|_{\Delta \cup \Sigma}$ is the identity on $\Delta \cup \Sigma$, because
\begin{align*}
\text{id}|_{\Delta \cup \Sigma} & = (h^{-1})^k|_{\Delta \cup \Sigma} \circ h^k|_{\Delta \cup \Sigma} \\
& = (h^{-1})^k|_{\Delta \cup \Sigma} \circ h^{(k+l)}|_{\Delta \cup \Sigma} \\
& = h^l|_{\Delta \cup \Sigma}
\end{align*}

There are now two possible cases.

\begin{enumerate}
	\item $T$ itself is a linear chain. Let us then look at the parent $p$ of $T$ in $P$. Again there are two possibilities:
	\begin{enumerate}
		\item $h^l(p)=p$: Since $h^l$ is a $\rho^l$-containment mapping from $(H,P)$ to $(\delta^l(H),P-T)$ and $T$ is redundant, we know that $T$ must be mapped to another subtree of $p$, $T'$, that is at least as deep as $T$. Hence, $T$ is an eliminable path. An illustration is given in Figure~\ref{redLemma_case1}. 
		\item $h^l(p)\neq p$: Then $p$ can only be an
		existential node. We now have two possibilities:
		\begin{enumerate}
			\item $T$ is the only subtree of $p$. We will show that the subtree $T'$, rooted in $p$ is redundant as well. Clearly we have the following containment relations:
			\begin{itemize}
			 	\item $h_1 = h^l$, a $\rho^l$-containment mapping from $(H,P)$ to $(\delta^l(H), P-T')$; and
				\item $h_2 = h^{-l}$, a $\rho^{-l}$-containment mapping from $(\delta^l(H),P-T')$ to $(H, P)$.
			\end{itemize} with $\delta$ and $\rho$ as above. 
			By Corollary~\ref{lemma_equivCont}, $T'$ is then a redundant subtree. This is in contraction with the assumption that $T$ is maximal. Hence, it is impossible that $p$ has only one subtree and $p$ is existential. An illustration of this is in given in Figure~\ref{redLemma_case2}.		
			\item $p$ has more than one subtree. Consider such another subtree $T'$. We will show that all subtrees $T'$ of $p$ consist entirely of existential nodes. Suppose a node $n \in T'$ is not an existential node. We then know that $h^l(n) = n$. However, since $h^l$ is a homomorphism and $p$ is an ancestor of $n$, $h^l(p)$ must be $p$. But this is in contradiction with the assumption that $h^l(p)\neq p$. So $T'$ must consist entirely of existential nodes. Hence this brings us to the second case where $T$ is not a linear chain. An illustration is given in Figure~\ref{redLemma_case3}.	
		\end{enumerate}
	\end{enumerate}

	\item $T$ is not a linear chain. An easy induction on the
	height of $T$, shows that any non-linear tree consisting
	entirely of existential nodes must contain an eliminable
	path. If the height of $T$ is $1$, there is an eliminable
	path of a single node: just choose one of the children of
	the root. If the height of $T$ is $n > 1$, consider the
	subtree $S$ of the root of $T$ with the smallest height,
	at most $n-1$. Then we have two possibilities: If $S$ is
	a linear chain, we found our eliminable path. And if $S$ is a non-linear chain we know by induction that $S$ will contain an eliminable path. Hence, $T$, and thus also $P$, contains an eliminable path as desired.
\end{enumerate}

\quad
\end{proof}

As we have seen in Section~\ref{innerLoop}, our algorithm
introduces existential nodes levelwise, one by one.  This makes the redundancy
test provided by the redundancy lemma particularly easy to perform.  Indeed,
if $(\Pi, \Sigma)$ is a parameterized tree patterns of which we already know it has no redundancies, and we make
one additional node $n$ existential, then it suffices to test whether $n$ thus
becomes part of a subtree $C$ as in the Redundancy Lemma.  If so, we will
prune the entire search at $\Pi \cup \{n\}$.

\subsubsection{Case B: Canonical forms}
\label{sec_can_form}

We may now assume that $P_{1}$ and $P_{2}$ do not contain redundancies, for if they would, they would have been dismissed
already.

Let us start by defining isomorphic parameterized tree patterns.
\paragraph*{Isomorphic Parametrized Tree Patterns}
We call two parameterized tree patterns $P_1$ and $P_2$ \emph{isomorphic} if there exists a homomorphism $f: P_1 \rightarrow P_2$ that is a bijection and that maps distinguished nodes to distinguished nodes, parameters to parameters and existential nodes to existential nodes. We call $f$ an \emph{isomorphism}.
Since we are working with trees, $f^{-1}$ is also a homomorphism.

For example, the two parameterized tree patterns in Figure~\ref{figisomorf} are indeed isomorphic with $f$ as follows: 

\begin{center}
\begin{tabular}{|c|c|}
\hline \multicolumn{2}{|c|}{$f$}  \\ \hline
$\sigma_{1}$ & $\sigma_{1}$ \\
$\exists$ & $\exists$ \\
$\sigma_{2}$ & $\sigma_{3}$ \\
$x_{1}$ & $x_{3}$ \\  
$x_{2}$ & $x_{1}$ \\ 
$\sigma_{3}$ & $\sigma_{2}$ \\
$x_{3}$ & $x_{2}$ \\ \hline
\end{tabular}\hspace*{1cm}
\end{center} Clearly, we have the following:

\begin{property}
\label{prop}
Any two isomorphic parameterized tree patterns $P_1$ and $P_2$ are equivalent. 
\end{property}
\begin{proof}
Using Corollary~\ref{lemma_equivCont} we have to show that the following exists:
\begin{enumerate}
	\item a bijective answer set correspondence $\delta: \Delta_1 \rightarrow \Delta_2$;
	\item a bijective parameter correspondence $\rho: \Sigma_1 \rightarrow \Sigma_2$;
	\item a $\rho$-containment mapping $f_1: (H_1, P_1) \rightarrow (\delta(H_1), P_2)$; and
	\item a $\rho^{-1}$-containment mapping $f_2: (\delta(H_1), P_2) \rightarrow (H_1, P_1)$.
\end{enumerate}
with $H_1$ the pure head for $P_1$.

Since $P_1$ and $P_2$ are isomorphic, there exists a homomorphism $f: P_1 \rightarrow P_2$ that is a bijection and that maps distinguished nodes to distinguished nodes, parameters to parameters and existential nodes to existential nodes. 

We now take $\delta = f|_{\Delta_1}$ and $\rho = f|_{\Sigma_1}$. Then $\delta$ and $\rho$ are bijections since $f$ is a bijection. 

For $(3)$ we will show that $f$ is $\rho$-containment mapping from $(H_1, P_1)$ to $(\delta(H_1), P_2)$, with $H_1$ the pure head for $P_1$:
\begin{itemize}
	\item $f$ is a homomorphism;
	\item $f$ maps distinguished nodes to distinguished nodes and $f|_{\Delta_1} = \delta$;
	\item $f$ maps parameters to parameters and $f|_{\Sigma_1} = \rho$; and
	\item $f(H_1) = \delta(H_1)$.
\end{itemize}

For $(4)$ we will show that $f^{-1}$ is $\rho^{-1}$-containment mapping from $(\delta(H_1), P_2)$ to $(H_1, P_1)$, with $H_1$ the pure head for $P_1$:
\begin{itemize}
	\item $f^{-1}$ is a homomorphism since $f$ is a bijection and we are working with trees;
	\item $f^{-1}$ maps distinguished nodes to distinguished nodes and $f^{-1}|_{\Delta_2} = \delta^{-1}$;
	\item $f^{-1}$ maps parameters to parameters and $f^{-1}|_{\Sigma_2} = \rho^{-1}$; and
	\item $f^{-1}(\delta(H_1)) = H_1$.
\end{itemize}

\quad
\end{proof}

The following lemma shows that two parameterized tree patterns without redundancies and with the same number of nodes can only be equivalent if they are isomorphic. 

\newtheorem*{isomorphismlemma}{Isomorphism Lemma}
\begin{isomorphismlemma}
Consider two parameterized tree patterns $P_1$ and $P_2$ without redundancies, and with the same number of nodes. Then $P_1$ and $P_2$ are equivalent if and only if $P_1$ and $P_2$ are isomorphic.
\end{isomorphismlemma}
\begin{proof}
We only need to show the only-if direction. 

Since $P_1$ and $P_2$ are equivalent we know that the following exists by Corollary~\ref{lemma_equivCont}:
\begin{enumerate}
	\item a bijective answer set correspondence $\delta: \Delta_1 \rightarrow \Delta_2$;
 	\item a bijective parameter correspondence $\rho: \Sigma_1 \rightarrow \Sigma_2$;
	\item a $\rho$-containment mapping $f_1: (H_1, P_1) \rightarrow (\delta(H_1), P_2)$; and
	\item a $\rho^{-1}$-containment mapping $f_2: (\delta(H_1), P_2) \rightarrow (H_1, P_1)$.
\end{enumerate}
with $H_1$ a pure head for $P_1$. 

We also know that $P_1$ and $P_2$ have the same number of existential nodes since $P_1$ and $P_2$ have the same number of nodes and $\rho$ and $\delta$ are bijections.

Hence to prove that $P_1$ and $P_2$ are isomorphic, we only need to show that:
\begin{enumerate}
	\item $f_1$ maps existential nodes to existential nodes; and
	\item $f_1|_{\Pi_1}$ is a bijection. 
\end{enumerate}

Thereto, it suffices to prove that $f_1$ is surjective on the existential nodes of $P_2$, because $f_1$ is already a bijection from $\Delta_1 \cup \Sigma_1$ to $\Delta_2 \cup \Sigma_2$.

Assume that $f_1|_{\Pi_1}$ is not surjective. Hence, there will be some existential nodes $p \in \Pi_2$ that are not in the range of $f_1$. Note that these existential nodes $p$ can never have descendants that are parameters or distinguished nodes since $f_1$ is a homomorphism and $\delta$ and $\rho$ bijections. Now fix some $p$ as high as possible in the tree. Then the entire subtree $R$ rooted in $p$ consists entirely of existential nodes, that are not in the range of $f_1$. We will now show that this subtree $R$ is a redundant subtree in $P_2$. We then have a contradiction since we assume that $P_2$ is redundancy free. 

Since the containment mappings $f_1$ and $f_2$ exist, we know that in particular the following containment mappings will exist:
\begin{enumerate}
	\item $g_1 = f_1$ a $\rho$-containment mapping from $(H_1, P_1)$ to $(\delta(H_1), P_2-R)$, and
	\item $g_2 = f_2|_{P_2-R}$ a $\rho^{-1}$-containment mapping from $(\delta(H_1), P_2-R)$ to $(H_1, P_1)$.
\end{enumerate}
with $\delta$ and $\rho$ as above.

Let us now look at the following mappings $h_1 = f_1 \circ g_2$ and $h_2 = g_1 \circ f_2$. By Lemma~\ref{lemma_comp}, $h_1$ and $h_2$ are identity-containment mappings. Using Corollary~\ref{lemma_equivCont}, we can now conclude that $P_2$ and $P_2-R$ are equivalent, and thus $R$ is a redundant subtree. 
\quad
\end{proof}

From the above lemma it follows that Case~B can only happen if $P_1$ and $P_2$ are actually isomorphic. In particular, $P_1$ and $P_2$ have the same underlying tree.

So, in our algorithm, we need an efficient way to avoid isomorphic parameterized tree patterns
based on the same tree $T$.

Fortunately, there is a standard way to do this, by working with
\emph{canonical forms} of parameterized tree patterns.  Consider a pair $(\Pi,\Sigma)$,
as in Section~\ref{innerLoop}.  We can view this pair as a labeling
of $T$: all nodes in $\Pi$ get the same generic label `$\exists$'; all nodes in
$\Sigma$ get `$\sigma$'; and all distinguished nodes get `$x$'.  We then observe that
the patterns $(\Pi_1,\Sigma_1)$ and $(\Pi_2,\Sigma_2)$ are isomorphic iff
there is a tree isomorphism between the corresponding labeled versions of $T$
that respects the labels.

In order to represent each pair $(\Pi,\Sigma)$ uniquely up to isomorphism, we
can rather straightforwardly refine the canonical ordering of the underlying
unlabeled tree $T$, which we already have (Section~\ref{outerLoop}), to take
into account the node labels.  Furthermore, the classical linear-time
algorithm to canonize a tree \cite{ahu} generalizes straightforwardly to
labeled trees. A nice review of these generalizations has been given by Chi,
Yang and Muntz \cite{muntz_trees}.

We will omit the details of the canonical form; in fact, there are several
ways to realize it.  All that is important is that we can check in linear time
whether a pair is canonical; that a pair can be canonized in linear time; and
that two pairs are isomorphic if and only if their canonical forms are
identical. 

\begin{example}
We can refine the level sequence introduced in
Section~\ref{outerLoop} to a \emph{refined level sequence} for
parameterized tree patterns $P$ as follows: if the tree pattern
$P$ consists of $n$ nodes, then the \emph{refined level sequence}
is now a sequence of $n$ elements, where the $i$th element is the
depth of the $i$th node in preorder in the pattern, followed by a
'd' if the node is distinguished; followed by a 'e' is the node
is existential and followed by a 'p' if the node is a parameter.
The canonical ordering of a parameterized tree pattern $P$, is
then the ordering of $P$ that yields the lexicographically maximal refined level sequence, among all orderings of $P$. Then the refined level sequences for the parameterized tree patterns in Figure~\ref{figisomorf} are:
\begin{itemize}
	\item[(a)] 0p1e2p2d1d2p2d 
	\item[(b)] 0p1d2d2p1e2d2p
\end{itemize}
and (a) is the canonical one. 
\end{example}

Armed by the canonical form, we are now in a position to describe how the
algorithm of Section~\ref{innerLoop} must be modified to avoid
equivalent parameterized tree patterns.  First of all, we only work with patterns $(\Pi,\Sigma)$ in
canonical form; the others are dismissed.  However, the problem then arises that a parent pattern $(\Pi',\Sigma')$, where we omit a variable from either $\Pi$ or $\Sigma$ as described in Section~\ref{innerLoop}, might be non-canonical.  In that case the frequency table for $(\Pi',\Sigma')$ will not exist.  We can solve this by
canonizing $(\Pi',\Sigma')$ to its canonical version $(\Pi'',\Sigma'')$, and remembering the renaming of variables this entails.  The table $\freqtab{\Pi''}{\Sigma''}$ can then serve in place of $\freqtab{\Pi'}{\Sigma'}$, after we have applied the inverse renaming to its column headings.

\begin{figure}
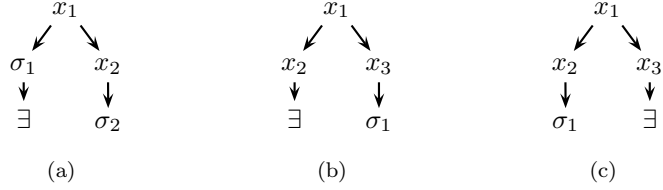

\begin{center}
\hspace*{\fill}
\subfigure[]{\label{figorder1}
\begin{tabular}{c}
\pstree{\TR{$x_1$}}
{
  \pstree{\TR{$\sigma_1$}}{\TR{$\exists$}}
  \pstree{\TR{$x_2$}}{\TR{$\sigma_2$}}
}
\end{tabular}
}
\hfill
\subfigure[]{\label{figorder2}
\begin{tabular}{c}
\pstree{\TR{$x_1$}}
{
  \pstree{\TR{$x_2$}}{\TR{$\exists$}}
  \pstree{\TR{$x_3$}}{\TR{$\sigma_1$}}
}
\end{tabular}
}
\hfill
\subfigure[]{\label{figorder3}
\begin{tabular}{c}
\pstree{\TR{$x_1$}}
{
  \pstree{\TR{$x_2$}}{\TR{$\sigma_1$}}
  \pstree{\TR{$x_3$}}{\TR{$\exists$}}
}
\end{tabular}
}
\hspace*{\fill}
\end{center}
\caption{The tree pattern in (b) is a parent of a tree pattern in (a), and (c) is the canonical version of (b).}
\label{figorder}
\end{figure}

This does not completely solve the problem, however. Indeed the frequency table of $(\Pi'', \Sigma'')$ might not yet have been computed. For example, consider the parameterized tree pattern in Figure~\ref{figorder1}, and one of its parents in Figure~\ref{figorder2}. The canonical version of this parent, using the canonical ordering from the previous example, is shown in Figure~\ref{figorder3}. Using the current order for computing the frequency tables as in Algorithm~\ref{figalgo}, the frequency table for the pattern in Figure~\ref{figorder3} is not yet computed, when we want to compute the frequency for the pattern in Figure~\ref{figorder1}. 

We can solve this by changing the order in which we compute the frequency tables. We work with increasing levels: in level $i$ we compute the frequency tables for all pairs $(\Pi, \Sigma)$, where $\# \Pi + \# \Sigma = i$. If we use this order, we are sure that when we compute the frequency table of a pair $(\Pi,\Sigma)$, all frequency tables of pairs $(\Pi', \Sigma')$ with $(\#\Pi' + \#\Sigma') < (\#\Pi + \#\Sigma)$, have been computed.

\subsubsection{The Algorithm}

The final algorithm is now given in Algorithm~\ref{algo_equiv}. The outline for canonizing a parameterized tree pattern in given in Function~\ref{algo_equiv_canon}.

\begin{algorithm}
\caption{Levelwise search for frequent tree patterns \newline with equivalence checking.}
\label{algo_equiv}
\begin{algorithmic}[1]
\FOR{each unordered, rooted tree $T$}
	\STATE level $:=$ number of nodes of $T$
	\STATE $i := 0$
	\STATE $C_0 := \{ (\emptyset, \emptyset) \}$; $F := \emptyset$
	\WHILE{$i \leq \text{level}$ AND $C_i \neq \emptyset$}
		\STATE \COMMENT{Candidate evaluation}
		\FOR{each pattern $(\Pi,\Sigma)$ in $C_i$}
			\IF{$\Sigma = \emptyset$}
				\STATE Compute $\freqtab \Pi \emptyset$ in SQL
			\ELSE
				\IF{($\#\Sigma = 1 $ AND $\#\Pi = 0$)}
					\STATE $\cantab \Pi \Sigma := \text{set of nodes of $G$}$
				\ELSE
					\STATE $\cantab \Pi \Sigma := \: \Join\{f^{-1}(\freqtab{\Pi''}{\Sigma''}) \mid (\Pi',\Sigma')$ parent of $(\Pi,\Sigma)$ 
					\STATE \hspace{3cm} $\text{ and } (f,(\Pi'', \Sigma''))=\text{\textbf{Canonize}}(\Pi',\Sigma') \}$
				\ENDIF
			\ENDIF
			\STATE Compute $\freqtab \Pi \Sigma$ in SQL
			\IF{($\freqtab \Pi \Sigma \neq \emptyset$)}
					\STATE $F = F \cup \{(\Pi, \Sigma)\}$
			\ENDIF
		\ENDFOR
		\STATE \COMMENT{Candidate generation}
		\STATE $C_{i+1} = \{(\Pi, \Sigma)|\ \text{all parents $(\Pi',\Sigma')$ of $(\Pi,\Sigma)$ are in $F$}\}$
		\STATE \COMMENT{Equivalence Check}
		\FOR{each pattern $(\Pi, \Sigma)$ in $C_{i+1}$}
			\IF{($(\Pi, \Sigma)$ contains a redundancy)}
				\STATE{remove $(\Pi,\Sigma)$ from $C_{i+1}$}
			\ELSIF{($(\Pi, \Sigma)$ is not canonical)}
				\STATE{remove $(\Pi,\Sigma)$ from $C_{i+1}$}
			\ENDIF
		\ENDFOR
		\STATE $i := i+1$
	\ENDWHILE
\ENDFOR
\end{algorithmic}
\end{algorithm}

\floatname{algorithm}{Function}
\begin{algorithm}
\caption{Canonize $(\Pi', \Sigma')$ based on $T$}
\label{algo_equiv_canon}
\begin{algorithmic}[1]
\STATE $(\Pi_c, \Sigma_c):=$ canonization of $(\Pi', \Sigma')$ based on $T$
\STATE $f:=$ isomorphism from $(\Pi_c, \Sigma_c)$ to $(\Pi', \Sigma')$
\RETURN $(f,(\Pi_c, \Sigma_c))$
\end{algorithmic}
\end{algorithm}

\subsubsection{Example run}
\label{sec_example_equiv}

In this Section we give an example run of the final algorithm in Algorithm~\ref{algo_equiv}. We use the same data graph $G$; unordered rooted tree $T$; and minimum support threshold, $3$, as in the example in Section~\ref{sec_example}. 

Note that there are two important differences between this run and the run in Section~\ref{sec_example}:
\begin{enumerate}
	\item duplicate work is avoided: equivalent tree patterns are not generated; and 
	\item the order for computing the tree patterns is different in the sense that here, the tree patterns are generated in levels, as explained in Section~\ref{sec_can_form}.
\end{enumerate}

The example run then looks as follows:

{\footnotesize
\begin{longtable}{|c|c|c|c|c|}
\hline \multicolumn{1}{|c|}{$\Pi$} & \multicolumn{1}{|c|}{$\Sigma$} & \multicolumn{1}{|c|}{$P$} & \multicolumn{1}{|c|}{${\it CanTab}$} & \multicolumn{1}{|c|}{${\it FreqTab}$}\\ \hline
\endfirsthead

\hline
\endlastfoot
\multicolumn{5}{|c|}{\textbf{Level 0}}
\\ \hline
$\emptyset$
&
$\emptyset$
&
\mystrut{\begin{tabular}{c}
\pstree{\TR{$x_1$}}
{
  \TR{$x_2$}
  \TR{$x_3$}
}
\end{tabular}}
&
&
\mystrut{\begin{tabular}{|c|}
\hline $\Freq$  \\ \hline
$15$ \\ \hline
\end{tabular}}
\\ \hline
\multicolumn{5}{|c|}{\textbf{Level 1}}
\\ \hline
$\emptyset$
&
$\{x_1\}$
&\mystrut{\begin{tabular}{c}
\pstree{\TR{$\sigma_1$}}
{
	\TR{$x_2$}
	\TR{$x_3$}
}
\end{tabular}}
&
\FreqEnv{\text{ nodes of }G}
&
\mystrut{\begin{tabular}{|c|c|}
\hline $\sigma_1$ & $\Freq$  \\ \hline
0 & 9 \\  
2 & 4 \\ \hline
\end{tabular}}
\\ \hline
$\emptyset$
&
$\{x_2\}$
&\mystrut{\begin{tabular}{c}
\pstree{\TR{$x_1$}}
{
\TR{$\sigma_2$}
\TR{$x_3$}
}
\end{tabular}}
&
\FreqEnv{\text{ nodes of }G}
&
\mystrut{\begin{tabular}{|c|c|}
\hline $\sigma_2$ & $\Freq$  \\ \hline
1 & 3 \\
2 & 3 \\
3 & 3 \\  
4 & 3 \\ \hline
\end{tabular}}
\\ \hline
$\emptyset$
&
$\{x_3\}$
&\mystrut{\begin{tabular}{c}
\pstree{\TR{$x_1$}}
{
\TR{$x_2$}
\TR{$\sigma_3$}
}
\end{tabular}}
&
\multicolumn{2}{|c|}{Equivalent with $(\emptyset, \{x_2\})$}
\\ \hline
$\{x_1\}$
&
$\emptyset$
&
\mystrut{
\begin{tabular}{c}
\pstree{\TR{$\exists$}}
{
\TR{$x_2$}
\TR{$x_3$}
}
\end{tabular}}
&
&
\mystrut{
\begin{tabular}{|c|}
\hline $\Freq$  \\ \hline
$14$ \\  \hline
\end{tabular}}
\\ \hline
$\{x_2\}$
&
$\emptyset$
&
\mystrut{\begin{tabular}{c}
\pstree{\TR{$x_1$}}
{
\TR{$\exists$}
\TR{$x_3$}
}
\end{tabular}}
&
\multicolumn{2}{|c|}{Redundancy}
\\ \hline
$\{x_3\}$
&
$\emptyset$
&
\mystrut{\begin{tabular}{c}
\pstree{\TR{$x_1$}}
{
\TR{$x_2$}
\TR{$\exists$}
}
\end{tabular}}
&
\multicolumn{2}{|c|}{Equivalent with $(\{x_2\}, \emptyset)$}
\\ \hline
\multicolumn{5}{|c|}{\textbf{Level 2}}
\\ \hline
$\emptyset$
&
$\{x_1, x_2\}$
&
\mystrut{\begin{tabular}{c}
\pstree{\TR{$\sigma_1$}}
{
\TR{$\sigma_2$}
\TR{$x_3$}
}
\end{tabular}}
&
\FreqEnv{\freqtab{\emptyset}{\{x_1\}} \\ \Join \freqtab{\emptyset}{\{x_2\}}}
&
\mystrut{\begin{tabular}{|c|c|c|}
\hline $\sigma_1$ & $\sigma_2$ & $\Freq$  \\ \hline
0 & 1 & 3 \\
0 & 2 & 3 \\
0 & 3 & 3 \\ \hline
\end{tabular}}
\\ \hline
$\emptyset$
&
$\{x_1, x_3\}$
&\mystrut{\begin{tabular}{c}
\pstree{\TR{$\sigma_1$}}
{
\TR{$x_2$}
\TR{$\sigma_3$}
}
\end{tabular}}
&
\multicolumn{2}{|c|}{Equivalent with $(\emptyset, \{x_1, x_2\})$}
\\ \hline
$\emptyset$
&
$\{x_2, x_3\}$
&\mystrut{\begin{tabular}{c}
\pstree{\TR{$x_1$}}
{
\TR{$\sigma_2$}
\TR{$\sigma_3$}
}
\end{tabular}}
&
\FreqEnv{\freqtab{\emptyset}{\{x_2\}} \\ \Join \freqtab{\emptyset}{\{x_3\}}}
&
$\emptyset$ 
\\ \hline
$\{x_1\}$ & $\{x_2\}$
&
\mystrut{\begin{tabular}{c}
\pstree{\TR{$\exists$}}
{
	\TR{$\sigma_2$}
	\TR{$x_3$}
}
\end{tabular}}
&
\FreqEnv{\freqtab{\emptyset}{\{ x_2 \}} \\ \Join \freqtab{\{x_1\}}{\{ \emptyset \}}}
&
\mystrut{\begin{tabular}{|c|c|}
\hline $\sigma_2$ & $\Freq$  \\ \hline
1 & 3 \\
2 & 3 \\
3 & 3 \\ \hline
\end{tabular}}
\\ \hline
$\{x_1\}$ & $\{x_3\}$
&
\mystrut{\begin{tabular}{c}
\pstree{\TR{$\exists$}}
{
\TR{$x_2$}
\TR{$\sigma_3$}
}
\end{tabular}}
&
\multicolumn{2}{|c|}{Equivalent with $(\{x_1\},\{x_2\})$}
\\ \hline
$\{x_2\}$
&
$\{x_1\}$
&
\mystrut{\begin{tabular}{c}
\pstree{\TR{$\sigma_1$}}
{
\TR{$\exists$}
\TR{$x_3$}
}
\end{tabular}}
&
\multicolumn{2}{|c|}{Redundancy}
\\ \hline
$\{x_2\}$
&
$\{x_3\}$
&
\mystrut{\begin{tabular}{c}
\pstree{\TR{$x_1$}}
{
\TR{$\exists$}
\TR{$\sigma_3$}
}
\end{tabular}}
&
\multicolumn{2}{|c|}{Redundancy}
\\ \hline
$\{x_3\}$
&
$\{x_1\}$
&
\mystrut{\begin{tabular}{c}
\pstree{\TR{$\sigma_1$}}
{
\TR{$x_2$}
\TR{$\exists$}
}
\end{tabular}}
&
\multicolumn{2}{|c|}{Equivalent with $(\{x_2\},\{x_1\})$}
\\ \hline
$\{x_3\}$
&
$\{x_2\}$
&
\mystrut{\begin{tabular}{c}
\pstree{\TR{$x_1$}}
{
\TR{$\sigma_2$}
\TR{$\exists$}
}
\end{tabular}}
&
\multicolumn{2}{|c|}{Equivalent with $(\{x_2\},\{x_3\})$}
\\ \hline
$\{x_1, x_2\}$
&
$\emptyset$
&
\mystrut{\begin{tabular}{c}
\pstree{\TR{$\exists$}}
{
\TR{$\exists$}
\TR{$x_3$}
}
\end{tabular}}
&
\multicolumn{2}{|c|}{Redundancy}
\\ \hline
$\{x_1, x_3\}$
&
$\emptyset$
&
\mystrut{\begin{tabular}{c}
\pstree{\TR{$\exists$}}
{
\TR{$x_2$}
\TR{$\exists$}
}
\end{tabular}}
&
\multicolumn{2}{|c|}{Equivalent with $(\{x_1, x_2\},\emptyset)$}
\\ \hline
$\{x_2, x_3\}$
&
$\emptyset$
&
\mystrut{\begin{tabular}{c}
\pstree{\TR{$x_1$}}
{
\TR{$\exists$}
\TR{$\exists$}
}
\end{tabular}}
&
\multicolumn{2}{|c|}{Redundancy}
\\ \hline
\multicolumn{5}{|c|}{\textbf{Level 3}}
\\ \hline
$\emptyset$
&
$\{x_1, x_2, x_3\}$
& 
\mystrut{
\begin{tabular}{c}
\pstree{\TR{$\sigma_1$}}
{
\TR{$\sigma_2$}
\TR{$\sigma_3$}
}
\end{tabular}}
&
\multicolumn{2}{|c|}{Pruned}
\\ \hline
$\{x_1\}$ & $\{x_2, x_3\}$
&
\mystrut{\begin{tabular}{c}
\pstree{\TR{$\exists$}}
{
\TR{$\sigma_2$}
\TR{$\sigma_3$}
}
\end{tabular}}
& 
\FreqEnv{\freqtab{\emptyset}{\{ x_2, x_3 \}} \\ \Join \freqtab{\{x_1\}}{\{ x_2 \}} \\ \Join \freqtab{\{x_1\}}{\{ x_3 \}}}
& 
$\emptyset$ 
\\ \hline
$\{x_2\}$
&
$\{x_1, x_3\}$
&
\mystrut{\begin{tabular}{c}
\pstree{\TR{$\sigma_1$}}
{
\TR{$\exists$}
\TR{$\sigma_3$}
}
\end{tabular}}
&
\multicolumn{2}{|c|}{Pruned}
\\ \hline
$\{x_3\}$
&
$\{x_1, x_2\}$
&\mystrut{\begin{tabular}{c}
\pstree{\TR{$\sigma_1$}}
{
\TR{$\sigma_2$}
\TR{$\exists$}
}
\end{tabular}}
&
\multicolumn{2}{|c|}{Equivalent with $(\{x_2\},\{x_1, x_3\})$}
\\ \hline
$\{x_1, x_2\}$
&
$\{x_3\}$
&
\mystrut{\begin{tabular}{c}
\pstree{\TR{$\exists$}}
{
\TR{$\exists$}
\TR{$\sigma_3$}
}
\end{tabular}}
&
\multicolumn{2}{|c|}{Redundancy}
\\ \hline
$\{x_1, x_3\}$
&
$\{x_2\}$
&
\mystrut{\begin{tabular}{c}
\pstree{\TR{$\exists$}}
{
\TR{$\sigma_2$}
\TR{$\exists$}
}
\end{tabular}}
&
\multicolumn{2}{|c|}{Equivalent with $(\{x_1,x_2\},\{x_3\})$}
\\ \hline
$\{x_2, x_3\}$
&
$\{x_1\}$
&
\mystrut{\begin{tabular}{c}
\pstree{\TR{$\sigma_1$}}
{
\TR{$\exists$}
\TR{$\exists$}
}
\end{tabular}}
&
\multicolumn{2}{|c|}{Redundancy}
\\ \hline
\end{longtable}
}

\subsection{Result Management: Pattern Database}
\label{sec_result_management}

When the algorithm is terminated, its final output consists of a set of frequency tables for each tree $T$ that was investigated. All frequency tables are kept in a relational database that we call the \emph{pattern database}. The pattern database is an ideal platform for an interactive tool for browsing the frequent queries. We developed such a browser called \emph{Certhia} and discuss it in Section~\ref{sec_certhia}.

The pattern database is also an ideal platform for tree query association mining as will be described in Section~\ref{sec_algo_associations}.

\section{Mining Tree-Query Associations}
\label{sec_algo_associations}

In this Section we present an algorithm for mining confident tree-query associations in a large data graph. Recall from Section~\ref{subsec_def_association} that a parameterized association rule (pAR) is something of the form $Q_1 \Rightarrow_{\rho} Q_2$, with $Q_1$ and $Q_2$ parameterized tree queries, $\rho: \Sigma_1 \rightarrow \Sigma_2$ a parameter correspondence, and $Q_2 \subseteq_{\rho} Q_1$. An instantiated association rule (iAR) is a pair $(Q_1 \Rightarrow_{\rho} Q_2,\alpha)$, with $Q_1 \Rightarrow_{\rho} Q_2$ a pAR and $\alpha: \Sigma_2 \rightarrow U$ a parameter assignment for $Q_1 \Rightarrow_{\rho} Q_2$. Also recall that the confidence of an iAR in a data graph $G$ is defined as $\Freq(\Qinst{2})/\Freq(Q_1^{\alpha_2 \circ \rho})$. 

The algorithm presented in this Section finds all iARs of the form $(\qleft \Rightarrow_{\rho} Q_{\text{right}}, \alpha)$ that are confident and frequent in a given data graph $G$ for a given lhs $\qleft$. Before presenting the algorithm we first show that we do not need to tackle the problem in its full generality.

\subsection{Problem Reduction}

In this section we show that, without loss of generality, we can focus on the case where the given lhs tree query $\qleft$ is pure in the sense that was defined in Section~\ref{sec_probred}. We will also show that this restriction can not be imposed on the rhs tree queries to be output. We also make a remark regarding ``free constants'' in the head of a tree query. 

\begin{figure}
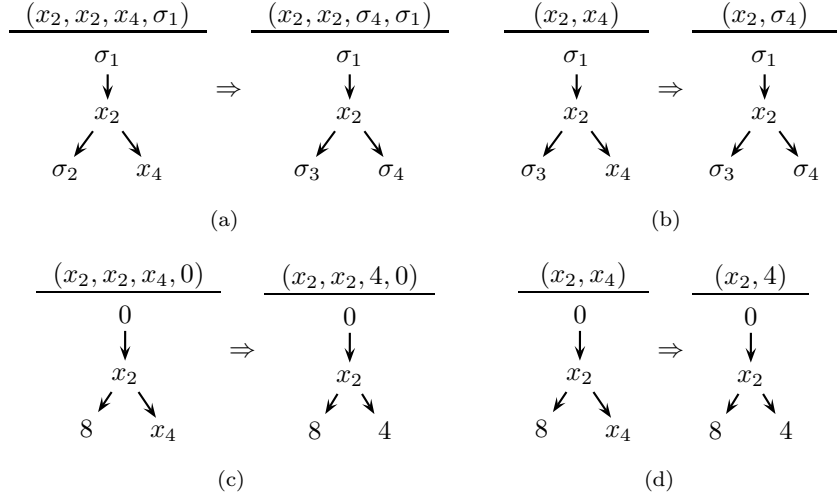

\hspace*{\fill}
\subfigure[]{\label{non_pure_AR}
\begin{tabular}{c}
$(x_2,x_2,x_4,\sigma_1)$\\
\hline
\pstree{\TR{$\sigma_1$}}
{
	\pstree{\TR{$x_2$}}
	{
		\TR{$\sigma_2$}
		\TR{$x_4$}
	}
}
\end{tabular}
$\Rightarrow$
\begin{tabular}{c}
$(x_2,x_2,\sigma_4,\sigma_1)$\\
\hline
\pstree{\TR{$\sigma_1$}}
{
	\pstree{\TR{$x_2$}}
	{
		\TR{$\sigma_3$}
		\TR{$\sigma_4$}
	}
}
\end{tabular}
}
\hfill
\subfigure[]{\label{pure_AR}
\begin{tabular}{c}
$(x_2,x_4)$\\
\hline
\pstree{\TR{$\sigma_1$}}
{
	\pstree{\TR{$x_2$}}
	{
		\TR{$\sigma_3$}
		\TR{$x_4$}
	}
}
\end{tabular}
$\Rightarrow$
\begin{tabular}{c}
$(x_2,\sigma_4)$\\
\hline
\pstree{\TR{$\sigma_1$}}
{
	\pstree{\TR{$x_2$}}
	{
		\TR{$\sigma_3$}
		\TR{$\sigma_4$}
	}
}
\end{tabular}
}
\hspace*{\fill}
\newline
\hspace*{\fill}
\subfigure[]{\label{non_pure_AR_inst}
\begin{tabular}{c}
$(x_2,x_2,x_4,0)$\\
\hline
\pstree{\TR{$0$}}
{
	\pstree{\TR{$x_2$}}
	{
		\TR{$8$}
		\TR{$x_4$}
	}
}
\end{tabular}
$\Rightarrow$
\begin{tabular}{c}
$(x_2,x_2,4,0)$\\
\hline
\pstree{\TR{$0$}}
{
	\pstree{\TR{$x_2$}}
	{
		\TR{$8$}
		\TR{$4$}
	}
}
\end{tabular}
}
\hfill
\subfigure[]{\label{pure_AR_inst}
\begin{tabular}{c}
$(x_2,x_4)$\\
\hline
\pstree{\TR{$0$}}
{
	\pstree{\TR{$x_2$}}
	{
		\TR{$8$}
		\TR{$x_4$}
	}
}
\end{tabular}
$\Rightarrow$
\begin{tabular}{c}
$(x_2,4)$\\
\hline
\pstree{\TR{$0$}}
{
	\pstree{\TR{$x_2$}}
	{
		\TR{$8$}
		\TR{$4$}
	}
}
\end{tabular}
}
\hspace*{\fill}
\caption{Rule (a) has a non-pure lhs. Rule (b) is the purification of rule (a), and expresses precisely the same information. Rules (c) and (d) are two example instantiations.}
\label{fig_pure_nonPure}
\end{figure}

\paragraph*{Pure lhs's} Assume that all possible variables (nodes of tree patterns) have been arranged in some fixed but arbitrary order. Recall then from Section~\ref{sec_probred} that we call a parameterized tree query $Q=(H,P)$ pure when $H$ consists of the enumeration in order and without repetitions of all distinguished variables of $P$. In particular $H$ can not contain parameters. We call $H$ the pure head for $P$. As an illustration, the lhs of rule~(a) of Figure~\ref{fig_pure_nonPure} is impure, while the lhs of rule~(b) is pure.

Consider the pARs in Figure~\ref{non_pure_AR} and Figure~\ref{pure_AR}, and their instantiations in Figure~\ref{non_pure_AR_inst} and Figure~\ref{pure_AR_inst}. The rules in Figure~\ref{non_pure_AR} and Figure~\ref{non_pure_AR_inst} have an impure lhs. If we apply the iARs in Figure~\ref{non_pure_AR_inst} and Figure~\ref{pure_AR_inst} to the data graph $G$ in Figure~\ref{figgraaf1}, both have the same frequency, namely $2$, and the same confidence, namely $33\%$. Indeed, since the frequency of a tree query is in fact the frequency of its body, repetitions of distinguished variables in the head and the occurrence of parameters in the head do not change the frequency of a tree query. In fact the pAR in Figure~\ref{pure_AR} is the purification of the pAR in Figure~\ref{non_pure_AR}: the repetition of the distinguished variable $x_2$ is removed from the heads, and the parameter $\sigma_1$ is removed from the heads.

Hence, a pAR with an impure lhs can always be rewritten to an equivalent pAR with a pure lhs, in such a way that all instantiations of the pAR with the impure lhs correspond to instantiations of the pAR with pure lhs, with the same confidence and frequency. Indeed, take a legal pAR $Q_1 \Rightarrow_{\rho} Q_2$ with $Q_1$ not pure. We know that
$Q_1$'s head is mapped to $Q_2$'s head by some $\rho$-containment mapping. Hence, we can purify $Q_1$ by removing all parameters and repetitions of distinguished variables from $Q_1$'s head, sort the head by the order on the variables, and
perform the corresponding actions on $Q_2$'s head as prescribed by the $\rho$-containment mapping. 

We can conclude that it is sufficient to only consider pARs with pure lhs's.
The rhs, however, need not be pure; impure rhs's are in fact interesting, as we will demonstrate next.

\begin{figure}
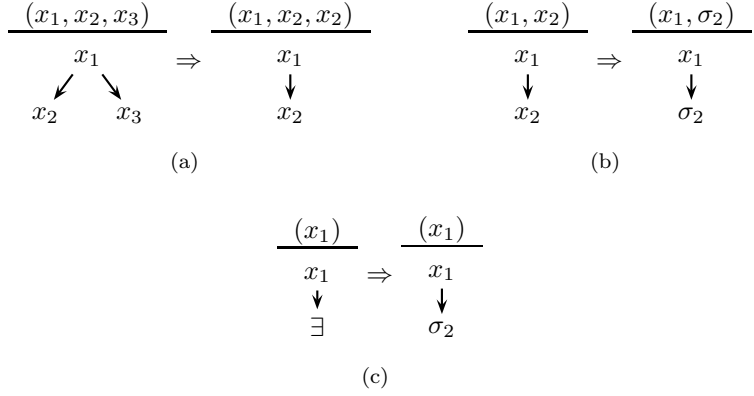

\hspace*{\fill}
\subfigure[]{\label{fig_not_pure_a}
\begin{tabular}{c}
$(x_1,x_2,x_3)$\\
\hline
\pstree{\TR{$x_1$}}
{
	\TR{$x_2$}
	\TR{$x_3$}
}
\end{tabular}
$\Rightarrow$
\begin{tabular}{c}
$(x_1,x_2,x_2)$\\
\hline
\pstree{\TR{$x_1$}}
{
	\TR{$x_2$}
}
\end{tabular}
}
\hfill
\subfigure[]{\label{fig_not_pure_b}
\begin{tabular}{c}
$(x_1,x_2)$\\
\hline
\pstree{\TR{$x_1$}}
{
	\TR{$x_2$}
}
\end{tabular}
$\Rightarrow$
\begin{tabular}{c}
$(x_1,\sigma_2)$\\
\hline
\pstree{\TR{$x_1$}}
{
	\TR{$\sigma_2$}
}
\end{tabular}
}
\hspace*{\fill}
\\
\begin{center}
\subfigure[]{\label{fig_not_pure_d}
\begin{tabular}{c}
$(x_1)$\\
\hline
\pstree{\TR{$x_1$}}
{
	\TR{$\exists$}
}
\end{tabular}
$\Rightarrow$
\begin{tabular}{c}
$(x_1)$\\
\hline
\pstree{\TR{$x_1$}}
{
	\TR{$\sigma_2$}
}
\end{tabular}
}
\end{center}
\caption{(a) and (b) are pARs with impure rhs. (c) is an
ill-advised attempt to purify (b) on the rhs.}
\label{fig_not_pure_rhs}
\end{figure}

\begin{figure}
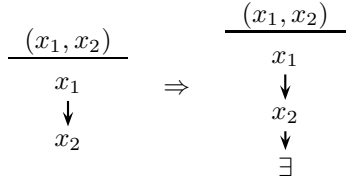

\centering
\begin{tabular}{c}
$(x_1,x_2)$\\
\hline
\pstree{\TR{$x_1$}}
{
	\TR{$x_2$}
}
\end{tabular}
\quad $\Rightarrow$ \quad
\begin{tabular}{c}
$(x_1,x_2)$\\
\hline
\pstree{\TR{$x_1$}}
{
	\pstree{\TR{$x_2$}}
	{
		\TR{$\exists$}
	}
}
\end{tabular}
\caption{A pAR with a pure rhs.}
\label{fig_pure_rhs}
\end{figure}

\paragraph*{Impure rhs's}
Consider the pAR in Figure~\ref{fig_not_pure_a}. The rhs is impure since $x_2$
appears twice in the head. The pAR expresses that a sufficient proportion of
the matchings of the lhs pattern, are also matchings of the rhs pattern, which
is the same as the lhs pattern except that $x_2$ is equal to
$x_3$. Since the pAR has no parameters, we can identify it with
its instantiation by the empty parameter assignment.  The confidence is then: $$\frac{m}{\sum_x \deg^2 x}$$ where $m$ is the number of edges, $x$ ranges over the nodes in the graph, and $\deg x$ is the outdegree
of (number of edges leaving) $x$.  Since  $m = \sum_x \deg x$, we show by an easy
calculation that this confidence is much larger than $1/m$: 
\begin{align*}
\frac{m}{\sum_x \deg^2 x} &= \frac{\sum_x \deg x}{\sum_x \deg^2 x} \\
&\geq \frac{\sum_x \deg x}{(\sum_x \deg x)^2}\\
&= \frac{1}{\sum_x \deg x}\\
&= \frac{1}{m}
\end{align*}
Hence, the sparser the graph (with the number of nodes remaining the same), the
higher the confidence, and thus the pAR is interesting in that it tells us
something about the sparsity of the graph.  As an illustration, on the graph
of Figure~\ref{figgraaf1} the confidence is $0.4$, but on the the graph of
Figure~\ref{figgraaf2}, it is $0.6$. 

Also consider the pAR in Figure~\ref{fig_not_pure_b}. Again the rhs is impure
since its head contains a parameter. Create an iAR for this pAR with $\alpha = \sigma_2 \mapsto 8$. With confidence $c$, this iAR then expresses that a fraction of $c$ of all edges point to node~8, which again would be an
interesting property of the graph.

The knowledge expressed by the above two example pARs cannot be expressed using
pARs with pure rhs's. To illustrate, the pAR of Figure~\ref{fig_not_pure_d}
may at first seem equivalent (and has a pure rhs) to that of
Figure~\ref{fig_not_pure_b}.  On second thought, however, it 
says nothing about the proportion of \emph{edges} pointing to $\sigma_2$, but only
about the proportion of \emph{nodes} with an edge to $\sigma_2$.

Of course, we are not implying that pARs with pure rhs's are uninteresting.
But all they can express are statements about the proportion of matchings of
the lhs that can be specialized or extended to a matching of the rhs (another
example is in Figure~\ref{fig_pure_rhs}, which says something about the
proportion of edges that can be extended); they cannot say anything about the
proportion of matchings of the lhs that satisfy certain equalities in the
distinguished variables.

\paragraph*{Free Constants}
Most treatments of conjunctive database queries \cite{cm, ahv_book, ullmanII} allow arbitrary constants in the head. In our treatment, a constant can only appear in the head as the value of a parameter. Fortunately this is enough. We do not need to consider ``free'' constants, i.e., constants not corresponding to a parameter value. To see this, first consider the possibility of free constants in the lhs. The same argument we already gave to assume that the lhs is pure can be used to dismiss this possibility. Next consider a constant in the rhs of an iAR $(Q_1 \Rightarrow_{\rho} Q_2,\alpha)$, with $Q_1=(H_1,P_1)$ and $Q_2=(H_2,P_2)$ and $Q_1$ already pure. Then there must be a $\rho$-containment mapping $f: Q_1 \rightarrow Q_2$, with $f(H_1) = H_2$, for the iAR to be legal. Hence, a constant $a$ can only appear in $H_2$ by one of the following two possibilities:
\begin{enumerate}
	\item $a = \alpha(f(\sigma)) = \alpha(\rho(\sigma))$, with $\sigma \in H_1$; or
	\item $a = \alpha(f(x))$, with $x$ a distinguished variable in $H_1$. 
\end{enumerate}

However, in both cases $a$ is not actually free, being equal to a parameter value.

\subsection{Overall Approach}
\label{sec_rules_overall}

Given the inputs: $G$; $\qleft = (\hleft, \pleft)$; \emph{minconf}; and \emph{minsup}, an outline of our algorithm for the association rule mining problem is that of four nested loops:

\begin{enumerate}

\item Generate, incrementally, all possible trees of increasing sizes. Avoid trees that are isomorphic to previously generated ones. The height of the generated trees must be at least the height of the
tree underlying $\pleft$.  (When enough trees have been generated, this loop can be terminated.)

\item For each new generated tree $T$, generate all frequent instantiated tree patterns $P^{\alpha}$ based on that tree.

\end{enumerate}
	
These first two loops are nothing but our algorithm for mining frequent tree queries as presented in Section~\ref{sec_algo_patterns}.

\begin{enumerate} \setcounter{enumi}{2}

\item For each parameterized tree pattern $P$, generate all containment mappings $f$ from $\pleft$ to $P$.  Here, a plain ``containment mapping'' is a $\rho$-containment mapping, as defined in Section~\ref{sec_treequery}, for some $\rho$. Note that $\rho$ then equals $f|_{\sigmaleft}$.

\item For each $f$, generate the parameterized tree query $Q_{\text{right}}=(f(\hleft), P)$, and all parameter assignments $\alpha$ such that $(\qleft \Rightarrow_{\rho} Q_{\text{right}}, \alpha)$ is frequent; and the confidence exceeds \emph{minconf}. The generation of all these $\alpha$'s happens in a parallel fashion.
\end{enumerate}

This approach is complete, i.e., it will output everything that must be output.  In proof, consider a legal, frequent and confident iAR $(\qleft \Rightarrow_{\rho_{0}} Q_{\text{right}}, \alpha_{0})$, with $Q_{\text{right}}=(H_{\text{right}},P_{\text{right}})$.  The tree $T$ is the underlying tree of $P_{\text{right}}$; $P_{\text{right}}$ is a tree pattern $P$ in loop~2; the containment mapping $f$ in loop~3 is the $\rho_{0}$-containment mapping that exists since the iAR is legal; $H_{\text{right}}$ is $f(\hleft)$; and $\alpha$ in loop~4 is $\alpha_{0}$.

The reader may wonder whether loop~3 cannot be organized in a levelwise fashion.  This is not obvious, however, since any two queries of the form $((f_1(\hleft), P), \alpha)$ and $((f_2(\hleft), P), \alpha)$ have exactly the same frequency, namely that of $P^\alpha$. Loop~4, however, is levelwise because it is based on loop~2 which is levelwise.

As already mentioned, these first two loops are nothing but our algorithm for mining frequent tree queries as presented in Section~\ref{sec_algo_patterns}. As already explained in Section~\ref{sec_result_management}, in loop 2 we build up a structured database containing all frequency tables for all trees in loop 1. We call this database the \emph{pattern database}. In fact these two loops should be regarded as a preprocessing step; once built up, this pattern database can be used to generate association rules. 

Hence, in practice an outline for our rule-mining algorithm is the following:

\begin{enumerate}

\item Preprocessing step: Generate a pattern database $D$ using the algorithm discussed in Section~\ref{sec_algo_patterns}. Halt this algorithm when enough patterns are generated. 

\item Consider, in a levelwise order, each parameterized tree pattern $P$ that has frequent instantiations in $D$, and such that the height of the underlying tree of $P$ is at least the height of the underlying tree of $\pleft$.

\item For each parameterized tree pattern $P$, generate all containment mappings $f$ from $\pleft$ to $P$ and let $\rho$ be $f|_{\sigmaleft}$.

\item For each $f$, generate the parameterized tree query $Q=(f(\hleft), P)$, and all parameter assignments $\alpha$ such that $(\qleft \Rightarrow_{\rho} Q, \alpha)$ is frequent; and the confidence exceeds \emph{minconf}. The generation of all these $\alpha$'s happens in a parallel fashion.
\end{enumerate}

We present loops $3$ and $4$ in detail in Sections \ref{subsec_contmap} and \ref{subsec_paras}. In Section \ref{subsec_equivrules}, we will show how our overall approach must be refined so that the generation of equivalent association rules is avoided.

\subsection{Generation of Containment Mappings}
\label{subsec_contmap}

In this section, we discuss loop $3$, the generation of all containment mappings $f$ from $\pleft$ to $P$. So, we need to solve the following problem: Given two parameterized tree patterns $P_1$ and $P_2$, find all containment mappings $f$ from $P_1$ to $P_2$.

Since the patterns are typically small, a naive algorithm suffices.  For a node $x_1$ of $P_1$ and a node $x_2$ of $P_2$, we say that $x_1$ ``matches'' $x_2$ if there is a containment mapping $f$ from the subpattern of $P_1$ rooted at $x_1$ to the subpattern of $P_2$ rooted at $x_2$ such that $f(x_1)=x_2$.  In a first phase, we determine for every node $y$ of $P_2$ separately whether the root $r_1$ of $P_1$ matches $y$.  While doing so, we also determine for every other node $x_1$ of $P_1$, and every node $x_2$ below $y$ at the same distance as $x_1$ is from $r_1$, whether $x_1$ matches $x_2$. We store all these boolean values in a two-dimensional matrix \emph{Map}. The function for filling in Map is given in function~\ref{algo_cm_fillin}. In line~$2$ of this function we mean by ``$x_1 \mapsto x_2$ is legal'', that if $x_1$ is a distinguished variable, then $x_2$ is a distinguished variable or a parameter; and if $x_1$ is a parameter then $x_2$ is a parameter, as prescribed by the definition of a $\rho$-containment mapping in Section~\ref{sec_treequery}.

\floatname{algorithm}{Function}
\begin{algorithm}
\caption{Function for filling in Map}
\label{algo_cm_fillin}
\begin{algorithmic}[1]
\STATE \textbf{bool FillInn}$\mathbf{(x_1 \in P_1,\ x_2 \in P_2)}$
\IF {$x_1 \mapsto x_2$ is legal}
\STATE Match $:=$ true;
\FOR {each child $c_1$ of $x_1$ from left to right}
\STATE MatchChild $:=$ false;
\FOR {each child $c_2$ of $x_2$ from left to right}
\STATE MatchChild $:=$ MatchChild OR $\text{FillIn}(c_1, c_2)$
\ENDFOR
\STATE Match $:=$ Match AND MatchChild;
\ENDFOR
\STATE $\text{Map}[x_1,x_2]:=\text{Match}$;
\RETURN Match;
\ELSE 
\STATE $\text{Map}[x_1,x_2]:=\text{false}$;
\RETURN false
\ENDIF 
\end{algorithmic}
\end{algorithm}
 
This first phase compares every possible pair $(x_1,x_2)$, with $x_1$ a node in $P_1$ and $x_2$ a node in $P_2$, at most once.  Indeed, if $x_1$ is at distance $d$ from $r_1$, then $x_1$ will be compared to $x_2$ only during the matching of $r_1$ with the node $y$ that is $d$ steps above $x_2$ in $P_2$ (if existing).  We thus have an $O(n_1 \times n_2)$ algorithm, where $n_1$ ($n_2$) is the number of nodes in $P_1$ ($P_2$).

In a second phase, we output all containment mappings.  Initially, by a synchronous preorder traversal of $P_1$ and $P_2$, we map each node of $P_1$ to the first matching node of $P_2$.  We store this first mapping in a one-dimensional matrix \emph{Cm}. In function~\ref{algo_cm_init} an outline for finding the initial containment mapping is given. 

In each subsequent step, we look for the last node $x_1$ (in preorder) of $P_1$, currently matched to some node $x_2$, with the property that $x_1$ can also be matched to a right sibling $x_3$ of $x_2$, and now map $x_1$ to the first such $x_3$.  The mappings of all nodes of $P_1$ coming after $x_1$ are reinitialized.  Every such step takes time that
is linear in $n_1$ and $n_2$.  Of course, the total number of different containment mappings may well be exponential in $n_1$. An outline of this step is given in Function~\ref{algo_cm_step}. 

\floatname{algorithm}{Function}
\begin{algorithm}
\caption{Function for finding the initial containment mapping}
\label{algo_cm_init}
\begin{algorithmic}[1]
\STATE \textbf{Init}$\mathbf{(x_1 \in P_1, x_2 \in P_2)}$
\STATE $\text{Cm}[x_1] := x_2$;
\FOR {each child $c_1$ of $x_1$ from left to right}
	\FOR {each child $c_2$ of $x_2$ from left to right}
		\IF {$\text{Map}[c_1,c_2]$}
			\STATE $\text{Init}(c_1, c_2)$;
			\STATE Break;
		\ENDIF	
	\ENDFOR
\ENDFOR
\end{algorithmic}
\end{algorithm}

\floatname{algorithm}{Function}
\begin{algorithm}
\caption{Function for finding the other containment mappings}
\label{algo_cm_step}
\begin{algorithmic}[1]
\STATE \textbf{bool Step}$\mathbf{(x \in P_1)}$
\STATE Found $:=$ false;
\FOR {each child $c$ from $x$ from right to left} 
	\IF{$\text{Step}(c)$}
		\STATE Found $:=$ true;
		\STATE Break;
	\ENDIF
\ENDFOR
\IF{Found}
	\FOR{each right-sibling $z$ of $c$ from left to right}
		\STATE $p_2 := \text{Cm}[x]$;
		\FOR{each child $c_2$ of $p_2$ from left to right}
			\IF{$\text{Map}[z,c_2]$}
				\STATE $\text{Init}(z,c_2)$
			\ENDIF
		\ENDFOR	
	\ENDFOR
	\RETURN true;
\ELSE	
	\IF{$x$ is the root of $P_1$}
		\RETURN false;
	\ELSE
		\STATE $m:= \text{Cm}[x]$;
		\FOR{each right-sibling $s$ of $m$ from left to right}
			\IF{$\text{Map}[x,s]$}
				\STATE $\text{Init}(x,s)$
				\STATE Break;
			\ENDIF
		\ENDFOR
		\RETURN true;
	\ENDIF
\ENDIF
\end{algorithmic}
\end{algorithm}

The complete outline for the generation of all containment mappings is given in Function~\ref{algo_cm}.

\floatname{algorithm}{Function}
\begin{algorithm}
\caption{Function for generating all containment mappings from $P_1$ to $P_2$}
\label{algo_cm}
\begin{algorithmic}[1]
\STATE \textbf{GenerateCm}$\mathbf{(P_1, P_2)}$
\STATE Initialize Map;
\STATE $r_1 :=$ root of $P_1$;
\FOR {each $x_2 \in P_2$ in preorder}
	\STATE FillIn$(r_1,x_2)$;
\ENDFOR
\FOR {each node $x_2 \in P_2$ in preorder}
	\IF{$\text{Map}[r_1,x_2]$}
		\STATE Initialize Cm;
		\STATE $\text{Init}(r_1,x_2)$
		\REPEAT
			\STATE Output Cm;
		\UNTIL{not $\text{Step}(r_1)$}
	\ENDIF
\ENDFOR
\end{algorithmic}
\end{algorithm}

We can thus easily generate all containment mappings $f$ from $\pleft$ to $P$ as required for loop~3 of our overall algorithm.  Note, however, that in loop~4 these mappings are used to produce the head $f(\hleft)$ of query $Q_{\text{right}}$.  For $Q_{\text{right}}$ to be a legal query, this head must contain all distinguished variables of $P$.  Hence, we only pass to loop~4 those $f$ whose image contains all distinguished variables of $P$.

\subsection{Generation of Parameter Assignments}
\label{subsec_paras}

In loop $4$, our task is the following. Given a containment mapping $f: \pleft \rightarrow P$, let $\rho = f|_{\sigmaleft}$, and generate all parameter assignments $\alpha$ such that $(\qleft \Rightarrow_{\rho} (f(\hleft),P) ,\alpha)$ is frequent and confident in $G$. We show how this can be done in a parallel database-oriented fashion. 

Recall from Section~\ref{sec_result_management} that the frequency tables for $\pleft$ and $P$ are available in a relational database. Our crucial observation is that we can compute precisely the required set of parameter assignments $\alpha$, together with the frequency and confidence of the corresponding association rules, by a single relational algebra expression. This expression has the following form: $$\proj_{\text{\emph{plist}}}\ \sel_{\frac{\freqtabPattern{P}.\text{freq}}{\freqtabPattern{\pleft}.\text{freq}} \geq \text{\emph{minconf}}} (\freqtabPattern{\pleft} \Join_{\theta} \freqtabPattern{P})$$ Here, $\proj$ denotes projection, $\sel$ denotes selection, and $\Join$ denotes join. The join condition $\theta$ and the projection list \emph{plist} are defined as follows. For $\theta$, we take the conjunction: $$\bigwedge_{\sigma \in \sigmaleft} \freqtabPattern{\pleft}.\sigma = \freqtabPattern{P}.\rho(\sigma)$$ Furthermore, \emph{plist} consists of all attributes $\pleft.\sigma_{\text{left}}$, with $\sigma_{\text{left}} \in \sigmaleft$; all attributes $P.\sigma$, with $\sigma \in \Sigma$; together with the attributes $\freqtabPattern{P}.\text{freq}$ and $\freqtabPattern{P}.\text{freq} / \freqtabPattern{\pleft}.\text{freq}$.

Referring back to our overall algorithm (Section~\ref{sec_rules_overall}), we thus generate, for each pattern $P$ from loop~2 and each containment mapping $f$ in loop~$3$, all association rules with the given $\qleft$ as lhs in parallel, by one relational database query (which can be implemented by a simple SQL select-statement).

\begin{example}
Consider $\qleft$ and $P=(\Pi,\Sigma)$ as shown in Figures \ref{ex_fig_qleft} and \ref{ex_fig_P}.
We have $\sigmaleft= \{\sigma_1, \sigma_4 \}$ and $\Pi_{\text{left}} = \{x_3, x_6 \}$, and $\Sigma=\{ \sigma_1, \sigma_4, \sigma_5\}$ and $\Pi = \{ x_3 \}$. Take the following containment mapping $f$ from $\pleft$ to $P$:

\begin{center}
\begin{tabular}{|c|c|}
\hline \multicolumn{2}{|c|}{$f$}  \\ \hline
$\sigma_{1}$ & $\sigma_{1}$ \\  
$x_1$ & $x_2$ \\ 
$\exists_3$ & $\exists$ \\ 
$\sigma_4$ & $\sigma_4$ \\ 
$x_5$ & $x_2$ \\ 
$\exists_6$ & $\exists$ \\ 
$x_7$ & $\sigma_{4}$ \\ \hline
\end{tabular}
\end{center}

Then the rhs query $Q_{\text{right}}$ equals $((x_2, x_2,
\sigma_4),P)$, and the relational algebra expression for
computing all parameter assignments and their corresponding
frequencies and confidences looks as follows: 

$$\proj_{\text{\emph{plist}}}\ \sel_{\frac{\freqtabPattern{P}.\text{freq}}{\freqtabPattern{\pleft}.\text{freq}} \geq \text{\emph{minconf}}}(\freqtabPattern{\pleft} \Join_{\theta} \freqtabPattern{P})$$
with \emph{plist} equal to
\begin{center}
$\freqtabPattern{\pleft}.\sigma_1$, $\freqtabPattern{\pleft}.\sigma_4$, $\freqtabPattern{P}.\sigma_1$, $\freqtabPattern{P}.\sigma_4$, $\freqtabPattern{P}.\sigma_5$, $\freqtabPattern{P}.\text{freq}$, $\freqtabPattern{P}.\text{freq} / \freqtabPattern{\pleft}.\text{freq}$
\end{center}
and $\theta$ equal to 
\begin{center} $\freqtabPattern{P}.\sigma_1 = \freqtabPattern{\pleft}.\sigma_1 \wedge \freqtabPattern{P}.\sigma_4=\freqtabPattern{\pleft}.\sigma_4$ 
\end{center}

In SQL, we get:
\begin{verbatim}
SELECT freqQleft.x1, freqQleft.x4, freqP.x1, freqP.x4, 
       freqP.x5, freqP.freq, freqP.freq/freqQleft.freq
FROM freqP, freqQleft
WHERE freqQleft.x1= freqP.x1 AND freqQleft.x4=freqP.x4
      AND freqP.freq/freqQleft.freq >= minconf
\end{verbatim}

\begin{figure}
\centering
\subfigure[]{\label{ex_fig_qleft}
\begin{tabular}{c}
$(x_2,x_5,x_7)$\\
\hline
\pstree{\TR{$\sigma_1$}}
{
  \pstree{\TR{$x_2$}}
  {
    \pstree{\TR{$\exists_3$}}
    {
    	\TR{$\sigma_4$}
    }
  }
  \pstree{\TR{$x_5$}}
  {
    \pstree{\TR{$\exists_6$}}
    {
    	\TR{$x_7$}
    }
  }
}
\end{tabular}
}
\quad
\quad
\quad
\quad
\subfigure[]{\label{ex_fig_P}
\begin{tabular}{c}
\pstree{\TR{$\sigma_1$}}
{
	\pstree{\TR{$x_2$}}
	{
		\pstree{\TR{$\exists$}}
		{
			\TR{$\sigma_4$}
			\TR{$\sigma_5$}
		}
	}
}
\end{tabular}
}
\caption{Example $\qleft$ and $P$.}
\end{figure}

\end{example}

\subsection{Example Run}
\label{sec_example_rules}


\begin{figure}
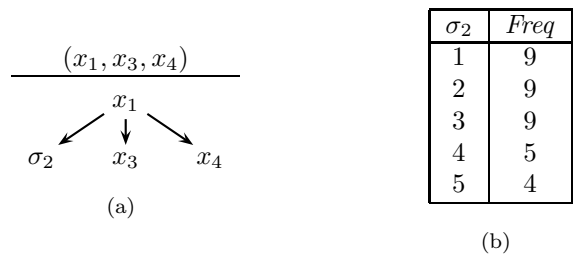

\hspace*{\fill}
\subfigure[]{\label{fig_examplerules_lhs_a}
\begin{tabular}{c}
$(x_1,x_3,x_4)$\\
\hline
\pstree{\TR{$x_1$}}
{
	\TR{$\sigma_2$}
	\TR{$x_3$}
	\TR{$x_4$}
}
\end{tabular}
}
\hfill
\subfigure[]{\label{fig_examplerules_lhs_b}
\begin{tabular}{|c|c|}
\hline $\sigma_2$ & $\Freq$ \\ \hline
$1$ & $9$ \\
$2$ & $9$ \\
$3$ & $9$ \\
$4$ & $5$ \\
$5$ & $4$ \\ \hline
\end{tabular}
}
\hspace*{\fill}
\caption{The fixed lhs and its frequency table for the example run in Section~\ref{sec_example_rules}.}
\end{figure}

In this Section we give an example run of the algorithm discussed in Section~\ref{sec_algo_associations}. We use the same data graph $G$, unordered rooted tree $T$, and minimum support threshold, $3$, as in the example run in Section~\ref{sec_example}. The fixed lhs tree query is given in Figure~\ref{fig_examplerules_lhs_a}, its corresponding frequency table in Figure~\ref{fig_examplerules_lhs_b}, and the minimum confidence threshold is $30\%$. All frequent tree patterns based on $T$ were already generated in the example run of Section~\ref{sec_example_equiv}.

The example run then looks as follows:

\begin{landscape}
\begin{longtable}{|c|c|c|c|}
\hline \multicolumn{1}{|c|}{$P$} & \multicolumn{1}{|c|}{Containment Mapping} & \multicolumn{1}{|c|}{$\qright$} & \multicolumn{1}{|c|}{${\it ConfTab}$}\\ \hline
\endfirsthead

\hline
\endlastfoot
\multicolumn{4}{|c|}{\textbf{Level 0}}
\\ \hline
$(\emptyset,\emptyset)$
&
\multicolumn{3}{|c|}{No Containment Mappings}
\\ \hline
\multicolumn{4}{|c|}{\textbf{Level 1}}
\\ \hline
$(\emptyset, \{x_1\})$
&
\multicolumn{3}{|c|}{No Containment Mappings}
\\ \hline
$(\emptyset, \{x_2\})$
&
\mystrut{\begin{tabular}{c}
\begin{pspicture}(0,-0.5)(6,1.5)
\rput(1,1){\rnode{A1}{$x_1$}}
\rput(0,0){\rnode{A2}{$\sigma_2$}}
\rput(1,0){\rnode{A3}{$x_3$}}
\rput(2,0){\rnode{A4}{$x_4$}}
\ncline{A1}{A2}
\ncline{A1}{A3}	
\ncline{A1}{A4}
\rput(5,1){\rnode{B1}{$x_1$}}
\rput(4,0){\rnode{B2}{$\sigma_2$}}
\rput(6,0){\rnode{B3}{$x_3$}}
\ncline{B1}{B2}
\ncline{B1}{B3}
\ncarc[arcangle=20,linestyle=dotted]{A1}{B1}
\ncarc[arcangle=20,linestyle=dotted]{A2}{B2}
\ncarc[arcangle=20,linestyle=dotted]{A3}{B3}
\ncarc[arcangle=20,linestyle=dotted]{A4}{B3}	
\end{pspicture}
\end{tabular}}
&
\mystrut{\begin{tabular}{c}
$(x_1,x_3,x_3)$\\
\hline
\pstree{\TR{$x_1$}}
{
	\TR{$\sigma_2$}
	\TR{$x_3$}
}
\end{tabular}}
&
\mystrut{\begin{tabular}{|c|c|c|c|}
\hline $\pleft.\sigma_2$ & $P.\sigma_2$ & $\Freq$ & $\Conf$ \\ \hline
$1$ & $1$ & $3$ & $33\%$\\
$2$ & $2$ & $3$ & $33\%$\\
$3$ & $3$ & $3$ & $33\%$\\
$4$ & $4$ & $3$ & $60\%$\\ \hline
\end{tabular}}
\\ \hline
$(\emptyset, \{x_2\})$
&
\mystrut{\begin{tabular}{c}
\begin{pspicture}(0,-0.5)(6,1.5)
\rput(1,1){\rnode{A1}{$x_1$}}
\rput(0,0){\rnode{A2}{$\sigma_2$}}
\rput(1,0){\rnode{A3}{$x_3$}}
\rput(2,0){\rnode{A4}{$x_4$}}
\ncline{A1}{A2}
\ncline{A1}{A3}	
\ncline{A1}{A4}
\rput(5,1){\rnode{B1}{$x_1$}}
\rput(4,0){\rnode{B2}{$\sigma_2$}}
\rput(6,0){\rnode{B3}{$x_3$}}
\ncline{B1}{B2}
\ncline{B1}{B3}
\ncarc[arcangle=20,linestyle=dotted]{A1}{B1}
\ncarc[arcangle=20,linestyle=dotted]{A2}{B2}
\ncarc[arcangle=20,linestyle=dotted]{A3}{B2}
\ncarc[arcangle=20,linestyle=dotted]{A4}{B3}	
\end{pspicture}
\end{tabular}}
&
\mystrut{\begin{tabular}{c}
$(x_1, \sigma_2, x_3)$\\
\hline
\pstree{\TR{$x_1$}}
{
	\TR{$\sigma_2$}
	\TR{$x_3$}
}
\end{tabular}}
&
\mystrut{\begin{tabular}{|c|c|c|c|}
\hline $\pleft.\sigma_2$ & $P.\sigma_2$ & $\Freq$ & $\Conf$ \\ \hline
$1$ & $1$ & $3$ & $33\%$\\
$2$ & $2$ & $3$ & $33\%$\\
$3$ & $3$ & $3$ & $33\%$\\
$4$ & $4$ & $3$ & $60\%$\\ \hline
\end{tabular}}
\\ \hline
$(\emptyset, \{x_2\})$
&
\mystrut{\begin{tabular}{c}
\begin{pspicture}(0,-0.5)(6,1.5)
\rput(1,1){\rnode{A1}{$x_1$}}
\rput(0,0){\rnode{A2}{$\sigma_2$}}
\rput(1,0){\rnode{A3}{$x_3$}}
\rput(2,0){\rnode{A4}{$x_4$}}
\ncline{A1}{A2}
\ncline{A1}{A3}	
\ncline{A1}{A4}
\rput(5,1){\rnode{B1}{$x_1$}}
\rput(4,0){\rnode{B2}{$\sigma_2$}}
\rput(6,0){\rnode{B3}{$x_3$}}
\ncline{B1}{B2}
\ncline{B1}{B3}
\ncarc[arcangle=20,linestyle=dotted]{A1}{B1}
\ncarc[arcangle=20,linestyle=dotted]{A2}{B2}
\ncarc[arcangle=20,linestyle=dotted]{A3}{B3}
\ncarc[arcangle=20,linestyle=dotted]{A4}{B2}	
\end{pspicture}
\end{tabular}}
&
\mystrut{\begin{tabular}{c}
$(x_1, x_3, \sigma_2)$\\
\hline
\pstree{\TR{$x_1$}}
{
	\TR{$\sigma_2$}
	\TR{$x_3$}
}
\end{tabular}}
&
\mystrut{\begin{tabular}{|c|c|c|c|}
\hline $\pleft.\sigma_2$ & $P.\sigma_2$ & $\Freq$ & $\Conf$ \\ \hline
$1$ & $1$ & $3$ & $33\%$\\
$2$ & $2$ & $3$ & $33\%$\\
$3$ & $3$ & $3$ & $33\%$\\
$4$ & $4$ & $3$ & $60\%$\\ \hline
\end{tabular}}
\\ \hline
$(\{x_1\}, \emptyset)$
&
\multicolumn{3}{|c|}{No containment mappings}
\\ \hline
\multicolumn{4}{|c|}{\textbf{Level 2}}
\\ \hline
$(\emptyset, \{x_1, x_2\})$
&
\mystrut{\begin{tabular}{c}
\begin{pspicture}(0,-0.5)(6,1.5)
\rput(1,1){\rnode{A1}{$x_1$}}
\rput(0,0){\rnode{A2}{$\sigma_2$}}
\rput(1,0){\rnode{A3}{$x_3$}}
\rput(2,0){\rnode{A4}{$x_4$}}
\ncline{A1}{A2}
\ncline{A1}{A3}	
\ncline{A1}{A4}
\rput(5,1){\rnode{B1}{$\sigma_1$}}
\rput(4,0){\rnode{B2}{$\sigma_2$}}
\rput(6,0){\rnode{B3}{$x_3$}}
\ncline{B1}{B2}
\ncline{B1}{B3}
\ncarc[arcangle=20,linestyle=dotted]{A1}{B1}
\ncarc[arcangle=20,linestyle=dotted]{A2}{B2}
\ncarc[arcangle=20,linestyle=dotted]{A3}{B3}
\ncarc[arcangle=20,linestyle=dotted]{A4}{B3}	
\end{pspicture}
\end{tabular}}
&
\mystrut{\begin{tabular}{c}
$(\sigma_1,x_3,x_3)$\\
\hline
\pstree{\TR{$\sigma_1$}}
{
	\TR{$\sigma_2$}
	\TR{$x_3$}
}
\end{tabular}}
&
\mystrut{\begin{tabular}{|c|c|c|c|c|}
\hline $\pleft.\sigma_2$ & $P.\sigma_1$ & $P.\sigma_2$ & $\Freq$ & $\Conf$ \\ \hline
$0$ & $1$ & $1$ & $3$ & $33\%$\\
$0$ & $2$ & $2$ & $3$ & $33\%$\\
$0$ & $3$ & $3$ & $3$ & $33\%$\\ \hline
\end{tabular}}
\\ \hline
$(\emptyset, \{x_1, x_2\})$
&
\mystrut{\begin{tabular}{c}
\begin{pspicture}(0,-0.5)(6,1.5)
\rput(1,1){\rnode{A1}{$x_1$}}
\rput(0,0){\rnode{A2}{$\sigma_2$}}
\rput(1,0){\rnode{A3}{$x_3$}}
\rput(2,0){\rnode{A4}{$x_4$}}
\ncline{A1}{A2}
\ncline{A1}{A3}	
\ncline{A1}{A4}
\rput(5,1){\rnode{B1}{$\sigma_1$}}
\rput(4,0){\rnode{B2}{$\sigma_2$}}
\rput(6,0){\rnode{B3}{$x_3$}}
\ncline{B1}{B2}
\ncline{B1}{B3}
\ncarc[arcangle=20,linestyle=dotted]{A1}{B1}
\ncarc[arcangle=20,linestyle=dotted]{A2}{B2}
\ncarc[arcangle=20,linestyle=dotted]{A3}{B2}
\ncarc[arcangle=20,linestyle=dotted]{A4}{B3}	
\end{pspicture}
\end{tabular}}
&
\mystrut{\begin{tabular}{c}
$(\sigma_1,\sigma_2,x_3)$\\
\hline
\pstree{\TR{$\sigma_1$}}
{
	\TR{$\sigma_2$}
	\TR{$x_3$}
}
\end{tabular}}
&
\mystrut{
\begin{tabular}{|c|c|c|c|c|}
\hline $\pleft.\sigma_2$ & $P.\sigma_1$ & $P.\sigma_2$ & $\Freq$ & $\Conf$ \\ \hline
$0$ & $1$ & $1$ & $3$ & $33\%$\\
$0$ & $2$ & $2$ & $3$ & $33\%$\\
$0$ & $3$ & $3$ & $3$ & $33\%$\\ \hline
\end{tabular}}
\\ \hline
$(\emptyset, \{x_1, x_2\})$
&
\mystrut{\begin{tabular}{c}
\begin{pspicture}(0,-0.5)(6,1.5)
\rput(1,1){\rnode{A1}{$x_1$}}
\rput(0,0){\rnode{A2}{$\sigma_2$}}
\rput(1,0){\rnode{A3}{$x_3$}}
\rput(2,0){\rnode{A4}{$x_4$}}
\ncline{A1}{A2}
\ncline{A1}{A3}	
\ncline{A1}{A4}
\rput(5,1){\rnode{B1}{$\sigma_1$}}
\rput(4,0){\rnode{B2}{$\sigma_2$}}
\rput(6,0){\rnode{B3}{$x_3$}}
\ncline{B1}{B2}
\ncline{B1}{B3}
\ncarc[arcangle=20,linestyle=dotted]{A1}{B1}
\ncarc[arcangle=20,linestyle=dotted]{A2}{B2}
\ncarc[arcangle=20,linestyle=dotted]{A3}{B3}
\ncarc[arcangle=20,linestyle=dotted]{A4}{B2}	
\end{pspicture}
\end{tabular}}
&
\mystrut{\begin{tabular}{c}
$(\sigma_1,x_3, \sigma_2)$\\
\hline
\pstree{\TR{$\sigma_1$}}
{
	\TR{$\sigma_2$}
	\TR{$x_3$}
}
\end{tabular}}
&
\mystrut{\begin{tabular}{|c|c|c|c|c|}
\hline $\pleft.\sigma_2$ & $P.\sigma_1$ & $P.\sigma_2$ & $\Freq$ & $\Conf$ \\ \hline
$0$ & $1$ & $1$ & $3$ & $33\%$\\
$0$ & $2$ & $2$ & $3$ & $33\%$\\
$0$ & $3$ & $3$ & $3$ & $33\%$\\ \hline
\end{tabular}}
\\ \hline
$(\{x_1\},\{x_2\})$
&
\multicolumn{3}{|c|}{No containment mappings}
\\ \hline
\multicolumn{4}{|c|}{\textbf{Level 3}}
\\ \hline
$(\{x_1\} , \{x_2, x_3\})$
&
\multicolumn{3}{|c|}{No containment mappings}
\\ \hline
\end{longtable}
\end{landscape}

\subsection{Equivalent Association Rules}
\label{subsec_equivrules}

In this section, we make a number of modifications to the algorithm described so far, so as to avoid duplicate work on equivalent rules. 

\begin{figure}
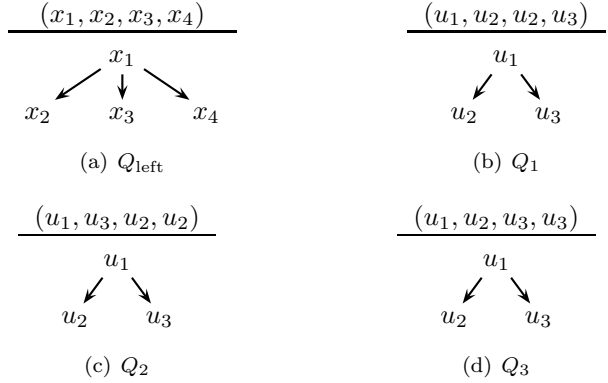

\hspace*{\fill}
\subfigure[$\qleft$]{\label{equiv_rules_qleft}
\begin{tabular}{c}
$(x_1,x_2,x_3,x_4)$\\
\hline
\pstree{\TR{$x_1$}}
{
	\TR{$x_2$}
	\TR{$x_3$}
	\TR{$x_4$}
}
\end{tabular}
}
\hfill
\subfigure[$Q_1$]{\label{equiv_rules_q1}
\begin{tabular}{c}
$(u_1,u_2,u_2,u_3)$\\
\hline
\pstree{\TR{$u_1$}}
{
	\TR{$u_2$}
	\TR{$u_3$}
}
\end{tabular}
}
\hspace*{\fill}
\\
\hspace*{\fill}
\subfigure[$Q_2$]{\label{equiv_rules_q2}
\begin{tabular}{c}
$(u_1,u_3,u_2,u_2)$\\
\hline
\pstree{\TR{$u_1$}}
{
	\TR{$u_2$}
	\TR{$u_3$}
}
\end{tabular}
}
\hfill
\subfigure[$Q_3$]{\label{equiv_rules_q3}
\begin{tabular}{c}
$(u_1,u_2,u_3,u_3)$\\
\hline
\pstree{\TR{$u_1$}}
{
	\TR{$u_2$}
	\TR{$u_3$}
}
\end{tabular}
}
\hspace*{\fill}
\caption{Queries to illustrate the duplicate work in the association mining algorithm}
\label{equiv_rules_example}
\end{figure}

Let us first look at an example of the duplicate work that the algorithm presented until now performs. Consider $\qleft$, $Q_1 = (f_1(\hleft), P)$, $Q_2 = (f_2(\hleft), P)$; and $Q_3 = (f_3(\hleft), P)$ in Figure~\ref{equiv_rules_example} with $f_1$, $f_2$ and $f_3$ as follows:

\begin{center}
\hspace*{\fill}
\begin{tabular}{|c|c|}
\hline \multicolumn{2}{|c|}{$f_1$}  \\ \hline
$x_{1}$ & $u_{1}$ \\  
$x_{2}$ & $u_{2}$ \\
$x_{3}$ & $u_{2}$ \\
$x_{4}$ & $u_{3}$ \\ \hline
\end{tabular}
\hfill
\begin{tabular}{|c|c|}
\hline \multicolumn{2}{|c|}{$f_2$}  \\ \hline
$x_{1}$ & $u_{1}$ \\  
$x_{2}$ & $u_{3}$ \\
$x_{3}$ & $u_{2}$ \\
$x_{4}$ & $u_{2}$ \\ \hline
\end{tabular}
\hfill
\begin{tabular}{|c|c|}
\hline \multicolumn{2}{|c|}{$f_3$}  \\ \hline
$x_{1}$ & $u_{1}$ \\  
$x_{2}$ & $u_{2}$ \\
$x_{3}$ & $u_{3}$ \\
$x_{4}$ & $u_{3}$ \\ \hline
\end{tabular}
\hspace*{\fill}
\end{center} Furthermore, consider pAR1: $\qleft \Rightarrow Q_1$; pAR2: $\qleft \Rightarrow Q_2$ and pAR3: $\qleft \Rightarrow Q_3$. 

The confidence of the first rule (pAR1) equals the proportion of tuples from the answer set of $\qleft$ where the values for variables $x_2$ and $x_3$ are equal (in the rhs those equal variables are represented by variable $u_2$, and the lhs variable $x_4$ is represented by the rhs variable $u_3$). Similarly, the confidence of the second rule (pAR2) equals the proportion of tuples from the answer set of $\qleft$ where the values for the variables $x_3$ and $x_4$ are equal (again the equal lhs variables $x_3$ and $x_4$ are represented by the rhs variable $u_2$, and the lhs variable $x_2$ is represented by the rhs variable $u_3$). Since, due to the symmetry in the lhs pattern, the columns for $x_2$, $x_3$ and $x_4$ are fully interchangeable in the answer set of $\qleft$, both rules convey precisely the same information: their confidences are equal. The third rule (pAR3) is yet another representation of the same association, but now the equal lhs variables $x_3$ and $x_4$ are represented by the rhs variable $u_3$. Again, it has the same confidence as pAR1 and pAR2.

It is important to note that the above pARs only differ in the containment mappings $f_1$, $f_2$ and $f_3$ that generate the rhs head. The algorithm discussed until now generates all these pARs, since we do not perform any check on the containment mappings generated in loop~3 of the overall approach (Section~\ref{sec_rules_overall}).  

In this Subsection, motivated by the above example, we consider the general problem of when two pARs $\qleft \Rightarrow_{\rho_1} Q_1$ and $\qleft \Rightarrow_{\rho_2} Q_2$ are equivalent, where $Q_1$ and $Q_2$ are of the form $(f_1(\hleft), P)$ and $(f_2(\hleft), P)$ for some common rhs pattern $P$, and containment mappings $f_1$ and $f_2$ from $\pleft$ to $P$. (Thus $\rho_1$ is $f_1|_{\sigmaleft}$ and $\rho_2$ is $f_2|_{\sigmaleft}$.) Since such two pARs differ only in $f_1$ and $f_2$ we can actually focus on $f_1$ and $f_2$.

It is important to remember for the rest of this Subsection that $\pleft$ and $P$ are arbitrary but fixed. Furthermore, without loss of generality we assume that the nodes of $\pleft$ and $P$ are disjoint. This assumption greatly simplifies the representation of containment mappings by graphs, as we will see shortly. 

\paragraph*{Equivalent Containment Mappings} Recall from Section~\ref{sec_can_form} that an \emph{isomorphism} from a parameterized tree pattern $P_1$ to a parameterized tree pattern $P_2$ is a homomorphism from $P_1$ to $P_2$ that is a bijection and that maps distinguished nodes to distinguished nodes, parameters to parameters and existential nodes to existential nodes. We now formalize equivalent containment mappings as follows:  Two containment mappings $f_1$ and $f_2$ are \emph{equivalent} if the structures $(\pleft, P, f_1)$ and $(\pleft, P, f_2)$ are isomorphic. Specifically, there must exist isomorphisms (actually automorphisms) $g: \pleft \rightarrow \pleft$ and $h: P \rightarrow P$ such that $f_2 \circ g = h \circ f_1$. 

Consider for instance $f_1$ and $f_3$ from the example above, then $h$ swaps $u_2$ and $u_3$, and $g$ is the cyclic permutation $u_2 \mapsto u_3 \mapsto u_4 \mapsto u_2$. 

\subsubsection{Testing for equivalence}
\label{subsubsec_testing_equivalence}

\begin{figure}
\begin{center}
\subfigure[]{
\begin{pspicture}(0,-0.5)(6,1.5)
\rput(1,1){\rnode{A1}{$x_1$}}
\rput(0,0){\rnode{A2}{$x_2$}}
\rput(1,0){\rnode{A3}{$x_3$}}
\rput(2,0){\rnode{A4}{$x_4$}}
\ncline{A1}{A2}
\ncline{A1}{A3}	
\ncline{A1}{A4}
\rput(5,1){\rnode{B1}{$u_1$}}
\rput(4,0){\rnode{B2}{$u_2$}}
\rput(6,0){\rnode{B3}{$u_3$}}
\ncline[linestyle=dashed]{B1}{B2}
\ncline[linestyle=dashed]{B1}{B3}
\ncarc[arcangle=20,linestyle=dotted]{A1}{B1}
\ncarc[arcangle=20,linestyle=dotted]{A2}{B2}
\ncarc[arcangle=20,linestyle=dotted]{A3}{B2}
\ncarc[arcangle=20,linestyle=dotted]{A4}{B3}	
\end{pspicture}
}
\end{center}

\begin{center}
\subfigure[]{
\begin{pspicture}(0,-0.5)(6,1.5)
\rput(1,1){\rnode{A1}{$x_1$}}
\rput(0,0){\rnode{A2}{$x_2$}}
\rput(1,0){\rnode{A3}{$x_3$}}
\rput(2,0){\rnode{A4}{$x_4$}}
\ncline{A1}{A2}
\ncline{A1}{A3}	
\ncline{A1}{A4}
\rput(5,1){\rnode{B1}{$u_1$}}
\rput(4,0){\rnode{B2}{$u_2$}}
\rput(6,0){\rnode{B3}{$u_3$}}
\ncline[linestyle=dashed]{B1}{B2}
\ncline[linestyle=dashed]{B1}{B3}
\ncarc[arcangle=20,linestyle=dotted]{A1}{B1}
\ncarc[arcangle=20,linestyle=dotted]{A2}{B2}
\ncarc[arcangle=20,linestyle=dotted]{A3}{B3}
\ncarc[arcangle=20,linestyle=dotted]{A4}{B3}	
\end{pspicture}
}
\end{center}
\caption{The graph representations of $f_1$ and $f_3$.}
\label{fig_graph_representation}
\end{figure}
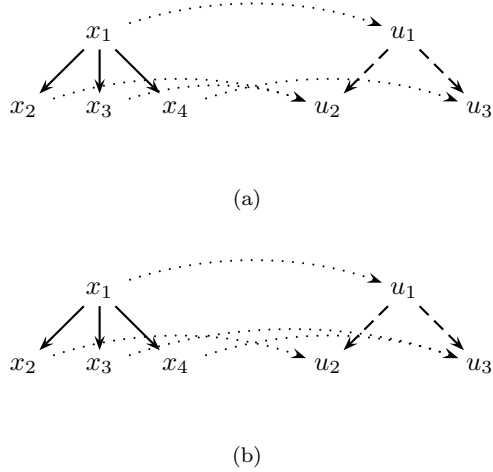

To test for equivalent containment mappings efficiently, we represent them using graphs.

\paragraph*{Graph representation of a containment mapping} The graph representation of a containment mapping $f: \pleft \rightarrow P$ is a directed, edge- and vertex-colored graph, with set of vertices $V_f = \text{Vertices}(\pleft) \cup \text{Vertices}(P)$ and set of edges $E_f = \text{Edges}(\pleft) \cup \text{Edges}(P) \cup \{(v,w) \mid f(v) = w\}$ (with the understanding that the edges of $\pleft$ and $P$ go from parent to child). We use different colors for the edges of $\pleft$, the edges of $P$ and the pairs in $f$, and we also use different colors for the distinguished nodes, the existential nodes and the parameters. 

As an illustration, Figure~\ref{fig_graph_representation} shows the graph representation of $f_1$ and $f_3$ from our example in the introduction above. 

\paragraph*{Graph Isomorphism} Two graphs $G_1=(V_1, E_1)$ and $G_2=(V_2, E_2)$ are \emph{colored isomorphic} if there exists a bijection $\varphi: V_1 \rightarrow V_2$, extended to edges $(v,w) \in E_1$ in a natural way by $\varphi(v,w) = (\varphi(v), \varphi(w))$, such that the colors of vertices and edges are preserved by $\varphi$, and such that $(v,w) \in E_1 \Leftrightarrow (\varphi(v), \varphi(w)) \in E_2$.   

The following Lemma shows then the utility of the colored graph representation of containment mappings. 

\begin{lemma}
Two containment mappings are equivalent if and only if their colored graph representations are isomorphic.
\end{lemma}
\begin{proof}
Let us start with the only-if direction. Consider two equivalent containment mappings $f_1$, $f_2$ from $\pleft$ to $P$. By definition of equivalent containment mappings, there exist isomorphisms $g: \pleft \rightarrow \pleft$ and $h: P \rightarrow P$ such that $f_2 \circ g = h \circ f_1$. Now take $\varphi = g \cup h$. Then, $\varphi$ is clearly a bijection from $V_{f_1}$ to $V_{f_2}$, and clearly preserves the colors of vertices and edges of $G_{f_1}$. Let $(v,w) \in E_{f_1}$. We show that $\varphi$ is indeed an isomorphism from $G_{f_1}$ to $G_{f_2}$. There are three possibilities:
\begin{enumerate}
	\item $(v,w) \in \text{Edges}(\pleft)$. Note that then also $g(v,w) \in \text{Edges}(\pleft)$. We have:
		\begin{align*}
			(v,w) \in E_{f_1} & \Leftrightarrow (v,w) \in \text{Edges}(\pleft) \\
			& \text{\quad\quad\quad\quad $g$ is an automorphism in $\pleft$} \\
			& \Leftrightarrow g(v,w) \in \text{Edges}(\pleft) \\
			& \text{\quad\quad\quad\quad $\varphi(v,w) = g(v,w)$} \\
			& \Leftrightarrow g(v,w) \in E_{f_2} \\
			& \text{\quad\quad\quad\quad $\varphi(v,w) \in E_{f_2}$} \\
			& \Leftrightarrow \varphi(v,w) \in E_{f_2}
		\end{align*}
	\item $(v,w) \in \text{Edges}(P)$. Note that then also $h(v,w) \in \text{Edges}(P)$. We have:
		\begin{align*}
			(v,w) \in E_{f_1} & \Leftrightarrow (v,w) \in \text{Edges}(P) \\
			& \text{\quad\quad\quad\quad $h$ is an automorphism in $P$} \\
			& \Leftrightarrow g(v,w) \in \text{Edges}(P) \\
			& \text{\quad\quad\quad\quad $\varphi(v,w) = h(v,w)$} \\
			& \Leftrightarrow h(v,w) \in E_{f_2} \\
			& \text{\quad\quad\quad\quad $\varphi(v,w) \in E_{f_2}$} \\
			& \Leftrightarrow \varphi(v,w) \in E_{f_2}
		\end{align*}
	\item $w = f_1(v)$. We have:
		\begin{align*}
			(v,w) \in E_{f_1} & \Leftrightarrow v = f_1(w) \\
			& \Leftrightarrow h(v) = h(f_1(w)) \\
			& \Leftrightarrow h(v) = f_2(g(w)) \\
			& \Leftrightarrow \varphi(v) = f_2(\varphi(w)) \\
			& \Leftrightarrow (\varphi(v),\varphi(w)) \in E_{f_2}
		\end{align*}
\end{enumerate}
So we can conclude that $G_{f_1}$ and $G_{f_2}$ are indeed colored isomorphic.

Let us now look at the if-direction. Let $\varphi$ be the given isomorphism from $G_{f_1}$ to $G_{f_2}$. Now take $g = \varphi|_{\text{Vertices}(\pleft)}$ and $h = \varphi|_{\text{Vertices}(P)}$. To prove that $f_1$ and $f_2$ are equivalent it suffices to show that:
\begin{enumerate}
	\item $g$ is an isomorphism from $\pleft$ to $\pleft$;
	\item $h$ is an isomorphism from $P$ to $P$; and
	\item $f_2 \circ g = h \circ f_1$.
\end{enumerate} 

Items 1 and 2 hold because $\varphi$ preserves the colors. For 3, let $v \in \pleft$. Since $\varphi$ is a graph isomorphism $f_2(\varphi(v))$ equals $\varphi(f_1(v))$. We then have:
\begin{align*}
f_2(g(v)) &= f_2(\varphi(v)) \\
 &= \varphi(f_1(v))\\
 &= h(f_1()v)
\end{align*}
\quad
\end{proof}

So, using graph isomorphism (to be precise, edge and vertex
colored directed graph isomorphism), we can test for equivalence.
Since our patterns are not very large, fast heuristics for graph
isomorphism can be used. We use the program Nauty \cite{nauty,
nautyUserGuide}, which is considered as the fastest heuristic for
graph isomorphism. Nauty is very efficient for small, dense
random graphs \cite{nautyPerformance}. Since our graph
representations are typically small (no more than $20$ vertices)
and dense, this works well in our case. 

Theoretically this situation is not entirely satisfying, as graph isomorphism is not known to be efficiently (polynomial-time) solvable in general.  We can show however that equivalence of our containment mappings is really as hard as the general graph isomorphism problem. This hardness argument is presented in the following Section~\ref{subsubsec_hardness}. As special case of the equivalence problem that is solvable in polynomial time is presented in Section~\ref{subsubsec_polynomial}

\subsubsection{Hardness argument}
\label{subsubsec_hardness}

First recall from graph theory that a graph $B = (V, E)$ is
\emph{bipartite} if $V$ can be split in two disjoint parts, $V =
V^a \cup V^b$ with $V^a \cap V^b = \emptyset$, such that for each
$(v,w) \in E$ then $v \in V^a$ and $w \in V^b$. The vertices in
$V^a$ are called lhs vertices and those in $V_b$ rhs vertices
(left-hand side, right-hand side).

We first reduce the problem of bipartite graph isomorphism to equivalence of our containment mapping.  Let $B_1 = (V_1, E_1)$ and $B_2 = (V_2, E_2)$ be bipartite graphs. We describe an efficient construction that produces from $B_1$ and $B_2$ two association rules $(\pleft,P,f_1)$ and $(\pleft,P,f_2)$ such that $B_1$ and $B_2$ are isomorphic if and only if the association rules are equivalent. This construction reduces the bipartite graph isomorphism problem to equivalence of containment mappings. 

Without loss of generality, we assume that $B_1$ and $B_2$ have precisely the same multiset of outdegrees (for vertices of $V^a_1$ and $V^a_2$), and precisely the same number of vertices in $V^b_1$ and $V^b_2$. Indeed, if these conditions are not satisfied, then $B_1$ and $B_2$ are never isomorphic and our reduction can output some arbitrary $\pleft$, $P$, $f_1$ and $f_2$ as long as $(\pleft,P,f_1)$ and $(\pleft,P,f_2)$ are not equivalent.

The construction in now as follows. By the premisses on $B_1$ and $B_2$, we may assume, without loss of generality, that $V^a_1 = V^a_2$ and $V^b_1 = V^b_2$. This can be accomplished by sorting the lhs vertices in each graph on their outdegrees an then numbering them arbitrarily (the rhs vertices can simply be numbered arbitrarily).
\begin{enumerate}
	\item Construction of $\pleft$. This is a tree with root called $r_{\text{left}}$ and as children of the root, all lhs vertices. Moreover, each lhs vertex $v$ has its own children as follows: if $v$ has outdegree $o$, then $v$ has $o$ children denoted by $[v,1]$, $[v,2]$, ..., $[v,o]$.
	\item Construction of $P$. This is a tree with root called $r_{\text{right}}$, and exactly one child of the root, called $c$. Moreover, $c$ has as children precisely all rhs vertices.
	\item Construction of $f_1$. We define $f(r_{\text{left}}) := r_{\text{right}}$, and define $f_1(v) := c$ for each lhs vertex $v$. Now for each such $v$, and all outgoing edges $(v,w_1)$, $(v,w_2)$,...,$(v,w_o)$ in $B_1$, listed in some arbitrary order, we define $f_1([v,i]) := w_i$, for $i=1,2,...,o$.
 	\item The construction of $f_2$ is analogous to that of
	$f_1$, but now we look at the outgoing edges in $B_2$.
\end{enumerate}

The construction is illustrated in Figure~\ref{fig_bipartite_pAR} for two bipartite graphs $B_1$ and $B_2$. 

\newpsobject{showgrid}{psgrid}{subgriddiv=1,griddots=10,gridlabels=6pt}
\psset{linewidth=0.8pt}

\begin{figure}
\hspace*{\fill}
\subfigure[$B_1$]{\label{fig_bipartite_pAR_a}
\scalebox{1}
{
\begin{pspicture}(-0.2,-0.2)(3.2,3.2)
\rput(0,0){\rnode{x4}{$x_4$}}
\rput(0,1){\rnode{x3}{$x_3$}}
\rput(0,2){\rnode{x2}{$x_2$}}
\rput(0,3){\rnode{x1}{$x_1$}}
\rput(3,0.50){\rnode{y3}{$y_3$}}
\rput(3,1.5){\rnode{y2}{$y_2$}}
\rput(3,2.5){\rnode{y1}{$y_1$}}
\ncarc{x1}{y1}
\ncarc{x1}{y2}
\ncarc{x2}{y1}
\ncarc{x2}{y2}
\ncarc{x4}{y3}
\end{pspicture}
}
}
\hfill
\subfigure[$B_2$]{\label{fig_bipartite_pAR_b}
\scalebox{1}
{
\begin{pspicture}(-0.2,-0.2)(3.2,3.2)
\rput(0,0){\rnode{q4}{$q_4$}}
\rput(0,1){\rnode{q3}{$q_3$}}
\rput(0,2){\rnode{q2}{$q_2$}}
\rput(0,3){\rnode{q1}{$q_1$}}
\rput(3,0.50){\rnode{p3}{$p_3$}}
\rput(3,1.5){\rnode{p2}{$p_2$}}
\rput(3,2.5){\rnode{p1}{$p_1$}}
\ncarc{q1}{p2}
\ncarc{q1}{p3}
\ncarc{q2}{p2}
\ncarc{q2}{p3}
\ncarc{q4}{p1}
\end{pspicture}
}
}
\hspace*{\fill}
\\
\hspace*{\fill}
\subfigure[$\pleft$]{\label{fig_bipartite_pAR_c}
\scalebox{1}
{
\begin{pspicture}(-0.2,-0.2)(5.2,2.2)
\rput(2.5,2){\rnode{rleft}{$r_{\text{left}}$}}
\rput(0.5,1){\rnode{v1}{$v_1$}}
\rput(2.5,1){\rnode{v2}{$v_2$}}
\rput(4,1){\rnode{v3}{$v_3$}}
\rput(5,1){\rnode{v4}{$v_4$}}
\rput(0,0){\rnode{v11}{$[v_1,1]$}}
\rput(1,0){\rnode{v12}{$[v_1,2]$}}
\rput(2,0){\rnode{v21}{$[v_2,1]$}}
\rput(3,0){\rnode{v22}{$[v_2,2]$}}
\rput(4,0){\rnode{v31}{$[v_3,1]$}}
\ncline{rleft}{v1}
\ncline{rleft}{v2}
\ncline{rleft}{v3}
\ncline{rleft}{v4}
\ncline{v1}{v11}
\ncline{v1}{v12}
\ncline{v2}{v21}
\ncline{v2}{v22}
\ncline{v3}{v31}
\end{pspicture}
}
}
\hfill
\subfigure[$P$]{\label{fig_bipartite_pAR_d}
\scalebox{1}
{
\begin{pspicture}(-0.2,-0.2)(2.2,2.2)
\rput(1,2){\rnode{rright}{$r_{\text{right}}$}}
\rput(1,1){\rnode{c}{$c$}}
\rput(0,0){\rnode{w1}{$w_1$}}
\rput(1,0){\rnode{w2}{$w_2$}}
\rput(2,0){\rnode{w3}{$w_3$}}
\ncline{rright}{c}
\ncline{c}{w1}
\ncline{c}{w2}
\ncline{c}{w3}
\end{pspicture}
}
}
\hspace*{\fill}
\\
\hspace*{\fill}
\subfigure[$f_1$]{\label{fig_bipartite_pAR_e}
\scalebox{1}
{
\begin{pspicture}(-0.2,-0.2)(9.2,2.5)
\rput(2.5,2){\rnode{rleft}{$r_{\text{left}}$}}
\rput(0.5,1){\rnode{v1}{$v_1$}}
\rput(2.5,1){\rnode{v2}{$v_2$}}
\rput(4,1){\rnode{v3}{$v_3$}}
\rput(5,1){\rnode{v4}{$v_4$}}
\rput(0,0){\rnode{v11}{$[v_1,1]$}}
\rput(1,0){\rnode{v12}{$[v_1,2]$}}
\rput(2,0){\rnode{v21}{$[v_2,1]$}}
\rput(3,0){\rnode{v22}{$[v_2,2]$}}
\rput(4,0){\rnode{v31}{$[v_3,1]$}}
\ncline{rleft}{v1}
\ncline{rleft}{v2}
\ncline{rleft}{v3}
\ncline{rleft}{v4}
\ncline{v1}{v11}
\ncline{v1}{v12}
\ncline{v2}{v21}
\ncline{v2}{v22}
\ncline{v3}{v31}
\rput(8,2){\rnode{rright}{$r_{\text{right}}$}}
\rput(8,1){\rnode{c}{$c$}}
\rput(7,0){\rnode{w1}{$w_1$}}
\rput(8,0){\rnode{w2}{$w_2$}}
\rput(9,0){\rnode{w3}{$w_3$}}
\ncline{rright}{c}
\ncline{c}{w1}
\ncline{c}{w2}
\ncline{c}{w3}
\ncarc[linestyle=dotted,arcangle=16]{rleft}{rright}
\ncarc[linestyle=dotted,arcangle=16,nodesepA=-3pt]{v11}{w1}
\ncarc[linestyle=dotted,arcangle=16,nodesepA=-3pt]{v12}{w2}
\ncarc[linestyle=dotted,arcangle=16,nodesepA=-3pt]{v21}{w1}
\ncarc[linestyle=dotted,arcangle=16,nodesepA=-3pt]{v22}{w2}
\ncarc[linestyle=dotted,arcangle=16,nodesepA=-3pt]{v31}{w3}
\ncarc[linestyle=dotted,arcangle=16]{v1}{c}
\ncarc[linestyle=dotted,arcangle=16]{v2}{c}
\ncarc[linestyle=dotted,arcangle=16]{v3}{c}
\ncarc[linestyle=dotted,arcangle=16]{v4}{c}
\end{pspicture}
}
}
\hspace*{\fill}
\\
\hspace*{\fill}
\subfigure[$f_2$]{\label{fig_bipartite_pAR_f}
\scalebox{1}
{
\begin{pspicture}(-0.2,-0.2)(9.2,2.5)
\rput(2.5,2){\rnode{rleft}{$r_{\text{left}}$}}
\rput(0.5,1){\rnode{v1}{$v_1$}}
\rput(2.5,1){\rnode{v2}{$v_2$}}
\rput(4,1){\rnode{v3}{$v_3$}}
\rput(5,1){\rnode{v4}{$v_4$}}
\rput(0,0){\rnode{v11}{$[v_1,1]$}}
\rput(1,0){\rnode{v12}{$[v_1,2]$}}
\rput(2,0){\rnode{v21}{$[v_2,1]$}}
\rput(3,0){\rnode{v22}{$[v_2,2]$}}
\rput(4,0){\rnode{v31}{$[v_3,1]$}}
\ncline{rleft}{v1}
\ncline{rleft}{v2}
\ncline{rleft}{v3}
\ncline{rleft}{v4}
\ncline{v1}{v11}
\ncline{v1}{v12}
\ncline{v2}{v21}
\ncline{v2}{v22}
\ncline{v3}{v31}
\rput(8,2){\rnode{rright}{$r_{\text{right}}$}}
\rput(8,1){\rnode{c}{$c$}}
\rput(7,0){\rnode{w1}{$w_1$}}
\rput(8,0){\rnode{w2}{$w_2$}}
\rput(9,0){\rnode{w3}{$w_3$}}
\ncline{rright}{c}
\ncline{c}{w1}
\ncline{c}{w2}
\ncline{c}{w3}
\ncarc[linestyle=dotted,arcangle=16]{rleft}{rright}
\ncarc[linestyle=dotted,arcangle=16,nodesepA=-3pt]{v11}{w2}
\ncarc[linestyle=dotted,arcangle=16,nodesepA=-3pt]{v12}{w3}
\ncarc[linestyle=dotted,arcangle=16,nodesepA=-3pt]{v21}{w2}
\ncarc[linestyle=dotted,arcangle=16,nodesepA=-3pt]{v22}{w3}
\ncarc[linestyle=dotted,arcangle=16,nodesepA=-3pt]{v31}{w1}
\ncarc[linestyle=dotted,arcangle=16]{v1}{c}
\ncarc[linestyle=dotted,arcangle=16]{v2}{c}
\ncarc[linestyle=dotted,arcangle=16]{v3}{c}
\ncarc[linestyle=dotted,arcangle=16]{v4}{c}
\end{pspicture}
}
}
\hspace*{\fill}
\caption{Illustration of the construction of pARs from bipartite graphs.}
\label{fig_bipartite_pAR}
\end{figure}
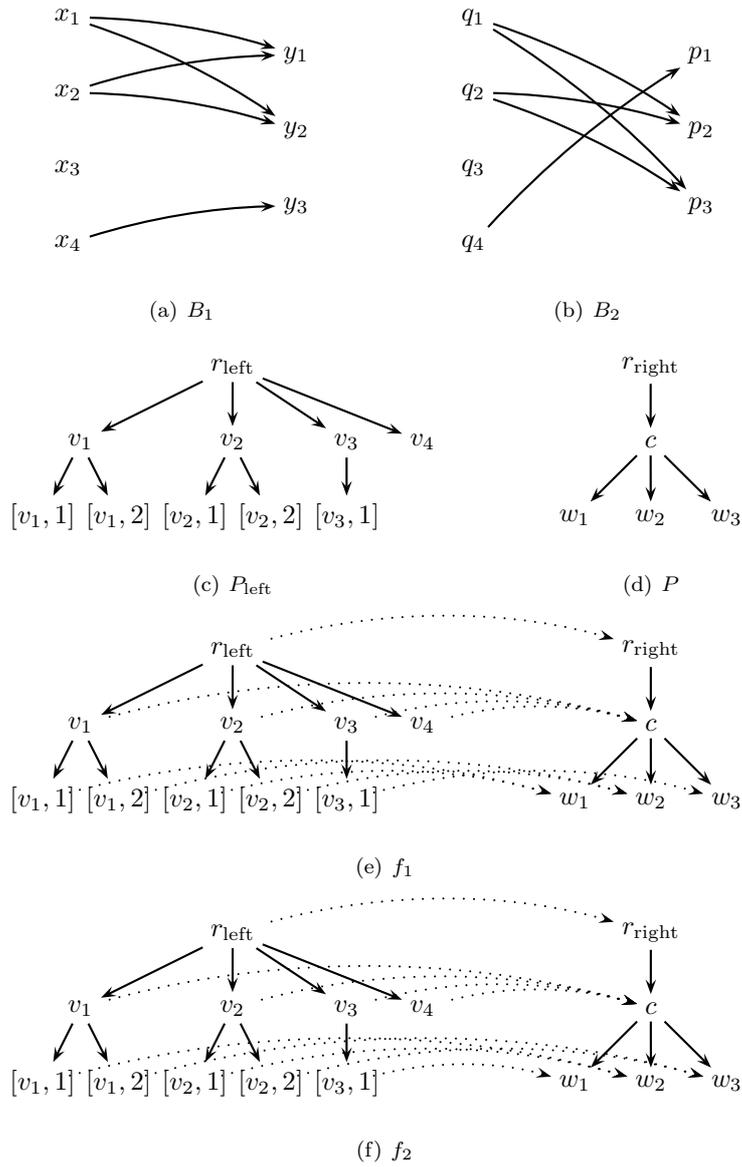

We now show the correctness of our reduction.

\newcommand{\rright}{r_{\text{right}}}
\newcommand{\rleft}{r_{\text{left}}}

\begin{lemma}
$B_1$ and $B_2$ are isomorphic if and only if $(\pleft,P,f_1)$ and $(\pleft,P,f_2)$ are isomorphic.
\end{lemma}
\begin{proof}
For the only-if direction, let $\psi$ be an isomorphism from $B_1$ to $B_2$. We define an isomorphism from $(\pleft,P, f_1)$ to $(\pleft,P, f_2)$ as follows:
\begin{itemize}
	\item $\varphi(\rleft) = \rleft$, $\varphi(\rright) = \rright$ and $\varphi(c) = c$;
	\item $\varphi(v) = \psi(v)$, for any vertex of $B_1$;
	\item for any lhs vertex $v$ of outdegree $o$, and any $i=1,2,...,o$, let $w$ be the rhs vertex such that $f_1([v,i])=w$. Then we define $\varphi([v,i]):=[\psi(v),j]$, where $j$ is such that $f_2([\psi(v),j]) = \psi(w)$.
\end{itemize}
To verify that $\varphi$ is indeed an isomorphism, we only check that $u=f_1([v,i]) \Leftrightarrow \psi(u)=f_2(\psi([v,i]))$. If $u=f_1([v,i])$ then $(v,u)$ is an edge in $B_1$ and thus $(\varphi(v),\varphi(w))=(\psi(v),\psi(w))$ is an edge in $B_2$. Hence there exists a $j$ such that, $\varphi(u)=f_2([\varphi(v),j])$, or equivalent, $\psi(u)=f_2([\psi(v),j])$. By definition of $\varphi$ we have $\varphi([v,i]) = [\psi(v),j]$ and thus $\varphi(u) = f_2(\varphi([v,i]))$ as desired. Conversely, suppose $\varphi(u) = f_2(\varphi([v,i]))$. By definition of $\varphi$, we have $\varphi([v,i])$ equals $[\psi(v),j]$ for some unique $j$, and $f_2([\psi(v),j])$ equals $\psi(f_1([v,i]))$. Hence, $\psi(u)=\varphi(u)=\psi(f_1([v,i]))$, and thus $u=f_1([v_i])$ as desired. 

For the if-direction, let $\varphi$ be an isomorphism from $(\pleft,P,f_1)$ to $(\pleft,P,f_2)$. We define an isomorphism $\psi$ from $B_1$ to $B_2$ as follows. Actually, $\psi$ is simple $\varphi$ restricted to the vertices of $B_1$. Indeed,
\begin{align*}
	(v,w) \in E_1 & \Leftrightarrow \exists i: f_1([v,i]) = w \\
	& \Leftrightarrow \exists i: f_2(\varphi([v,i])) = \varphi(w) \\
	& \Leftrightarrow \exists j: f_2(\varphi(v),j) = \varphi(w) \\
	& \Leftrightarrow (\varphi(v), \varphi(w)) \in E_2
\end{align*}
\quad
\end{proof}

We can already conclude from this reduction that equivalence of pARs is really as hard as isomorphism of bipartite directed graphs. The latter problem, however, is well-known to be as hard as isomorphism of general directed graphs. Indeed, any directed graph $G=(V,E)$ can be transformed into the bipartite directed graph $B(G):=(V \cup E, \{(v,(v,w)) \mid (v,w) \in E\} \cup \{((v,w),w) \mid (v,w) \in E\})$, and it is easily verified that $G_1$ and $G_2$ are isomorphic if and only if $B(G_1)$ and $B(G_2)$ are isomorphic.

So, we can now conclude that equivalence of our pARs is really as hard as the general graph isomorphism problem. But as we show next, we can still capture an important special case in polynomial time, so that the
general graph isomorphism heuristics only have to be applied on instances not captured by the special case.

\subsubsection{Polynomial case} 
\label{subsubsec_polynomial}

The special efficient case is to check whether
$(\pleft,P,f_1)$ and $(\pleft,P,f_2)$ are already isomorphic with $g$ the
identity, i.e., whether the structures $(P,f_1)$ and $(P,f_2)$ are already isomorphic.
So, we look for an automorphism $h$ of $P$ such that $f_2 = h \circ f_1$.
This can be solved efficiently by a reduction to node-labeled tree isomorphism.
As explained in Section~\ref{innerLoop}, if we know the tree $T$ underlying $P$, then $P$ is
characterized by the pair $(\Pi,\Sigma)$, and thus $(P,f)$ is characterized by
$(\Pi,\Sigma,f)$.  We can view this triple as a labeling of $T$, as follows.
We label every node $y$ of $P$ with a triple $(b_\Pi,b_\Sigma,f^{-1}(y))$,
where $b_\Pi$ is a bit that is 1 iff $y \in \Pi$; $b_\Sigma$ is a bit that is
defined likewise; and $f^{-1}(y)$ is the set of nodes of $\pleft$ that are
mapped by $f$ to $y$.  Then $(P,f_1)$ and $(P,f_2)$ are isomorphic if and only
if the corresponding node-labeled trees are isomorphic, and the latter can be checked
in linear time using canonical ordering
\cite{ahu,muntz_trees}.

\subsubsection{The Algorithm} 

We are now in a position to describe how our general algorithm
must be modified to avoid equivalent
association rules.  There is only extra checking to be done in loop~3 (recall Sections~\ref{sec_rules_overall} and \ref{subsec_contmap}).
For each new containment mapping $f$ from $\pleft$ to $P$,
we canonize the corresponding node-labeled tree and we check if the
canonical form is identical to an earlier generated canonical form;
 if so, $f$ is dismissed.
We can keep track of the canonical forms seen so far efficiently using a trie
data structure.
If the canonical form was not yet seen, we can either let $f$ through to
loop~4, if the
presence of duplicates in the output is tolerable for the application at hand,
or we can perform the colored graph isomorphism check of Section~\ref{subsubsec_testing_equivalence} with
the containment mappings previously seen, to be absolutely sure we will not
generate a duplicate.

\section{Certhia: Pattern and Association Browsing}
\label{sec_certhia}

In this Section we introduce an interactive tool, called \emph{Certhia}, for browsing the frequent tree patterns, and generating association rules.

As already noted in Section~\ref{sec_result_management}, the result of our tree query mining algorithm in Section~\ref{sec_algo_patterns} is a structured database, called a Pattern Database, containing all frequency tables for each tree $T$ that was investigated. This pattern database is an ideal platform for an interactive tool for browsing the frequent queries. However, this pattern database is also an ideal platform for generating association rules as explained in Section~\ref{sec_rules_overall}, since the first two loops of association rule algorithm are exactly our tree query mining algorithm. 

In a typical scenario for Certhia, the user draws a tree shape, marks some nodes as existential, marks some others as parameters, instantiates some parameters by constants, but possibly also leaves some parameters open. The browser then returns, by consulting the appropriate frequency table in the database, all instantiations of the free parameters that make the pattern frequent, together with the frequency.  The user can then select one of these instantiations, set a minconf value, and ask the browser to return all rhs's that form a confident association with the selected pure tree query as lhs. In another scenario the user lets the browser suggest some frequent tree patterns to choose from as an lhs. 

\begin{figure}
\begin{center}
\resizebox{0.8\columnwidth}{!}{
\includegraphics{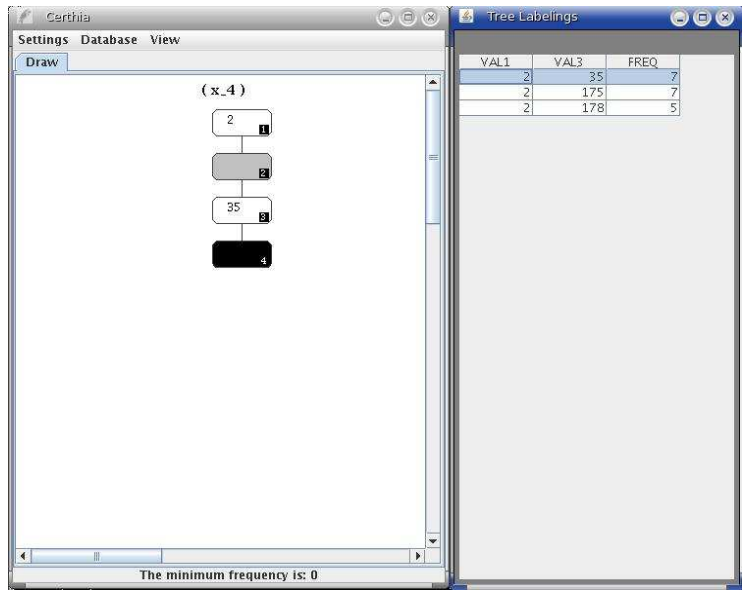}}
\end{center}
\caption{Screenshot of Certhia: the user draws a pattern and asks for the instantiations.}
\label{fig_certhia_a}
\end{figure}

\begin{figure}
\begin{center}
\resizebox{0.8\columnwidth}{!}{
\includegraphics{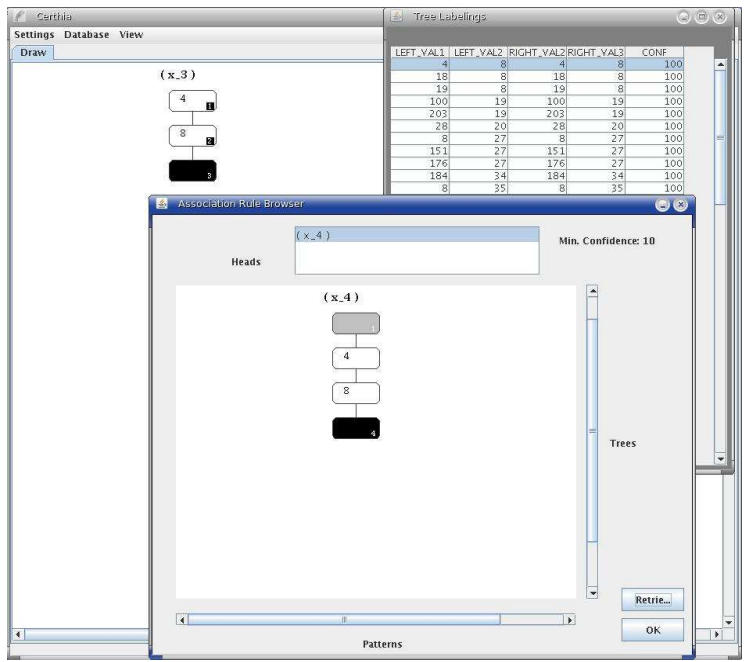}}
\end{center}
\caption{Screenshot of Certhia: the user asks for all association rules for this lhs.}
\label{fig_certhia_b}
\end{figure}

\begin{figure}
\begin{center}
\resizebox{0.8\columnwidth}{!}{
\includegraphics{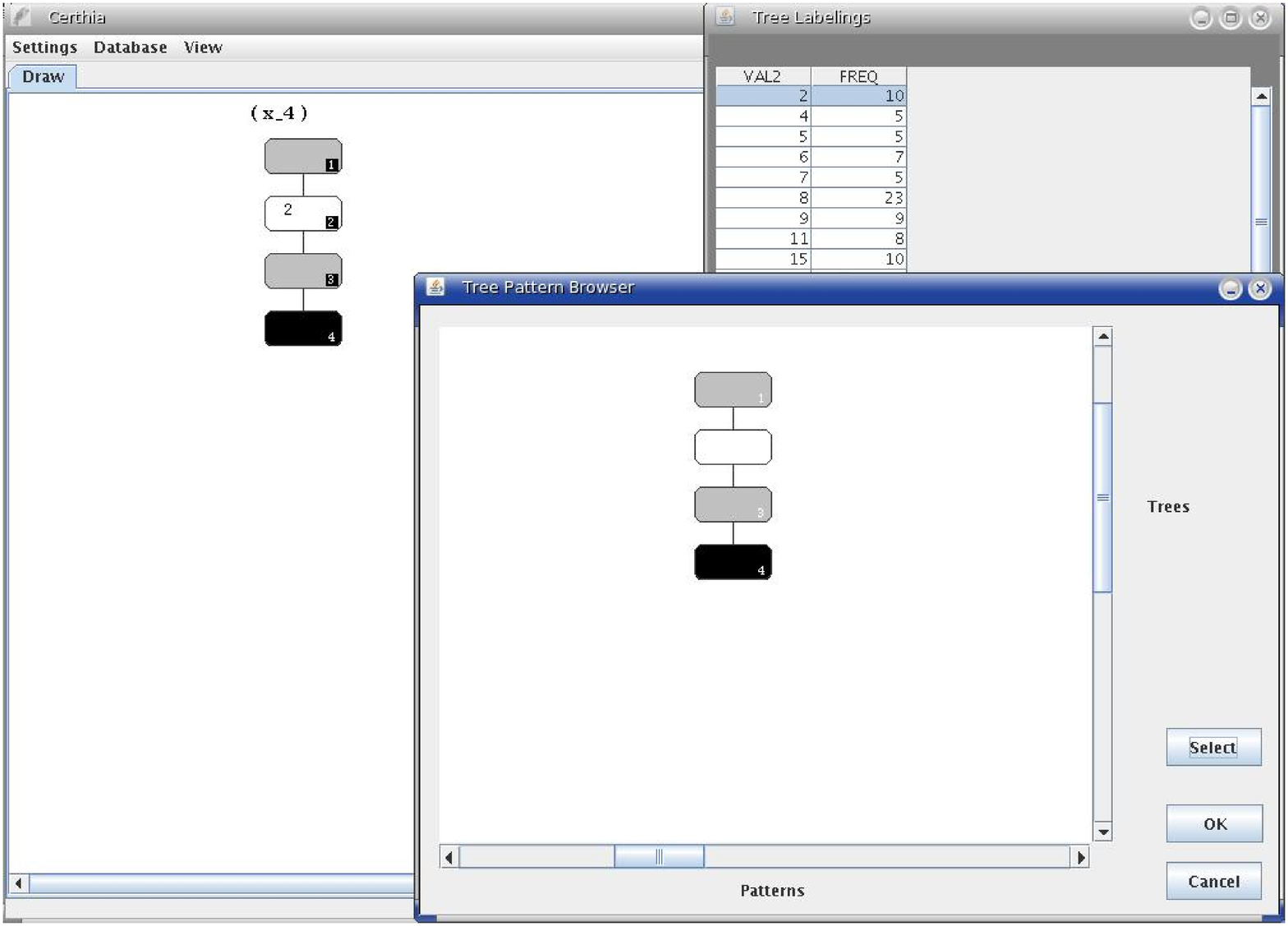}}
\end{center}
\caption{Screenshot of Certhia: the browser suggests frequent tree patterns.}
\label{fig_certhia_c}
\end{figure}

Some screenshots of Certhia are given in Figure~\ref{fig_certhia_a}, Figure~\ref{fig_certhia_b} and Figure~\ref{fig_certhia_c}.
\begin{itemize}
	\item In Figure~\ref{fig_certhia_a}, the user draws a
	tree, marks some nodes nodes as existential, some others
	as parameters, instantiates some parameters with
	constants, and asks the browser to return all possible
	instantiations of the remaining parameters and the
	corresponding frequencies. 
	\item In Figure~\ref{fig_certhia_b}, the user asks the
	browser to return all association rules for a fixed lhs.
	The user selects a rhs in the dialog box and asks the
	browser to return the instantiations and the
	corresponding frequencies. 
	\item In Figure~\ref{fig_certhia_c}, the browser suggests some frequent tree patterns where the user can choose from.
\end{itemize}

\paragraph*{Efficiency}

The preprocessing step, i.e., the building up of the pattern database with frequent tree patterns, is of course a hugely intensively task. First because the large datagraph must be accessed intensively, and secondly because the number of frequent patterns is huge. In Section~\ref{sec_exp_random} we show that this preprocessing step can be implemented with satisfactory performance. Also, in scientific discovery applications it is no problem, indeed typical, if a preprocessing step takes a few hours, as long as after that the interactive exploration of the found results can happen very fast. And indeed we found that the actual generation of association rules is very fast. This is also shown in Section~\ref{sec_exp_random}.

\section{Experimental Results}
\label{sec_exp}

In this section, we report on some experiments performed using our prototype implementation applied to both real-life and synthetic datasets to show that our approach is indeed workable. 

\subsection{Real-life datasets}
\label{sec_exp_real}

We have worked with a food web, a protein interactions graph, and a citation
graph.  For each dataset we built up a pattern database using the following
parameters:  

\begin{center}
\begin{tabular}{r|rrrr}
& \#nodes & \#edges & $k$ & size  \\
\hline
food web & 154 & 370 & 25 & 6
\\
proteins & 2114 & 4480 & 10 & 5
\\
citations & 2500 & 350000 & 5 & 4
\end{tabular}
\end{center} As we set rather generous limits on the maximum size of trees, or on the
minimum frequency threshold, each run took several hours.

The \textbf{food web} \cite{silwood} comprises 154 species that are all directly or indirectly dependent on the Scotch Broom (a kind of shrub). One of the patterns that was mined with frequency 176 is the following:

\begin{center}
\pstree[treemode=R,levelsep=3em]{\TR{$\exists$}}
{
  \pstree{\TR{$x_1$}}
  {
    \pstree{\TR{$\exists$}}
    {
      \TR{20}
      \TR{$x_2$}
    }
  }
  \TR{20}
}
\end{center}
This is really a rather arbitrary example, just to give an idea of the kind of complex
patterns that can be mined.  Note also that, thanks to the constant 20 appearing twice, this is really a non-tree shaped pattern: we could equally well draw both arrows to a single node labeled 20.

While we were thus browsing through the results, we quickly noticed that the constant 20 actually occurs quite predominantly, in many different frequent patterns.  This constant denotes the species \emph{Orthotylus adenocarpi}, an omnivorous plant bug.  To confirm our hypothesis that this species plays a central role in the food web, we asked for all association rules with the following left-hand side:

\begin{center}
\begin{tabular}{c}
$(x_1,x_2)$\\
\hline
\pstree{\TR{$x_1$}}
{
\pstree{\TR{$\exists$}}
{
\pstree{\TR{$\exists$}}
{
\pstree{\TR{$\exists$}}
{
\TR{$x_2$}
}}}}
\end{tabular}
\quad $\Rightarrow$ \quad
\begin{tabular}{c}
$(x_1,x_2)$\\
\hline
\pstree{\TR{$x_1$}}
{
\pstree{\TR{$\exists$}}
{
\pstree{\TR{20}}
{
\pstree{\TR{$\exists$}}
{
\TR{$x_2$}
}}}}
\end{tabular}
\end{center}
Indeed, the rule shown above turned up with 89\% confidence! For 89\% of all pairs of species that are linked by a path of length four, \emph{Orthotylus adenocarpi} is involved in between.

Two other rules we discovered are:

\begin{center}
\begin{tabular}{c}
	$(x_1,x_2,x_3, x_4, x_5)$\\
	\hline
	\pstree{\TR{$x_1$}}
	{
	  \pstree{\TR{$x_2$}}
	  {
	  	\pstree{\TR{$x_3$}}
		{
			\pstree{\TR{$x_4$}}
			{
				\TR{$x_5$}
			}
		}
	  }
	}
\end{tabular}
\begin{tabular}{c}
$45 \%$ \\
$\Rightarrow$ 
\end{tabular}
\begin{tabular}{c}
	$(0,x_2,x_3, x_4, x_5)$\\
	\hline
	\pstree{\TR{$0$}}
	{
	  \pstree{\TR{$x_2$}}
	  {
	  	\pstree{\TR{$x_3$}}
		{
			\pstree{\TR{$x_4$}}
			{
				\TR{$x_5$}
			}
		}
	  }
	}
\end{tabular}
\end{center}

\begin{center}
\begin{tabular}{c}
	$(x_1,x_2,x_3, x_4, x_5)$\\
	\hline
	\pstree{\TR{$x_1$}}
	{
	  \pstree{\TR{$x_2$}}
	  {
	  	\pstree{\TR{$x_3$}}
		{
			\pstree{\TR{$x_4$}}
			{
				\TR{$x_5$}
			}
		}
	  }
	}
\end{tabular}
\begin{tabular}{c}
$55 \%$ \\
$\Rightarrow$ 
\end{tabular}
\begin{tabular}{c}
	$(x_1,x_2,x_3, x_4, x_5)$\\
	\hline
	\pstree{\TR{$0$}}
	{
		\pstree{\TR{$x_1$}}
		{
		  \pstree{\TR{$x_2$}}
		  {
		  	\pstree{\TR{$x_3$}}
			{
				\pstree{\TR{$x_4$}}
				{
					\TR{$x_5$}
				}
			}
		  }
		}
	}
\end{tabular}
\end{center}
Since $45\%+55\%=100\%$, these rules together say that each path of length $5$ either starts in $0$, or one beneath $0$.  This tells us that the depth of the food web equals 6. Constant $0$ turns out to denote the Scotch Broom itself, which is the root of the food web.

Another rule we mined, just to give a rather arbitrary example of the kind of rules we find with our algorithm, is the following:

\begin{center}
\begin{tabular}{c}
	$(x_1,x_2,x_3, x_4, x_5)$\\
	\hline
	\pstree{\TR{$x_1$}}
	{
	  \pstree{\TR{$x_2$}}
	  {
	  	\TR{$x_3$}
	  }
	  \pstree{\TR{$x_4$}}
	  {
	  	\TR{$x_5$}
	  }
	}
\end{tabular}
\begin{tabular}{c}
$11 \%$ \\
$\Rightarrow$ 
\end{tabular}
\begin{tabular}{c}
	$(x_1,x_2,x_4, x_2, x_5)$\\
	\hline
	\pstree{\TR{$x_1$}}
	{
	  \pstree{\TR{$x_2$}}
	  {
	  	\TR{$101$}
		\TR{$x_4$}
		  \TR{$x_5$}
	  }
	}
\end{tabular}
\end{center}

The \textbf{protein interaction graph} \cite{proteinetwork} comprises
molecular interactions (symmetric) among 1870 proteins occurring in the yeast
\emph{Saccharomyces cerevisiae}.  In such interaction networks, typically a
small number of highly connected nodes occur.  Indeed, we discovered the
following association rule with 10\% confidence, indicating that protein \#224
is highly connected:

\begin{center}
\begin{tabular}{c}
\pstree{\TR{$x_1$}}
{
\pstree{\TR{$\exists$}}
{
\TR{$x_2$}
}}
\end{tabular}
\quad $\Rightarrow$ \quad
\begin{tabular}{c}
\pstree{\TR{$x_1$}}
{
\pstree{\TR{224}}
{
\TR{$x_2$}
}}
\end{tabular}
\end{center}

We also found the following rule:

\begin{center}
\begin{tabular}{c}
	$(x_1,x_2)$\\
	\hline
	\pstree{\TR{$x_1$}}
	{
	  \pstree{\TR{$x_2$}}
	  {
	  	\TR{$746$}
	  }
	}
\end{tabular}
\begin{tabular}{c}
$90 \%$ \\
$\Rightarrow$ 
\end{tabular}
\begin{tabular}{c}
	$(x_1,x_2)$\\
	\hline
	\pstree{\TR{$x_1$}}
	{
	  \pstree{\TR{$x_2$}}
	  {
	  	\TR{$746$}
		\TR{$376$}
	  }
	}
\end{tabular}
\end{center}
This rule expresses that almost all interactions that link to protein~746 also
link to protein~376, which unveils a close relationship between these two
proteins.

The \textbf{citation graph} comes from the KDD cup 2003, and contains around 2500 papers about high-energy physics taken from arXiv.org, with around $350\,000$ cross-references.  One of the discovered patterns is the following, with frequency $1655$, showing two papers that are frequently cited together (by 6\% of all papers).

\begin{center}
\pstree{\TR{$x_1$}}{\TR{9711200}\TR{9802150}}
\end{center}

One of the discovered rules is the following:

\begin{center}
\begin{tabular}{c}
	$(x_1,x_2)$\\
	\hline
	\pstree{\TR{$x_1$}}
	{
	  \pstree{\TR{$\exists$}}
	  {
	  	\TR{$\exists$}
	  }
	  \TR{$x_2$}
	}
\end{tabular}
\begin{tabular}{c}
$15 \%$ \\
$\Rightarrow$ 
\end{tabular}
\begin{tabular}{c}
	$(x_1,x_2)$\\
	\hline
	\pstree{\TR{$x_1$}}
	{
	  \pstree{\TR{$x_2$}}
	  {
	  	\TR{$9503124$}
	  }
	}
\end{tabular}
\end{center}
This rule shows that paper~9503124 is an important paper.  In 15\% of all
``non-trivial'' citations (meaning that the citing paper cites at least one
paper that also cites a paper), the cited paper cites 9503124.

\subsection{Performance}
\label{sec_exp_random}

While our prototype implementations are not tuned for
performance, we still conducted some preliminary performance measurements,
with encouraging results.  The experiments were performed on a Pentium IV (2.8GHz) architecture with 1GB of internal memory, running under Linux 2.6. The program was written in C++ with embedded SQL, with DB2 UDB v8.2 as the relational database system.

We have used two types of synthetic datasets.

\paragraph*{Random Web graphs}  Naturally occurring graphs (as found in
biology, sociology, or the WWW) have a number of typical characteristics, such
as sparseness and a skewed degree distribution \cite{newman_networks}.
Various random graph models have been proposed in this respect, of which we
have used the ``copy model'' for Web graphs
\cite{copymodel_focs,henzinger_wholinkswhom}.  We use degree
5 and probability $\alpha=10\%$ to link to a random node (thus 90\% to copy a
  link).

\begin{figure*}
\centering
\resizebox{\columnwidth}{!}{
\includegraphics{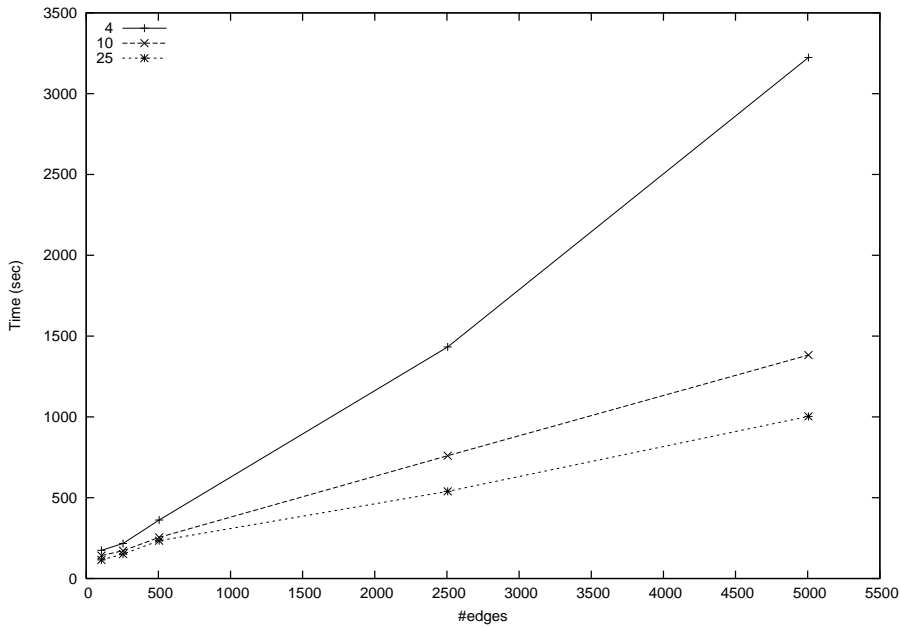}
}
\caption{Performance on Web graphs.}
\label{webgraphs}
\end{figure*}

On these graphs, we have measured the total running time of the tree query mining algorithm as a function of the size (number of edges) of the graph, where we mine up to tree size 5, with varying minimum frequency thresholds of 4, 10, and 25.  The results, depicted in Figure~\ref{webgraphs}, show that the performance of these runs
is quite adequate.

\paragraph*{Uniform random graphs} We have also experimented with the
well-known Erd\"os--R\'enyi random graphs, where one specifies a number $n$
of nodes and gives each of the possible $n^2$ edges a uniform probability (we
used 10\%) of actually belonging to the graph.  In contrast to random Web
graphs, these graphs are quite dense and uniform, and they serve well as a
worst-case scenario to measure the performance of the tree query mining algorithm as a function of the number of discovered patterns, which will be huge.

\begin{figure*}
\centering
\resizebox{\columnwidth}{!}{
\includegraphics{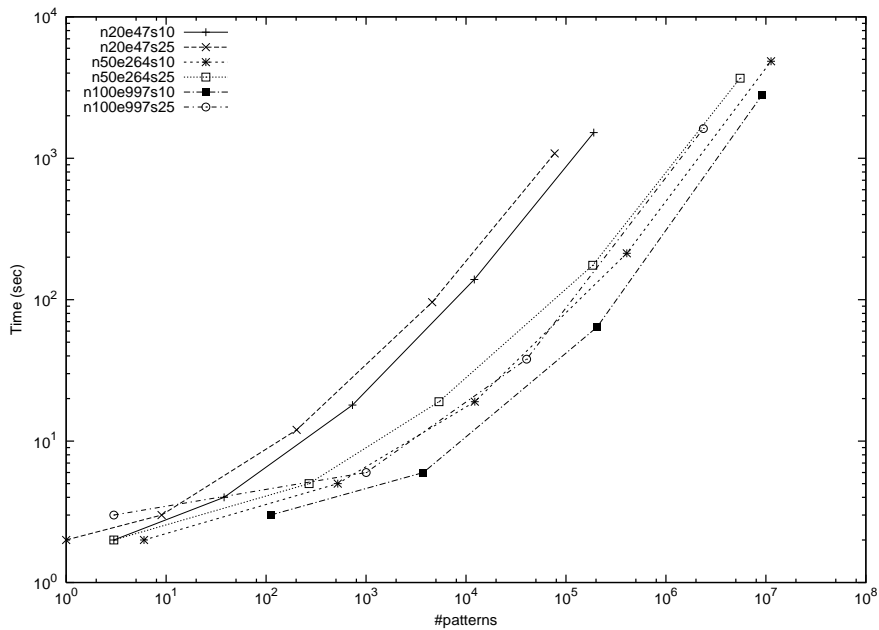}
}
\caption{Performance in terms of number of discovered patterns.}
\label{ER}
\end{figure*}

We have run on graphs with 47, 264, and 997 edges, with minimum frequency
thresholds of 10 and 25.  The results, depicted in Figure~\ref{ER}, show, 
first, that huge numbers of patterns are mined within a reasonable time, and
second, that the overhead per discovered pattern is constant (all six lines
have the same slope).

\begin{figure}
\centering
\resizebox{\columnwidth}{!}{
\includegraphics{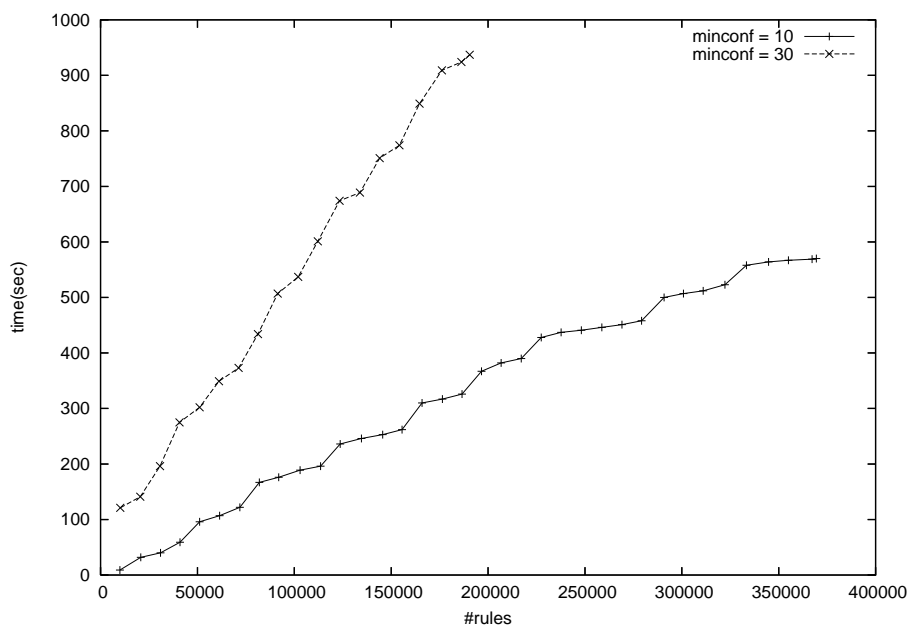}
}
\caption{Performance in terms of number of discovered rules.}
\label{fig_performance}
\end{figure}

On these uniform random graphs we also conducted some experiments to check the performance of the association rule mining algorithm. We found the actual generation of association rules (i.e., loops 3 and 4, assuming that a pattern database is already build up) to be very fast.  For instance, Figure~\ref{fig_performance} shows the
performance of generating association rules for two different (absolute)
values of minconf, against a frequency table database built up for a random
graph with 33 nodes and 113 edges, an absolute minsup of 25, and all trees
up to size 7.  We see that associations are generated with constant
overhead, i.e., in linear-output time.  The coefficient is larger for the
larger minconf, because in this experiment we have counted instantiated
rhs's, and per rhs query less instantiations satisfy the confidence
threshold for larger such thresholds.  Had we simply counted rhs's
regardless of the number of confident instantiations, the two lines would
have had the same slope.

\paragraph*{Performance issues} One major performance issue that we have not addressed in the present study is that some of the SQL queries that are performed due to pattern generation take a very long time (in order of hours) to answer by the database system. This happens in those cases where the data graph is large (5000 edges or more) with many cycles, and the candidate patterns are large (6 nodes are more). Certainly, some SQL queries can be hand-optimized (or replaced by a combination of simpler queries), to alleviate these performance problems, but we leave this issue to future research.

\section*{Acknowledgment}

We thank Bart Goethals for his contributions in the initial conception of the ideas presented in this work. We also thank Jan Hidders and Dries Van Dyck for their help with our results on graph isomorphism.

\bibliographystyle{plain}
\bibliography{datamining}

\newpage

\section*{Appendix}
\label{sec_appendix}

\begin{longtable}{|c|l|}
\hline \multicolumn{1}{|c|}{Notation} & \multicolumn{1}{|l|}{Interpretation} \\ \hline

$U$ & set of data constants \\ \hline
$T$ & ordered rooted tree \\ \hline 
$G$ & data graph \\ \hline
$P$ & parameterized tree pattern \\ \hline
$\Pi$ & set of existential nodes \\ \hline
$\Delta$ & set of distinguished nodes \\ \hline
$\Sigma$ & set of parameters \\ \hline
$\exists$ & existential node \\ \hline
$\sigma$ & parameter \\ \hline
$x$ & distinguished node\\  \hline
$\alpha$ & parameter assignment \\ \hline
$P^{\alpha}$, $(P,\alpha)$ & instantiated tree pattern\\ \hline
$P^{\alpha}(G)$ & $\{\mu|_\Delta:\ \mu \text{ is a matching of } P^{\alpha} \text{ in } G\}$ \\ \hline
$\mathit{\text{minsup}}$ & the frequency threshold\\ \hline
$Q=(H,P)$ & parameterized tree query with $H$ the head and $P$ the body \\ \hline
$Q^{\alpha}$, $(Q, \alpha)$ & instantiated tree query \\ \hline
$Q^{\alpha}(G)$ & answer set of the instantiated tree query $Q^{\alpha}$ in $G$ \\ \hline
$\rho$ & parameter correspondence \\ \hline
$Q_2 \subseteq_{\rho} Q_1$ & $Q_2$ is $\rho$-contained in $Q_1$ \\ \hline
$\text{freeze}_{\beta}(P)$ & the freezing of a tree pattern $P$\\ \hline
pAR & parameterized association rule \\ \hline
iAR & instantiated association rule \\ \hline
$Q_1 \Rightarrow_{\rho} Q_2$ & pAR from $Q_1$ to $Q_2$ \\ \hline
$(Q_1 \Rightarrow_{\rho} Q_2, \alpha)$ & iAR from $Q_1$ to $Q_2$ \\ \hline
$\mathit{\text{minconf}}$ & the confidence threshold \\ \hline
$\Freq(P^{\alpha})$ & the frequency of $P^{\alpha}$ in $G$ \\ \hline
$(\Pi, \Sigma)$ & a parameterized tree pattern $P$ based on a fixed tree $T$ \\ \hline
$(\Pi, \Sigma, \alpha)$ &  an instantiated tree pattern $P$ based on a fixed tree $T$ \\ \hline
$\cantab \Pi \Sigma $ &  $\{\alpha \mid \text{$\Pinst{}$ is a candidate instantiated tree pattern}\}$ \\ \hline
$\freqtab \Pi \Sigma $ & $\{\alpha \mid \text{$\Pinst{}$ is a frequent instantiated tree pattern}\}$ \\ \hline
$\delta$ & answer set correspondence \\ \hline
$P_1 \equiv_{\rho}^{\delta} P_2$ & $P_1$ is $(\delta, \rho)$-equivalent with $P_2$ \\ \hline
$P_2^{\alpha_{2}}(G) \circ \delta$ & $\{f \circ \delta :\ f \in \Pinst{2}(G)\}$ \\ \hline
$P_1 \cong P_2$ & $P_1$ and $P_2$ are isomorphic\\ \hline
\end{longtable}

\end{document}